\pdfoutput=1
\newcommand*{\ATLASLATEXPATH}{}
\documentclass[atlasdraft=false, texlive=2016, UKenglish, cernpreprint, PAPER]{\ATLASLATEXPATH atlasdoc}
 
\usepackage[full,subcaption]{\ATLASLATEXPATH atlaspackage}
\usepackage[backref=false]{\ATLASLATEXPATH atlasbiblatex}
 
\usepackage[jetetmiss]{\ATLASLATEXPATH atlasphysics}
 
\graphicspath{{figures/}{logos/}}
 
\usepackage{ANA-SUSY-2018-16-PAPER-defs}
\usepackage{tablefootnote}
 
\usepackage{multirow}
\usepackage{adjustbox}

\AtlasTitle{Searches for electroweak production of supersymmetric particles with compressed mass spectra in $\sqrt{s}=13$ TeV $pp$ collisions with the ATLAS detector }
 
\AtlasAbstract{
This paper presents results of searches for electroweak production of supersymmetric particles in models with compressed mass spectra.
The searches use 139~$\mbox{fb\(^{-1}\)}$ of $\sqrt{s}=13~\ensuremath{\text{Te\kern -0.1em V}}$ proton--proton collision data collected by the ATLAS experiment at the Large Hadron Collider.
Events with missing transverse momentum and two same-flavor, oppositely charged, low transverse momentum leptons are selected, and are further categorized by the presence of hadronic activity from initial-state radiation or a topology compatible with vector-boson fusion processes.
The data are found to be consistent with predictions from the Standard Model. The results are interpreted using simplified models of $R$-parity-conserving supersymmetry in which the lightest supersymmetric partner is a neutralino with
a mass similar to the lightest chargino, the second-to-lightest neutralino or the slepton.
Lower limits on the masses of charginos in different simplified models range from 193~${\text{Ge\kern -0.1em V}}$ to 240~${\text{Ge\kern -0.1em V}}$ for moderate mass splittings, and extend down to mass splittings of 1.5~${\text{Ge\kern -0.1em V}}$ to 2.4~${\text{Ge\kern -0.1em V}}$ at the LEP chargino bounds (92.4~${\text{Ge\kern -0.1em V}}$).  Similar lower limits on degenerate light-flavor sleptons extend up to masses of 251~${\text{Ge\kern -0.1em V}}$ and down to mass splittings of 550~${\text{Me\kern -0.1em V}}$.  Constraints on vector-boson fusion production of electroweak SUSY states are also presented.
}
 
\AtlasJournal{\PRD}
\AtlasJournalRef{Phys. Rev. D 101 (2020) 052005}
\AtlasDOI{10.1103/PhysRevD.101.052005}
 
\author{The ATLAS Collaboration}
 
\AtlasRefCode{SUSY-2018-16}
 
\PreprintIdNumber{CERN-EP-2019-242}

\AtlasCoverSupportingNote{Internal Note}{https://cds.cern.ch/record/2308395}
 
\AtlasCoverCommentsDeadline{31 October 2019}
 
\AtlasCoverAnalysisTeam{Hass Abouzeid (editor), Shun Akatsuka, Andrew Aukerman, Moritz Backes, Alan Barr, Ben Carlson, Shion Chen, Julia Gonski, Mike Hance (editor), Michael Holzbock, Tae Min Hong, Joe Kroll, Tommaso Lari, Elliot Lipeles, Jesse Liu, Alexander Mann, Joana Machado Migu\'{e}ns (editor), Masahiro Morii, Andreas Redelbach, Joey Reichert, Elodie Resseguie, Lorenzo Rossini, Jeff Shahinian, Toshi Sumida, Stefano Zambito (editor)}

\AtlasCoverEdBoardMember{Anna Lipniacka~(chair)}
\AtlasCoverEdBoardMember{Alaettin Serhan Mete}
\AtlasCoverEdBoardMember{Javier Montejo Berlingen}
 
\AtlasCoverEgroupEditors{atlas-susy-2018-16-editors@cern.ch}
 
\AtlasCoverEgroupEdBoard{atlas-susy-2018-16-editorial-board@cern.ch}

\hypersetup{pdftitle={ATLAS document},pdfauthor={The ATLAS Collaboration}}
 
\begin{document}
 
\maketitle

\section{Introduction}
\label{sec:intro}
Extensions of the Standard Model (SM) that include new states with nearly degenerate masses can help to resolve
open issues in particle physics while evading constraints from experiments at high-energy colliders.
The mass spectra of such new states are referred to in this paper as ``compressed''.
Supersymmetry (SUSY)~\cite{ourGolfand:1971iw,Volkov:1973ix,Wess:1974tw,Wess:1974jb,Ferrara:1974pu,Salam:1974ig} predicts new particles that have identical quantum numbers to their SM partners with the exception of spin, with SM fermions having bosonic partners and SM bosons having fermionic partners. The neutralinos $\widetilde{\chi}^{0}_{1,2,3,4}$ and charginos $\widetilde{\chi}^\pm_{1,2}$ are collectively
referred to as electroweakinos, where the subscripts indicate increasing electroweakino mass. If the \chioz{} is stable, e.g.~as the lightest SUSY partner (LSP) in $R$-parity-conserving SUSY models~\cite{Farrar:1978xj},
then it is a viable dark-matter candidate~\cite{Goldberg:1983nd,Ellis:1983ew}.
In the compressed SUSY models considered in this paper, the \chioz{} is close in mass to a heavier SUSY partner such as a
chargino (\chiopm), second-lightest neutralino (\chitz), or slepton (\slep, the SM lepton partner).
 
This paper presents searches for physics beyond the SM in signatures sensitive to models with compressed mass spectra.
Simplified SUSY models~\cite{Alwall:2008ve,Alwall:2008ag,Alves:2011wf} are used to optimize the searches and interpret the results.
The searches use $13~\TeV$
$pp$ collision data corresponding to \FullRunTwoLumi{} of integrated luminosity,
collected by the ATLAS experiment~\cite{PERF-2007-01} from 2015 to 2018 at the CERN Large Hadron Collider (LHC).
 
All searches assume pair production of SUSY particles via electroweak interactions, with subsequent decays into the \chioz{}
and SM particles.
The electroweakino mass eigenstates are a mixture of wino, bino, and higgsino fields,\footnote{In the minimal supersymmetric extension of the SM,
the Higgs sector is extended to contain two Higgs doublets and $\tan(\beta)$ is the ratio of the vacuum expectation values
of the two Higgs doublets.} which form the SUSY partners of the SM $W$, $\gamma/Z$, and Higgs fields, respectively.
In the minimal supersymmetric extension of the SM (MSSM)~\cite{Fayet:1976et,Fayet:1977yc} the masses of the bino, wino,
and higgsino states are parameterized in terms of $M_1$, $M_2$, and $\mu$, respectively.
For large values of $\tan(\beta)$, these three parameters drive the phenomenology of the electroweakinos.
 
Four SUSY scenarios are considered in the interpretation of the searches.
In the first scenario, the lightest SUSY partners are assumed to be a triplet of higgsino-like states ($\chioz,\chiopm,\chitz$),
in which the mass-splitting between the states is partially determined by the magnitude of $M_1$ or $M_2$ relative to $|\mu|$.
Such a scenario, referred to here as higgsino models, is motivated by naturalness arguments~\cite{Barbieri:1987fn,deCarlos:1993yy},
which suggest that $|\mu|$ should be near the weak scale~\cite{Barbieri:2009ev,Baer:2011ec,Papucci:2011wy,Baer:2012up} while $M_1$
and/or $M_2$ can be larger.
 
\begin{figure}[tbp]
\centering
\begin{subfigure}[b]{0.32\columnwidth}
\centering
\includegraphics[width=\columnwidth]{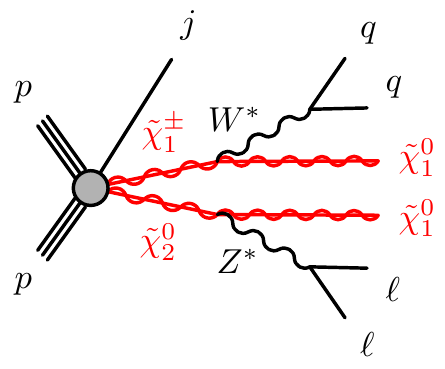}
\caption{}
\label{fig:intro:feynman:ewkino}
\end{subfigure}
\begin{subfigure}[b]{0.32\columnwidth}
\centering
\includegraphics[width=\columnwidth]{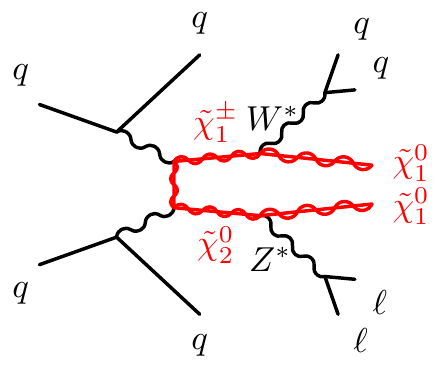}
\caption{}
\label{fig:intro:feynman:vbf}
\end{subfigure}
\begin{subfigure}[b]{0.32\columnwidth}
\centering
\includegraphics[width=\columnwidth]{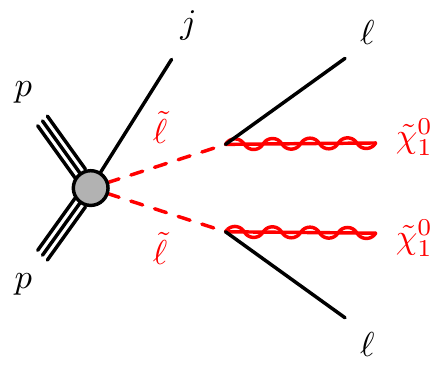}
\caption{}
\label{fig:intro:feynman:slepton}
\end{subfigure}
\caption{\label{fig:intro:feynman} Diagrams representing the two-lepton final state of (a) production of electroweakinos $\chitz\chipm$ with initial-state radiation ($j$), (b) VBF production of electroweakinos $\chitz\chipm$,
and (c) slepton pair ($\widetilde{\ell}\widetilde{\ell}$) production in association with initial-state radiation ($j$).
The higgsino simplified model also considers \chitz\chioz\ and \chip\chim\ production.}
\end{figure}

The second scenario features a similar particle spectrum to the first, except with $|M_1| < |M_2| \ll |\mu|$, so that the
produced electroweakinos have wino and/or bino nature.
In such wino/bino scenarios the LSP can be a thermal-relic dark-matter candidate that was depleted in the early universe through
coannihilation processes to match the observed dark-matter density~\cite{Griest:1990kh,Edsjo:1997bg}.
The production cross-section in such scenarios is typically larger than in the first scenario. They are also poorly constrained by dark-matter direct-detection experiments, and collider searches constitute the only direct probe for $|\mu| > 800~\GeV$~\cite{Profumo:2017ntc}.
Diagrams representing the production mode for the first two scenarios are shown in Figure~\ref{fig:intro:feynman:ewkino}.
A \chitz{} produced in either scenario can decay into a dilepton pair via an off-shell $Z$ boson ($Z^*$), such that the dilepton
invariant mass $\mll$ is kinematically restricted to be smaller than the mass-splitting between the \chitz{} and \chioz.
Hadronic initial-state radiation (ISR) is also required to boost the system as a way of enhancing the sensitivity of the search.
 
The third scenario is similar to the previous two, but it instead assumes that the pair production of the electroweakinos proceeds via vector-boson fusion (VBF) processes,
in which SM weak bosons exchange an electroweakino in a $t$-channel process to produce two electroweakinos and a pair of forward jets.
Such scenarios typically have very low cross-sections, but can complement the sensitivity of $q\bar{q}$ annihilation modes
that dominate the inclusive higgsino and wino/bino cross-sections, especially for LSP masses above a
few hundred \GeV~\cite{Nelson:2015jza}.  An example of such a process is illustrated in Figure~\ref{fig:intro:feynman:vbf}.
The kinematic cutoff of the \mll{} distribution is also used as the primary discriminant in this scenario, along with the
presence of two forward jets consistent with a VBF production mode.
 
The fourth scenario assumes the presence of scalar partners of the SM leptons (slepton, \slepton) that are slightly heavier
than a bino-like LSP.
Such models can explain dark-matter thermal-relic densities through coannihilation channels, as well as the
muon $g-2$ anomaly~\cite{Bennett:2006fi,Belyaev:2018vkl}.  This process is illustrated in Figure~\ref{fig:intro:feynman:slepton}.
This scenario exploits the relationship between the lepton momenta and the missing transverse momentum through the stransverse mass,
$\mtt$~\cite{Lester:1999tx,Barr:2003rg},
which exhibits a kinematic endpoint similar to that for \mll{} in electroweakino decays.

Events with two same-flavor opposite-charge leptons (electrons or muons), significant missing transverse momentum of size \met,
and hadronic activity are selected for all scenarios.
Signal regions (SRs) are defined by placing additional requirements on a number of kinematic variables.
The dominant SM backgrounds are either estimated with \emph{in situ} techniques or constrained using data control regions (CRs)
that enter into a simultaneous likelihood fit with the SRs.
The fit is performed in bins of either the \mll{} distribution (for electroweakinos) or the \mtt{} distribution (for sleptons).
 
Constraints on these compressed scenarios were first established at
LEP~\cite{LEPlimits,LEPlimitsSlepton,Heister:2001nk,Heister:2002mn,Heister:2002jca,Heister:2003zk,Abdallah:2003xe,Abdallah:2003xe,Acciarri:2000wy,Achard:2003ge,Abbiendi:2003ji,Abbiendi:2002vz}.
The lower bounds on direct chargino production from these results correspond to $m(\chiopm) > 103.5$~\GeV{} for
$\Delta m(\chiopm, \chioz) > 3$~\GeV{} and $m(\chiopm) > 92.4$~\GeV{} for smaller mass differences,
although the lower bound on the chargino mass weakens to around 75~\GeV{} for models with additional new scalars and
higgsino-like cross-sections~\cite{Egana-Ugrinovic:2018roi}.
For sleptons, conservative lower limits on the mass of the scalar partner of the right-handed muon, denoted $\smuon_R$,
are approximately $m(\smuon_R) \gtrsim 94.6$~\GeV{} for mass splittings down to $m(\smuon_R) - m(\chioz) \gtrsim 2$~\GeV.
For the scalar partner of the right-handed electron, denoted $\selectron_R$, LEP established a universal lower bound of
$m(\selectron_R) \gtrsim 73$~\GeV{} that is independent of $\Delta m(\selectron_R, \chioz)$~\cite{Heister:2002jca}.
Recent papers from the CMS~\cite{Sirunyan:2018iwl,Sirunyan:2018ubx,CMS:VBF:2019} and ATLAS~\cite{Aaboud:2017leg} collaborations
have extended the LEP limits for a range of mass splittings.
 
This paper extends previous LHC results by increasing the integrated luminosity, extending the search with additional channels,
and exploiting improvements in detector calibration and performance. The dedicated search for production via VBF is also added and
the event selection was reoptimized and uses techniques based on Recursive Jigsaw Reconstruction~\cite{Jackson:2017gcy},
which improve the separation of the SUSY signal from the SM backgrounds.


\section{ATLAS detector}
The ATLAS experiment is a general-purpose particle detector
that surrounds the interaction point with nearly $4\pi$ solid angle
coverage.\footnote{\AtlasCoordFootnote}
It comprises an inner detector, calorimeter systems, and a muon
spectrometer.
The inner detector provides precision tracking of charged particles in the pseudorapidity
region $|\eta| < 2.5$, consisting of pixel and microstrip silicon subsystems within a
transition radiation tracker.
The innermost pixel detector layer, the insertable B-layer~\cite{PIX-2018-001,ATLAS-TDR-2010-19}, was
added for $\sqrt{s} = 13$~TeV data-taking to improve tracking performance.
The inner detector is immersed in a 2~T axial magnetic field provided by a superconducting
solenoid.
High-granularity lead/liquid-argon electromagnetic sampling calorimeters are
used for $|\eta| < 3.2$. Hadronic energy deposits are measured in a
steel/scintillator tile barrel calorimeter in the $|\eta| < 1.7$ region. Forward
calorimeters cover the region $3.2 <|\eta|<4.9$ for both
electromagnetic and hadronic measurements.
The muon
spectrometer comprises trigger and high-precision tracking chambers spanning $|\eta| <
2.4$ and $|\eta|<2.7$, respectively, with a magnetic field provided by three large superconducting
toroidal magnets.
Events of interest are selected using a \mbox{two-level} trigger
system~\cite{TRIG-2016-01}, consisting of a first-level trigger implemented in
hardware, which is followed by a software-based high-level trigger.

\section{Data and simulated event samples}
\label{sec:samples}
 
Events were selected with a \met\ trigger, employing varied trigger thresholds as a function of the data-taking periods.
The trigger is $>95\%$ efficient for offline \met\ values above 200~\GeV{} for all periods.
The dataset used corresponds to \FullRunTwoLumi\ of $\sqrt{s}=13$~\TeV{} $pp$ collision data, where the uncertainty in the integrated luminosity is 1.7\%~\cite{ATLAS-CONF-2019-021}, obtained using the LUCID-2 detector \cite{LUCID2} for the primary luminosity measurements. The average number of interactions per bunch-crossing was 33.7.

Samples of Monte Carlo (MC) simulated events are used to estimate the signal yields, and for estimating the background from processes with prompt leptons, as well as in the determination of systematic uncertainties.
 
For the first signal scenario introduced in Section~\ref{sec:intro}, samples were generated for a simplified model of higgsino LSPs, including production of $\widetilde{\chi}^{-}_{1}\widetilde{\chi}^{+}_{1}$, $\widetilde{\chi}^{0}_{2}\widetilde{\chi}^\pm_{1}$ and $\widetilde{\chi}^0_{2}\widetilde{\chi}^0_{1}$.
The masses of the neutralinos ($\widetilde{\chi}^0_{1,2}$) were varied while the chargino mass was set to $\widetilde{\chi}^\pm_{1} = \frac{1}{2}\left [ m(\widetilde{\chi}^0_{1}) + m(\widetilde{\chi}^0_{2}) \right ]$.
Mass splittings in the case of pure higgsinos are generated by radiative corrections, and are of the order of hundreds of \MeV~\cite{Thomas:1998wy}.
Mass splittings of the order of tens of \GeV\ can be obtained by introducing mixing with wino or bino states.
In this simplified model, mass differences ranging from $1~\GeV$ to $60~\GeV$ are considered, but the calculated cross-sections
assume electroweakino mixing matrices corresponding to pure higgsino \chitz{}, \chiopm{} and \chioz{} states,
and all other SUSY particles are decoupled.
Typical values of cross-sections for $m(\chitz) = 110~\GeV$ and $m(\chioz) = 100~\GeV$ are $4.3\pm 0.1$~pb for $\chitz\chiopm$
production, $2.73\pm 0.07$~pb for $\chitz\chioz$ production, and $2.52 \pm 0.08$~pb for $\chiop\chiom$ production.
The samples were generated at leading order (LO) with MG5\_aMC@NLO 2.6.1~\cite{Alwall:2014hca} using the NNPDF23LO~\cite{Ball:2012cx}
parton distribution function (PDF) set and included up to two extra partons in the matrix element (ME).
The electroweakinos were decayed with \textsc{MadSpin}~\cite{Artoisenet:2012st}. The events were then interfaced with
\textsc{Pythia 8.212}~\cite{Sjostrand:2014zea} to model the parton shower (PS), hadronization, and underlying event (UE)
using the A14 set of tuned parameters (tune)~\cite{ATL-PHYS-PUB-2014-021}.
The ME--PS matching was performed using the CKKW-L scheme~\cite{Loennblad2012} with the merging scale set to 15~\GeV.
To enforce an ISR topology, at least one parton in the final state was required to have a transverse momentum ($\pT$)
greater than 50~\GeV{}. Possible diagrams including colored SUSY particles were excluded from the generation.
 
In the wino/bino scenario, the generated process is $pp \to \widetilde{\chi}^{0}_{2}\widetilde{\chi}^\pm_{1}$.
The $\widetilde{\chi}^{0}_{1}$ is a pure bino state, with the $\widetilde{\chi}^{0}_{2}$ and $\widetilde{\chi}^{\pm}_{1}$ states
forming degenerate pure wino states.
The generator configurations
are consistent with those used for the higgsino samples.
A typical value of the $\chitz\chiopm$ production cross-section is $16.0\pm0.5$~pb for $m(\chitz) = m(\chiopm) = 110~\GeV$.

Additional samples were generated for the third scenario of pair production of electroweakinos produced via VBF.
These were generated with the same decay, PS, hadronization, and UE configuration as the higgsino simplified model samples.
The ME generation was the same as in the higgsino case, but used an updated version of MG5\_aMC@NLO (version 2.6.2).
In order to select uniquely the VBF topologies, the number of QCD vertices was set to zero.
An additional filter was applied to select events with exactly two parton emissions in the ME.
The invariant mass of the two partons is required to be at least $200~\GeV{}$, while the minimum transverse momentum of
each parton is $12~\GeV{}$.
Typical values of LO cross-sections with these requirements for $m(\chitz) = 100~\GeV$ and $m(\chioz) = 90~\GeV$ are $16\pm 1$~fb and
$47 \pm 4$~fb, for the higgsino and wino/bino models, respectively.
For higgsino masses smaller than half of the Higgs boson mass, the cross-sections include contributions from VBF Higgs production
with decays $h\to\chitz\chioz$.

The electroweakino searches exploit the kinematic endpoint in the dilepton invariant mass spectrum from the decay chain
$\widetilde{\chi}^0_{2} \to Z^* \widetilde{\chi}^0_{1},\ Z^* \to \ell\ell$.
Therefore, processes that involve the production of a $\widetilde{\chi}^0_{2}$ neutralino dominate the sensitivity of the search.
The branching ratios for the processes
$\widetilde{\chi}^0_{2}\to Z^* \widetilde{\chi}^0_{1}$ and $\widetilde{\chi}^\pm_{1}\to W^{*} \widetilde{\chi}^0_{1}$ were fixed to 100\%
for all the scenarios given above.
The branching ratios of $Z^* \to \ell\ell$ and $W^{\pm *} \to \ell \nu$ depend on the invariant mass of the off-shell vector boson.
For both the higgsino and wino/bino models, the branching ratios were computed with SUSY-HIT 1.5a~\cite{Djouadi:2006bz},
which accounts for finite $b$-quark and $\tau$-lepton masses.
At $\Delta m \left ( \widetilde{\chi}^0_{2}, \widetilde{\chi}^0_{1}\right ) = 40$~\GeV{} the $Z^*\to \ell\ell$ branching ratio
to electrons or muons is 3.5\%.
This increases to 5.3\% and 5.0\%, respectively, at
$\Delta m \left ( \widetilde{\chi}^0_{2}, \widetilde{\chi}^0_{1}\right ) = 1$~\GeV{} as decays into heavier quarks or $\tau$-leptons become kinematically inaccessible.
Similarly, for $W^*\to \ell\nu$ the branching ratios to electrons or muons are both 11\% at a mass splitting of 40~\GeV{},
but increase to 20\% and 17\%, respectively, for $\Delta m \left ( \widetilde{\chi}^0_{2}, \widetilde{\chi}^0_{1}\right ) = 1$~\GeV.
 
\begin{figure}[tbp]
\centering
\includegraphics[width=0.7\columnwidth]{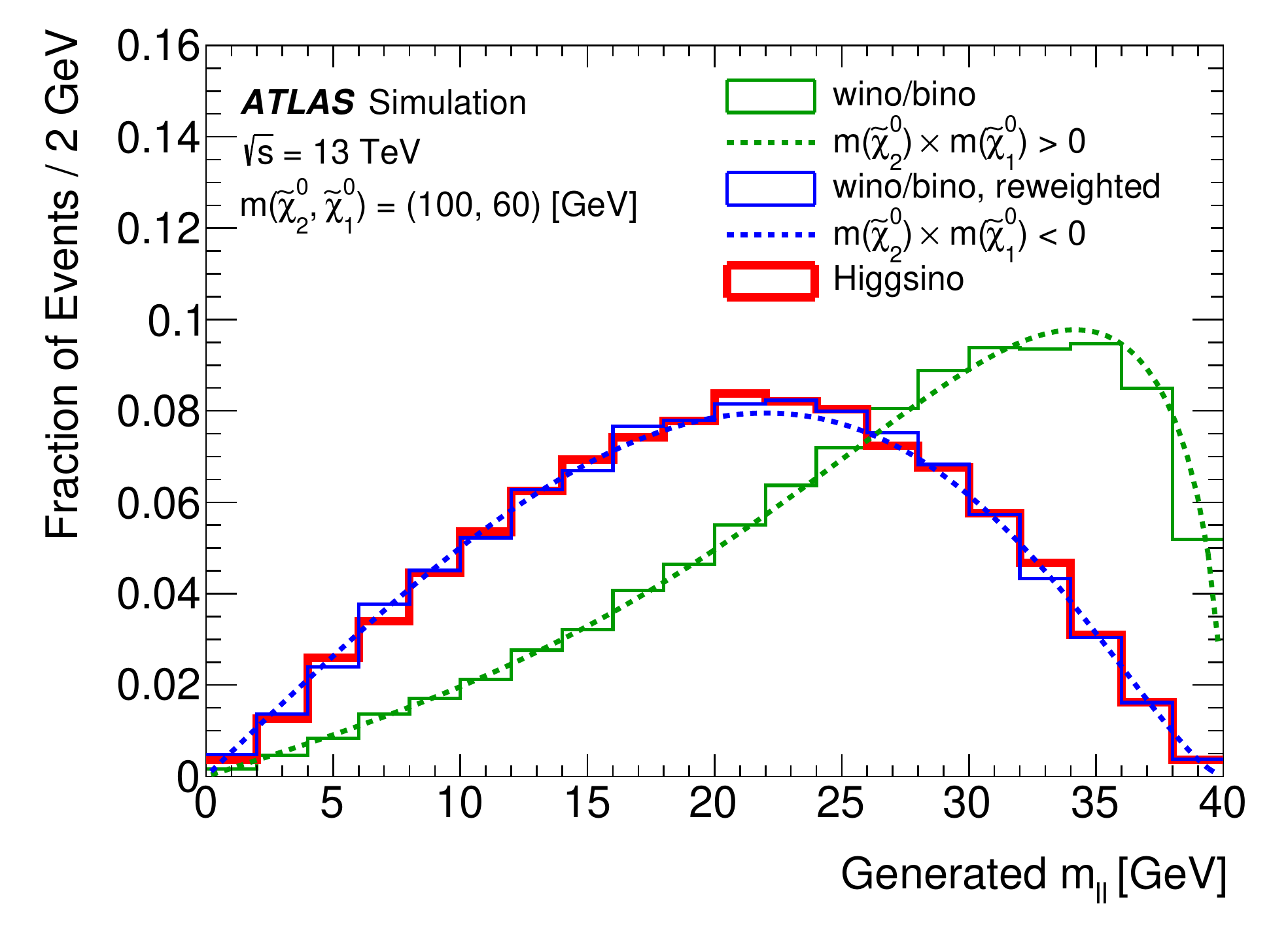}
\caption{\label{fig:samples:lineshape}Dilepton invariant mass for higgsino and wino/bino simplified models. The endpoint of the distribution is determined by the difference between the masses of the $\widetilde{\chi}^0_{2}$ and $\widetilde{\chi}^0_{1}$. The results from simulation (histograms) are compared with analytic calculations of the expected lineshape (dashed lines) presented in Ref.~\cite{DeSanctis:2007yoa}. The product of the signed mass eigenvalues $\left ( m\left ( \widetilde{\chi}^0_{2} \right ) \times  m \left ( \widetilde{\chi}^0_{1} \right ) \right )$ is negative for the higgsino model and can be either negative or positive for wino/bino scenarios.}
\end{figure}

The distribution of the dilepton invariant mass from the decay of the virtual $Z^*$~\cite{DeSanctis:2007yoa} depends on the relative sign of
the $\chioz$ and $\chitz$ mass parameters.
In a pure higgsino model the product of the signed mass eigenvalues
$\left ( m\left ( \widetilde{\chi}^0_{2} \right ) \times  m \left ( \widetilde{\chi}^0_{1} \right ) \right )$ can only be negative,
while for the wino/bino case either positive or negative products are allowed.\footnote{The mixing matrix used to diagonalize the neutral electroweakino states is forced to be a real matrix in the SLHA2 format~\cite{Allanach:2008qq}.  A consequence of this choice is a negative sign given to one or more mass eigenvalues, determined in part by the relative fractions of wino, bino, or higgsino content of the physical states.  For additional discussion of this, see Ref.~\cite{Fuks:2017rio}.}
The generated wino/bino process assumes the product of the signed mass eigenvalues is positive, and the analytical description of
the expected lineshape is used to reweight the \mll\ distribution to the case of the product being negative.
The difference between wino/bino and higgsino lineshapes, as well as the level of agreement between the reweighted distribution and the expected
lineshape, is shown in Figure~\ref{fig:samples:lineshape}. The two possible wino/bino \mll\ distributions are used to provide two separate model-dependent interpretations of the results. With the exception of the signal modeling, the interpretations for higgsino and both wino/bino samples are otherwise conducted identically, and use the same search regions as defined in Section~\ref{sec:signalreg}.

For the fourth scenario, samples with direct production of selectrons $\widetilde{e}_{L,R}$ or smuons $\widetilde{\mu}_{L,R}$ were generated.
The $L,R$ subscripts denote left- or right-handed chirality of the corresponding SM lepton partners.
All sleptons flavors and chirality contributions are assumed to be degenerate in mass.
A typical value of the slepton production cross-section is $0.55\pm0.01$~pb for $m(\slepton_{L,R}) = 110~\GeV$.
These particles decay with a 100\% branching ratio into their corresponding SM partner lepton and a pure bino neutralino,
$\widetilde{\chi}^{0}_{1}$. The slepton samples were generated with MG5\_aMC@NLO 2.6.1 and interfaced with \textsc{Pythia 8.230}.
The PDF set used was NNPDF2.3LO with the A14 tune.  Similar to the higgsino and wino/bino samples, CKKW-L merging~\cite{Loennblad2012} was used
for the ME--PS matching, with the merging scale set to a quarter of the slepton mass.

Cross-sections for all but the VBF signal scenarios are calculated with
\textsc{Resummino} 2.0.1 at NLO+NLL precision~\cite{Beenakker:1999xh,Debove:2010kf,Fuks:2012qx,Fuks:2013vua,Fiaschi:2018hgm,Bozzi:2007qr,Fuks:2013lya,Fiaschi:2018xdm}.
The VBF cross-sections are computed at LO precision with MG5\_aMC@NLO 2.6.2.
The evaluation of the cross-sections and corresponding uncertainty are taken from an envelope of cross-section predictions
using different PDF sets, and varied factorization and renormalization scales.
This procedure is described in Ref.~\cite{Borschensky:2014cia}, and is the same procedure as used in the
previous search~\cite{Aaboud:2017leg}.
 
The SM background processes are estimated from a combination of MC simulation as well as data-driven approaches.
The latter are described in Section~\ref{sec:backgrounds}.
\textsc{Sherpa} 2.2.1 and 2.2.2~\cite{Gleisberg:2008ta} were used to model the $V+$jets ($V= W,Z,\gamma^*$) samples involving
leptonically decaying vector bosons, as well as diboson ($WW$, $ZZ$ and $WZ$, collectively referred to as $VV$),
and fully leptonic triboson processes.
The $Z^{(*)}/\gamma^*$ + jets and $VV$ samples provide coverage of dilepton invariant masses down to 0.5~\GeV{} for $Z^{(*)}/\gamma^*\to e^+e^-/\mu^+\mu^-$, and 3.8~\GeV{} for $Z^{(*)}/\gamma^*\to\tau^+\tau^-$.
A separate set of $Z(\to\mu\mu)$+jets samples were generated using MG5\_aMC@NLO using the same configuration as for the signal
samples described above in order to evaluate initial- and final-state radiation modeling in signal samples.
Gluon--gluon fusion (ggF) and VBF single-Higgs production were generated with \textsc{Powheg-Box}~\cite{Alioli:2010xd},
while Higgs production in association with a massive vector boson was generated with \textsc{Pythia 8.186}, and $\ttbar h$ production was generated with MG5\_aMC@NLO 2.2.3. 
\textsc{Powheg-Box} was used to generate $t\bar{t}$~\cite{Frixione:2007nw,Nason:2004rx,Frixione:2007vw,Alioli:2010xd}, single top~\cite{Frederix:2012dh} and top quarks produced in association with $W$ bosons~\cite{Re:2010bp}.
Rarer top-quark processes all used MG5\_aMC@NLO (versions 2.2.2/2.3.3).
Matrix elements, excluding those generated with \textsc{Pythia} or \textsc{Sherpa}, were then interfaced with \textsc{Pythia~8}
using the ME+PS prescription.
Further details on the configuration of the simulation of SM processes can be
found in
Refs.~\cite{ATL-PHYS-PUB-2017-005,ATL-PHYS-PUB-2017-006,ATL-PHYS-PUB-2017-007,ATL-PHYS-PUB-2017-008,ATL-PHYS-PUB-2016-005}.
A summary of the generator configurations, including the PDF sets and the order of the cross-section calculations used for normalization, is given
in Table~\ref{table:samples:BGs}.
 
\begin{table}[tbp]
\centering
\caption{Simulated SM background processes. The PDF set refers to that used for the matrix element.}
\resizebox{\textwidth}{!}{
\begin{tabular}{l l l l l}
\toprule
Process &  Matrix element & Parton shower & PDF set & Cross-section \\
\midrule
$V$+jets & \multicolumn{2}{c}{\textsc{Sherpa 2.2.1}}       & NNPDF 3.0 NNLO~\cite{Ball:2014uwa} & NNLO~\cite{Anastasiou:2003ds} \\
$VV$  & \multicolumn{2}{c}{\textsc{Sherpa 2.2.1/2.2.2}} & NNPDF 3.0 NNLO                     & Generator NLO \\
Triboson & \multicolumn{2}{c}{\textsc{Sherpa 2.2.1}}       & NNPDF 3.0 NNLO                     & Generator LO, NLO \\
\midrule
$h$ (ggF)          & \textsc{Powheg-Box}                       & \textsc{Pythia 8.212} & NLO CTEQ6L1~\cite{ourPumplin:2002vw} & $\mathrm{N^3LO}$~\cite{ourdeFlorian:2016spz} \\
$h$ (VBF)          & \textsc{Powheg-Box}                       & \textsc{Pythia 8.186} & NLO CTEQ6L1~\cite{ourPumplin:2002vw} & NNLO + NLO~\cite{ourdeFlorian:2016spz}  \\
$h+W/Z$      & \multicolumn{2}{c}{\textsc{Pythia 8.186}}                               & NNPDF 2.3 LO~\cite{Ball:2012cx}   & NNLO + NLO~\cite{ourdeFlorian:2016spz} \\
$h+t\bar{t}$ & MG5\_aMC@NLO 2.2.3                        & \textsc{Pythia 8.210}       & NNPDF 2.3 LO                      & NLO~\cite{ourdeFlorian:2016spz} \\
\midrule
$t\bar{t}$        & \textsc{Powheg-Box} & \textsc{Pythia 8.230} & NNPDF 2.3 LO & NNLO+NNLL~\cite{Cacciari:2011hy,Czakon:2012zr,Czakon:2012pz,Czakon:2013goa,Czakon:2011xx} \\
$t$ ($s$-channel) & \textsc{Powheg-Box} & \textsc{Pythia 8.230} & NNPDF 2.3 LO & NNLO+NNLL~\cite{Kidonakis:2010tc} \\
$t$ ($t$-channel) & \textsc{Powheg-Box} & \textsc{Pythia 8.230} & NNPDF 2.3 LO & NNLO+NNLL~\cite{Kidonakis:2011wy,Frederix:2012dh} \\
$t + W$           & \textsc{Powheg-Box} & \textsc{Pythia 8.230} & NNPDF 2.3 LO & NNLO+NNLL~\cite{Kidonakis:2010ux} \\
\midrule
$t+Z$                   & MG5\_aMC@NLO 2.3.3 & \textsc{Pythia 8.212}       & NNPDF 2.3 LO & NLO~\cite{Alwall:2014hca} \\
$t\bar{t}WW$            & MG5\_aMC@NLO 2.2.2 & \textsc{Pythia 8.186}       & NNPDF 2.3 LO & NLO~\cite{Alwall:2014hca} \\
$t\bar{t}+Z/W/\gamma^*$ & MG5\_aMC@NLO 2.3.3 & \textsc{Pythia 8.210/8.212} & NNPDF 2.3 LO & NLO~\cite{ourdeFlorian:2016spz} \\
$t + WZ$                & MG5\_aMC@NLO 2.3.3 & \textsc{Pythia 8.212}       & NNPDF 2.3 LO & NLO~\cite{Alwall:2014hca} \\
$t+t\bar{t}$            & MG5\_aMC@NLO 2.2.2 & \textsc{Pythia 8.186}       & NNPDF 2.3 LO & LO~\cite{Alwall:2014hca} \\
$tt\bar{t}\bar{t}$      & MG5\_aMC@NLO 2.2.2 & \textsc{Pythia 8.186}       & NNPDF 2.3 LO & NLO~\cite{Alwall:2014hca} \\
\bottomrule
\end{tabular}
}
\label{table:samples:BGs}
\end{table}

To simulate the effects of additional $pp$ collisions, referred to as pileup, in the same and neighboring bunch crossings,
additional interactions were generated using the soft QCD processes of \textsc{Pythia} 8.186 with the
A3 tune~\cite{ATL-PHYS-PUB-2016-017} and the MSTW2008LO PDF set~\cite{Martin:2009iq}, and were overlaid onto each
simulated hard-scatter event.
The MC events were reweighted to match the pileup distribution observed in the data.
 
Background and signal samples made use of \textsc{EvtGen} 1.6.0 and 1.2.0~\cite{Lange:2001uf} to model the decay of bottom
and charm quarks, with the exception of the background samples modeled with \textsc{Sherpa}. All MC simulated samples were processed through the ATLAS simulation framework~\cite{SOFT-2010-01} in
\textsc{Geant4}~\cite{Agostinelli:2002hh}.
The samples for the signal scenarios made use of the ATLAS fast simulation, which parameterizes the response of the calorimeters.

\section{Event reconstruction}
\label{sec:selection}
Events are required to have at least one reconstructed $pp$ interaction vertex with a minimum of two associated tracks
with $\pt > 500~\MeV$.
In events with multiple vertices, the primary vertex is defined as the one with the highest $\sum\pt^2$ of associated tracks.
To reject events with detector noise or noncollision backgrounds, a set of basic quality criteria~\cite{DAPR-2012-01} are applied.
 
Leptons, jets and tracks are ``preselected'' using loose identification criteria, and must survive tighter ``signal''
identification requirements in order to be selected for the search regions.
Preselected leptons and jets are used in fake/nonprompt (FNP) lepton background estimates, as well as in resolving ambiguities
between tracks and clusters associated with multiple lepton and jet candidates.
 
Isolation criteria are used in the definition of signal leptons, and are based on tracking information, calorimeter clusters, or both.  Isolation energies are computed as a $\sum\pt$ of nearby activity, excluding the contributions from nearby leptons, and are effective in reducing contributions from semileptonic heavy-flavor hadron decays and jets faking prompt leptons.  The isolation requirements used in this analysis are based on those described in Refs.~\cite{PERF-2015-10} and~\cite{EGAM-2018-01}, with updates to improve their performance under the increased pileup conditions encountered in the 2017 and 2018 data samples.
 
Electrons are required to have $\pt > 4.5~\GeV$ and $|\eta| < 2.47$.
Preselected electrons are further required to pass the calorimeter- and tracking-based \emph{VeryLoose} likelihood
identification~\cite{EGAM-2018-01}, and to have a longitudinal impact parameter $z_0$ relative to the primary vertex
that satisfies $|z_0 \sin \theta| < 0.5$~mm.
Signal electrons must satisfy the \emph{Medium} identification criterion~\cite{EGAM-2018-01}, and be compatible
with originating from the primary vertex, with the significance of the transverse impact parameter defined relative
to the beam position satisfying \mbox{$|d_0|/\sigma(d_0) < 5$}.
Signal electrons are further refined using the \emph{Gradient} isolation working point~\cite{EGAM-2018-01},
which uses both tracking and calorimeter information.
 
Muons are required to satisfy $\pt > 3~\GeV$ and $|\eta|<2.5$.
Preselected muons are identified using the \emph{LowPt} criterion~\cite{ATL-PHYS-PUB-2020-002}, a reoptimized selection similar to those defined
in Ref.~\cite{PERF-2015-10} but with improved signal efficiency and background rejection for $\pt<10~\GeV$ muon candidates.
The \emph{LowPt} working point has improved efficiency for muons with $\pt<4~\GeV$ traversing the central detector region, which can lose enough energy in the calorimeters that they do not reach the second station of precision muon tracking chambers.  The \emph{LowPt} selection accepts candidates composed of track segments in the inner detector matched to track segments from a single station of the muon spectrometer.  Misidentified muon candidates originating from in-flight hadron decays are rejected by requirements on the significance of a change in trajectory along the track, and by requiring that the momentum measurements in the inner tracker and in the muon spectrometer are compatible with each other. For prompt muons with $3<\pt < 6~\GeV$, the \emph{LowPt} criterion recovers approximately 20\% of the identification efficiency in the $|\eta|<1.2$ region, while maintaining an average misidentification probability comparable to the \emph{Medium} selection described in Ref.~\cite{PERF-2015-10}.
 
Preselected muons must also satisfy $|z_0 \sin \theta| < 0.5$~mm.
From the remaining preselected muons, signal muons must satisfy \mbox{$|d_0|/\sigma(d_0) < 3$}.
Finally, signal muons are required to pass the \emph{FCTightTrackOnly} isolation working point~\cite{PERF-2015-10}, which uses only tracking information.
 
Preselected jets are reconstructed from calorimeter topological energy clusters~\cite{PERF-2014-07} in the
region $|\eta| < 4.5$ using the anti-$k_t$ algorithm~\cite{Cacciari:2008gp,Fastjet} with radius parameter $R = 0.4$.
The jets are required to have $\pt>20$~\GeV{} after being calibrated in accord with Ref.~\cite{PERF-2016-04} and
having the expected energy contribution from pileup subtracted according to the jet area~\cite{PERF-2014-03}.
In order to suppress jets due to pileup, jets with $\pt < 120~\GeV$ and $|\eta| < 2.5$ are required to satisfy
the \emph{Medium} working point of the jet vertex tagger~\cite{PERF-2014-03}, which uses information from the tracks
associated with the jet.
The \emph{Loose} working point of the forward jet vertex tagger~\cite{ATL-PHYS-PUB-2015-034} is in turn used to
suppress pileup in jets with $\pt < 50~\GeV$ and $|\eta| >2.5$.
From the sample of preselected jets, signal jets are selected if they satisfy $\pt > 30~\GeV$ and $|\eta| < 2.8$.
The VBF search uses a modified version of signal jets, labeled VBF jets, satisfying $\pt > 30~\GeV$ and $|\eta| < 4.5$.
 
Jets identified as containing $b$-hadron decays, referred to as $b$-tagged jets, are identified from preselected
jets within $|\eta| < 2.5$ using the \emph{MV2c10} algorithm~\cite{FTAG-2018-01}.
The $\pt > 20~\GeV$ requirement is maintained to maximize the rejection of the \ttbar background.
The $b$-tagging algorithm working point is chosen so that $b$-jets from simulated $t\bar{t}$ events are identified
with an 85\% efficiency, with rejection factors of 2.7 for charm-quark jets and 25 for light-quark and gluon jets.
 
The following procedure is used to resolve ambiguities between the reconstructed leptons and jets.
It employs the distance measure $\Delta R_y = \sqrt{(\Delta y)^2 + (\Delta \phi)^2}$, where $y$ is the rapidity.
Electrons that share an inner detector track with a muon candidate are discarded to remove bremsstrahlung from muons
followed by a photon conversion. Non-$b$-tagged jets that are separated from the remaining electrons by $\Delta R_y < 0.2$ are removed.
Jets containing a muon candidate within $\Delta R_y < 0.4$ and with fewer than three tracks with $\pt>500~\MeV$ are removed to
suppress muon bremsstrahlung. Electrons or muons with $\Delta R_y < 0.4$ from surviving jet candidates are removed to
suppress bottom and charm hadron decays.
 
Signal regions based on a signal lepton and an isolated low-\pt\ track are used to increase the efficiency for electroweakino signals with the lowest mass splittings, where the lepton \pt\ can be very low.
For these regions the track is selected to be matched to a reconstructed electron or muon candidate with no identification requirements, including muons reconstructed with the \emph{CaloTagged} and \emph{SegmentTagged} algorithms described in Ref.~\cite{PERF-2015-10}.
Preselected tracks with $\pt> 500~\MeV$ and $\eta<2.5$ are selected using the \emph{Tight-Primary} working point defined in
Ref.~{\cite{ATL-PHYS-PUB-2015-051}.
Signal tracks are required to be within $\DeltaR = 0.01$ of a reconstructed electron or muon candidate.
Electron~(muon) candidates can be reconstructed with transverse momenta as low as 1~(2)~\GeV, and are required to fail the
signal lepton requirements defined above to avoid any overlap.
Signal tracks with $\pt$ that differs from the transverse momentum of the matched lepton by more than $20\%$ are rejected.
The track--lepton matching allows the tracks to be identified as electron or muon tracks, reducing backgrounds from
tracks not originating from the leptonic decay of a SUSY particle.
Signal tracks must also satisfy dedicated isolation criteria: they are required to be separated from preselected jets
by at least $\Delta R > 0.5$, and the $\sum\pt$ of preselected tracks within $\DeltaR = 0.3$ of signal tracks, excluding
the contributions from nearby leptons, is required to be smaller than $0.5~\GeV$.
Finally, signal tracks must satisfy $\pt> 1~\GeV$, $|z_0 \sin \theta| < 0.5$~mm and \mbox{$|d_0|/\sigma(d_0) < 3$}.
 
The missing transverse momentum \ptmiss, with magnitude \met, is defined as the negative vector sum of the transverse momenta
of all preselected objects (electrons, muons, jets, and photons~\cite{EGAM-2018-01}), and an additional soft term that is constructed from all tracks that are not associated with any lepton or jet, but that are associated with the primary vertex.
A dedicated overlap removal procedure is used to resolve ambiguities between the reconstructed objects~\cite{ATLAS-CONF-2018-023}.
In this way, \met\ is adjusted for the best calibration of jets and leptons, while maintaining pileup independence in the
soft term~\mbox{\cite{PERF-2016-07}}.
 
Small scale factors are applied to the efficiencies of reconstructed electrons, muons, $b$-tagged jets, and tracks in the simulated samples to
match the reconstruction efficiencies in data.
The scale factors for $b$-tagged jets account for the differences between data and simulated samples in the identification
efficiencies for jets, including $b$-hadron decays, as well as misidentification rates of jets initiated from charm quarks, light-flavor quarks, or gluons.
The scale factors for low-momentum leptons are obtained from $J/\psi\rightarrow ee/\mu\mu$ events with the same tag-and-probe
methods as used for higher-\pt{} electrons~\cite{EGAM-2018-01} and muons~\cite{PERF-2015-10}.
The scale factors used to account for track--lepton matching efficiency differences between data and simulation are derived
from events with a $J/\psi$ meson decaying into a low-$\pt$ signal lepton and a preselected track.
The track-isolation scale factors are measured using events with a $Z$ boson decaying into a signal lepton and a track matched to a
reconstructed lepton candidate. All track scale factors are found to be compatible with 1.
 
After all lepton selection criteria and efficiency scale factors are applied, the efficiency for reconstructing and identifying
signal electrons within the detector acceptance in the higgsino and slepton signal samples ranges from 20\% for $\pt=4.5~\GeV$ to
over 75\% for $\pt>30~\GeV$.
The corresponding efficiency for signal muons ranges from approximately 50\% at $\pt=3~\GeV$ to 90\% for $\pt>30~\GeV$.
The efficiency of selecting signal tracks for electroweakino events peaks at 78\% for tracks with $\pt=2.5~\GeV$,
with lower efficiencies at lower \pt{} due to track selection criteria and at higher \pt{} due to increasing electron and
muon efficiencies.
The efficiency for signal electrons, muons, and isolated tracks in a mix of slepton and higgsino samples is shown
in Figure~\ref{fig:leptoneffs} as a function of lepton \pt.
 
\begin{figure}[tbp]
\centering
\includegraphics[width=0.7\columnwidth]{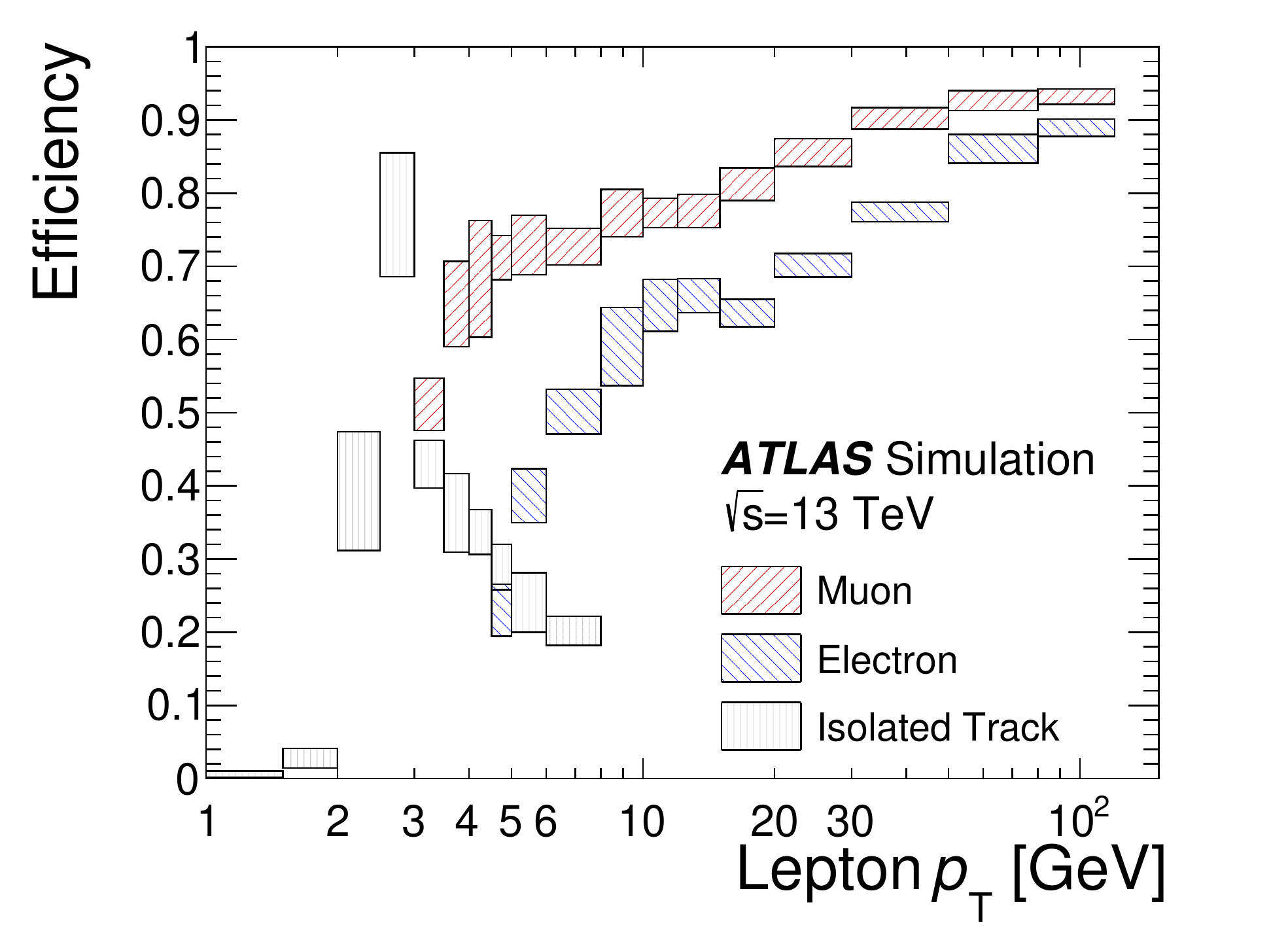}
\caption{Signal lepton efficiencies for electrons, muons, and isolated tracks in a mix of slepton and higgsino samples. Combined reconstruction, identification, isolation and vertex association efficiencies are shown for leptons within detector acceptance, and with lepton \pt{} within a factor of 3 of $\Delta m(\slep,\chioz)$ for sleptons or of $\Delta m(\chitz,\chioz)/2$ for higgsinos. The efficiencies for isolated tracks include  track reconstruction and vertex association efficiencies~\cite{ATL-PHYS-PUB-2015-051},
as well as the efficiencies for track--lepton matching and track isolation, which are specific to this search. Scale factors are applied to match reconstruction efficiencies in data.  The average number of interactions per crossing in the MC samples is $33.7$; the number of pileup interactions match the distribution in data in spread and mean value. Uncertainty bands represent the range of efficiencies observed across all signal samples used for the given \pt{} bin. 
}
\label{fig:leptoneffs}
\end{figure}
 
Dedicated scale factors are also used to re-weight MC events to properly model the trigger efficiency observed in data.
These scale factors are measured in events selected with single-muon triggers, passing kinematic selections similar to the
ones used to define the SRs.
They are parameterized as a function of \met\ and found to vary between 0.85 and 1 in the \met\ range of interest.
The uncertainty in the parameterization of the scale factors is negligible. An uncertainty of 5\% is assigned to the scale factors
to cover their dependence on other kinematic quantities of interest, such as $\mll$ and $\mtt$.
Additional uncertainties of at most 4\% are assigned due to differences between the trigger efficiencies determined with MC events
for the different signal and background processes.

\section{Signal regions}
\label{sec:signalreg}
Events entering into all SRs share a common preselection, with requirements listed in Table~\ref{table:signalreg:preselectioncuts}.
The $2\ell$ channels require exactly two opposite-charge (OS) signal leptons of the same flavor, while
the \oneleponetrack{} channel requires exactly one signal lepton and at least one OS signal track of the same flavor.
In events where more than one OS same-flavor signal track is present, the candidate with the highest \pt{} is used to define the \oneleponetrack{} system.
In regions with two leptons, the higher-\pt{} lepton is referred to as the ``leading'' lepton ($\ell_1$) while the lower-\pt{} lepton
is the ``subleading'' lepton ($\ell_2$).
 
\begin{table}[tbp]
\centering
\caption{Preselection requirements applied to all events entering into electroweakino, slepton and VBF search regions. Requirements marked with $\dagger$ are not applied to VBF search regions. Requirements on jets are applied to VBF jets (satisfying $|\eta| < 4.5$) in the VBF channel. }
\resizebox{\textwidth}{!}{
\begin{tabular}{l l l}
\toprule
\toprule
&\multicolumn{2}{c}{Preselection requirements}\\
\cline{2-3}\\[-1.0em]
Variable   		                     & $2\ell$  					     & \oneleponetrack{}\\
\midrule
Number of leptons (tracks)           & $=2$ leptons 								 & $=1$~lepton~and~$\geq1$~track\\
Lepton \pt{}$~[\GeV]$      & $\ptlone>5$                                   & $\ptl<10$				\\
\DRll                          	     & $\DRee>0.30$, $\DRmm>0.05$, $\DRem>0.2$       & $0.05<\Delta R_{\ell\mathrm{track}}<1.5$\\
Lepton (track) charge and flavor     & $e^{\pm}e^{\mp}$ or $\mu^{\pm}\mu^{\mp}$      & $e^{\pm}e^{\mp}$ or $\mu^{\pm}\mu^{\mp}$ \\
Lepton (track) invariant mass~[\GeV] & $3<m_{ee}<60$, $1<m_{\mu\mu}<60$              		 & $0.5<\mlt<5$\\
$J/\psi$ invariant mass~[\GeV]       & veto $3<\mll<3.2$                             & veto $3<\mlt<3.2$\\
$\mtautau$~[\GeV]                    & $<0$ or $>160$                                & no requirement \\
\met~[\GeV]                          & $>120$                                        & $>120$ \\
Number of jets                       & $\geq 1$                                      & $\geq 1$ \\
Number of $b$-tagged jets      	     & $=0$                                          & no requirement \\
Leading jet \pt~[\GeV]               & $\geq 100$                                    & $\geq 100$\\
\mindphijetsmet                	     & $>0.4$                                        & $>0.4$ \\
\dphijetmet~$^{\dagger}$                          & $\geq 2.0$                                    & $\geq 2.0$ \\
\bottomrule
\bottomrule
\end{tabular}
}
\label{table:signalreg:preselectioncuts}
\end{table}


Preselection requirements are employed to reduce backgrounds and form a basis for SRs and CRs used in the simultaneous fit.
The leading lepton is required to have $\pt{}>5$~\GeV{}, which reduces backgrounds from FNP leptons.
Pairs of muons are required to be separated by $\DRmm>0.05$, while pairs of electrons are required to be separated by $\DRee>0.3$ to avoid reconstruction inefficiencies due to overlapping
electron showers in the EM calorimeter. Electrons and muons are likewise required to be separated
by $\DeltaR_{e\mu}>0.2$ to avoid energy deposits from muons spoiling electron shower shapes.
An additional requirement that $\mll$ be outside of $[3.0,3.2]$~\GeV{} removes contributions from $J/\psi$ decays,
while requiring $\mll<60$~\GeV{} reduces contributions from on-shell $Z$-boson decays.
Contributions from other hadronic resonances, e.g. $\Upsilon$ states, are expected to be negligible in the search regions
and are not explicitly vetoed.
Requirements on the minimum angular separation between the lepton candidates ($\DRll$) and invariant mass ($\mll$)
remove events in which an energetic photon produces collinear lepton pairs.
 
The \mtautau{} variable~\cite{Han:2014kaa,Baer:2014kya,Barr:2015eva} approximates the invariant mass of a leptonically decaying $\tau$-lepton pair if both $\tau$-leptons
are sufficiently boosted so that the neutrinos from each $\tau$ decay are collinear with the visible lepton momentum.
It is defined as
\mbox{$m_{\tau\tau} = \text{sign}\left(m_{\tau\tau}^2\right)\sqrt{\left|m_{\tau\tau}^2\right|}$}, which is the
signed square root of $m_{\tau\tau}^2 \equiv 2p_{\ell_1} \cdot p_{\ell_2}(1+\xi_1)(1+\xi_2)$, where
$p_{\ell_1}$ and $p_{\ell_2}$ are the lepton four-momenta, while the parameters $\xi_1$ and $\xi_2$ are determined by
solving $\mathbf{p}_\text{T}^\textrm{miss} = \xi_1 \mathbf{p}_\textrm{T}^{\ell_1} + \xi_2 \mathbf{p}_\mathrm{T}^{\ell_2}$.
It can be less than zero in events where one of the lepton momenta has a smaller magnitude than
the $\met$ and points in the hemisphere opposite to the \ptmiss{} vector.
Events with $0<\mtautau<160$~\GeV{} are rejected, which reduces backgrounds from $Z\to\tau\tau$ and
has an efficiency greater than $80\%$ for the signals considered.
 
The reconstructed \met{} is required to be greater than 120~\GeV{} in preselection, with higher thresholds applied in some SRs.
For SUSY events in which much of the invisible momentum is carried by the \chioz{} pair, these requirements on \met{} suggest that
the SUSY system is recoiling against additional hadronic activity, either in the form of ISR or the forward jets in VBF processes.
All events are therefore required to have at least one jet with $\pt>100~\GeV$. Additional jets in the event are also required to
be separated from the \ptmiss{} by $\mindphijetsmet>0.4$ in order to suppress the impact of jet energy mismeasurement on \met.
For searches involving ISR, the leading jet is required to be separated from the \ptmiss{} by at least 2.0 radians in $\phi$.
In the 2$\ell$ channel, events with one or more $b$-tagged jets with $\pt>20~\GeV$ (\nbtagtwenty) are vetoed to reduce
backgrounds from $\ttbar$ production.
 
After applying the preselection requirements above, SRs are further optimized for specific SUSY scenarios.
Three categories of SRs, labeled SR--E, SR--VBF and SR--S, are constructed: the first for electroweakinos recoiling against
ISR (or simply electroweakinos), the second for electroweakinos produced through VBF, and the last targeting sleptons
recoiling against ISR.
 
The SRs designed for optimal sensitivity to electroweakinos are defined in Table~\ref{table:signalreg:ewkinocuts}.
High-\met\ regions, labeled SR--E--high and SR--E--\oneleponetrack{}, require $\met>200~\GeV{}$, where the online \met{} triggers are fully efficient
for the SUSY signal. Low-\met\ regions are constructed using events with $120~\GeV<\met<200~\GeV$: SR--E--med targets
electroweakinos with small mass splittings and SR--E--low targets mass splittings larger than $\sim$10~\GeV{}.
 
\begin{table}[tbp]
\centering
\caption{Requirements applied to events entering into the four signal regions used for electroweakino searches. The $\oneleponetrack$ preselection requirements from Table~\ref{table:signalreg:preselectioncuts} are implied for SR--E--\oneleponetrack{}, while the $2\ell$ ones are implied for the other SRs.}
\resizebox{\columnwidth}{!}{
\begin{tabular}{l l l l l }
\toprule
\toprule
&\multicolumn{4}{c}{Electroweakino SR Requirements}                                                                                 \\
\cline{2-5}\\[-1.0em]
Variable                           &SR--E--low           & SR--E--med            &SR--E--high                             &SR--E--\oneleponetrack{} \\
\midrule
\met~[\GeV]                        &$[120, 200]$         &$[120, 200]$           &$>200$                                  & $>200$                  \\
\met/\htlep                        &$<10$                &$>10$                  &--                                      & $>30$                \\
\dphilepmet	                       &--				   	 &--				     &--					                  & $<1.0$               \\
Lepton or track \pt{}~[\GeV]       &$\ptltwo> 5 + \mll/4$&--                     &$\ptltwo>\min(10, 2 + \mll/3)$          & $\ptt<5$       \\
\MTS~[\GeV]                        &--                   &$<50$                  &--                                      &--                   \\
\mtlone~[\GeV]                     &$[10,60]$            &--                     &$<60$                                   &--                   \\
\RISR                              &$[0.8,1.0]$          &--                     &$[\max(0.85,0.98-0.02\times\mll),~1.0]$ &--                   \\
\bottomrule
\bottomrule
\end{tabular}
}
\label{table:signalreg:ewkinocuts}
\end{table}
 
The \pt{} threshold for the subleading lepton is defined with sliding cuts that retain efficiency for soft leptons from
low-$\dm$ signals, while reducing backgrounds from FNP leptons in events with larger values of \mll.
The sliding requirement was optimized using a significance metric separately in each SR, considering signal models with a variety of masses and mass splittings. The significance was calculated following the profile likelihood method of Ref.~\cite{2008NIMPA.595..480C}, under the assumption the observation in each SR matches the expected number of signal plus background events.
 
The transverse mass of the leading lepton and \met{} is defined as
$\mtlone =\sqrt{2(E_\mathrm{T}^{\ell_1}\met -\mathbf{p}_\mathrm{T}^{\ell_1}\cdot\ptmiss)}$ and is used in
the SR--E--low and SR--E--high regions to reduce contributions from fake and nonprompt leptons.
 
In events with high-\pt{} ISR jets, the axis of maximum back-to-back \pt{}, referred to here as the thrust axis,
approximates the direction of the recoil of the ISR activity against the sparticle pair.
The recursive jigsaw reconstruction (RJR) technique \cite{Jackson:2017gcy} is used to divide each event into two
hemispheres perpendicular to the thrust axis: a supersymmetric-particles hemisphere $S$, expected to contain the decay
products of the electroweakinos or slepton pair and therefore the \met{}; and an ISR hemisphere, containing hadronic activity.
This bisection allows the calculation of two discriminating variables that are useful in isolating events with
ISR-induced \met{} topologies: $\RISR$, the ratio of the \met{} to the transverse momentum of the ISR system, and $\MTS$,
the transverse mass of the $S$ system.
The \RISR{} variable in particular is sensitive to the mass splitting, with values near 1.0 for the most compressed SUSY events.
Figure~\ref{fig:signalreg:RISR} shows the relationship between \RISR{} and \mll{} and \mtth, which is exploited in
SR--E--high and SR--S--high (\mtth\ and SR--S--high are defined below) through sliding requirements on \RISR.

\begin{figure}[tbp]
\centering
\includegraphics[width=0.49\columnwidth]{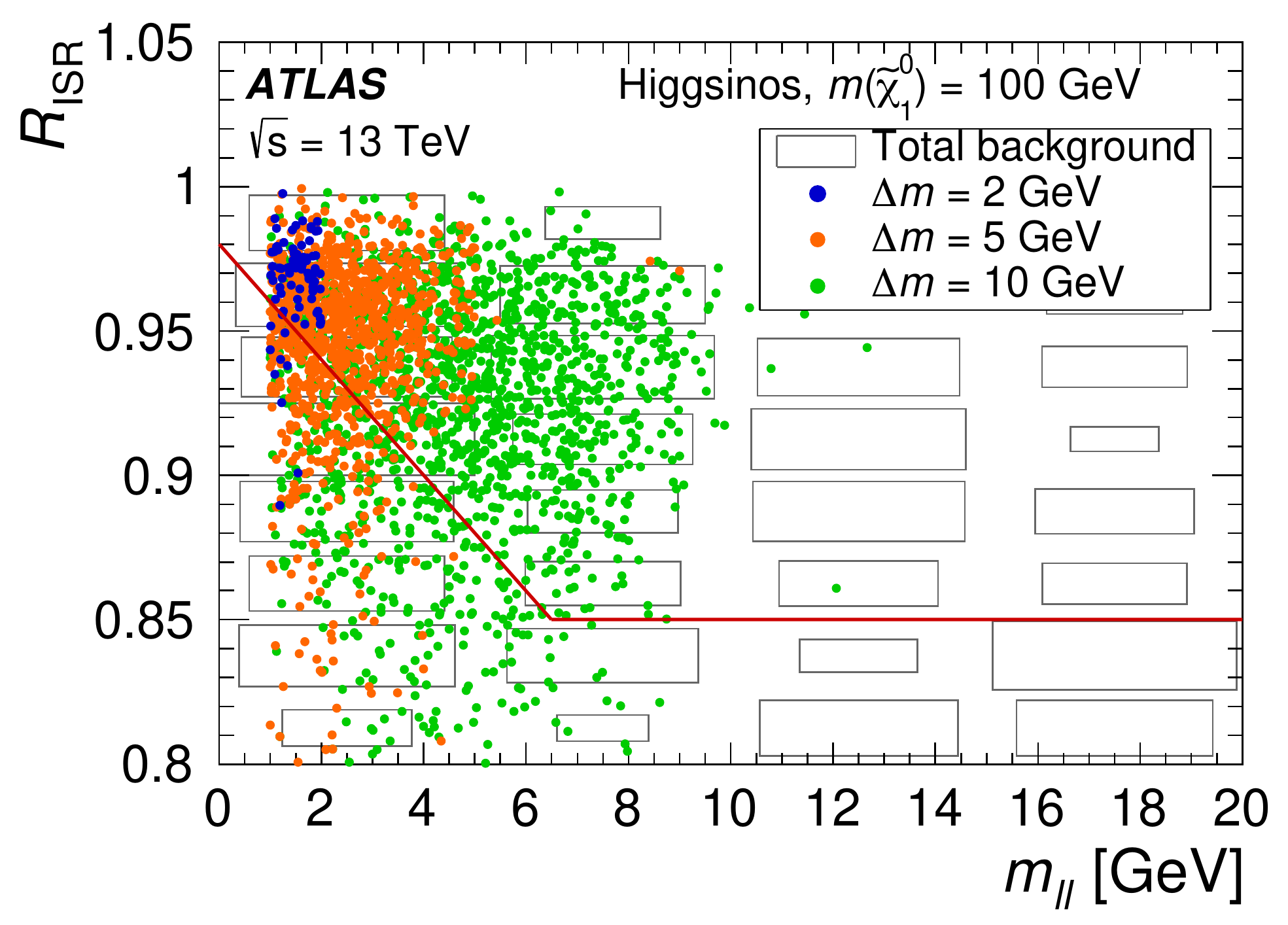}
\includegraphics[width=0.49\columnwidth]{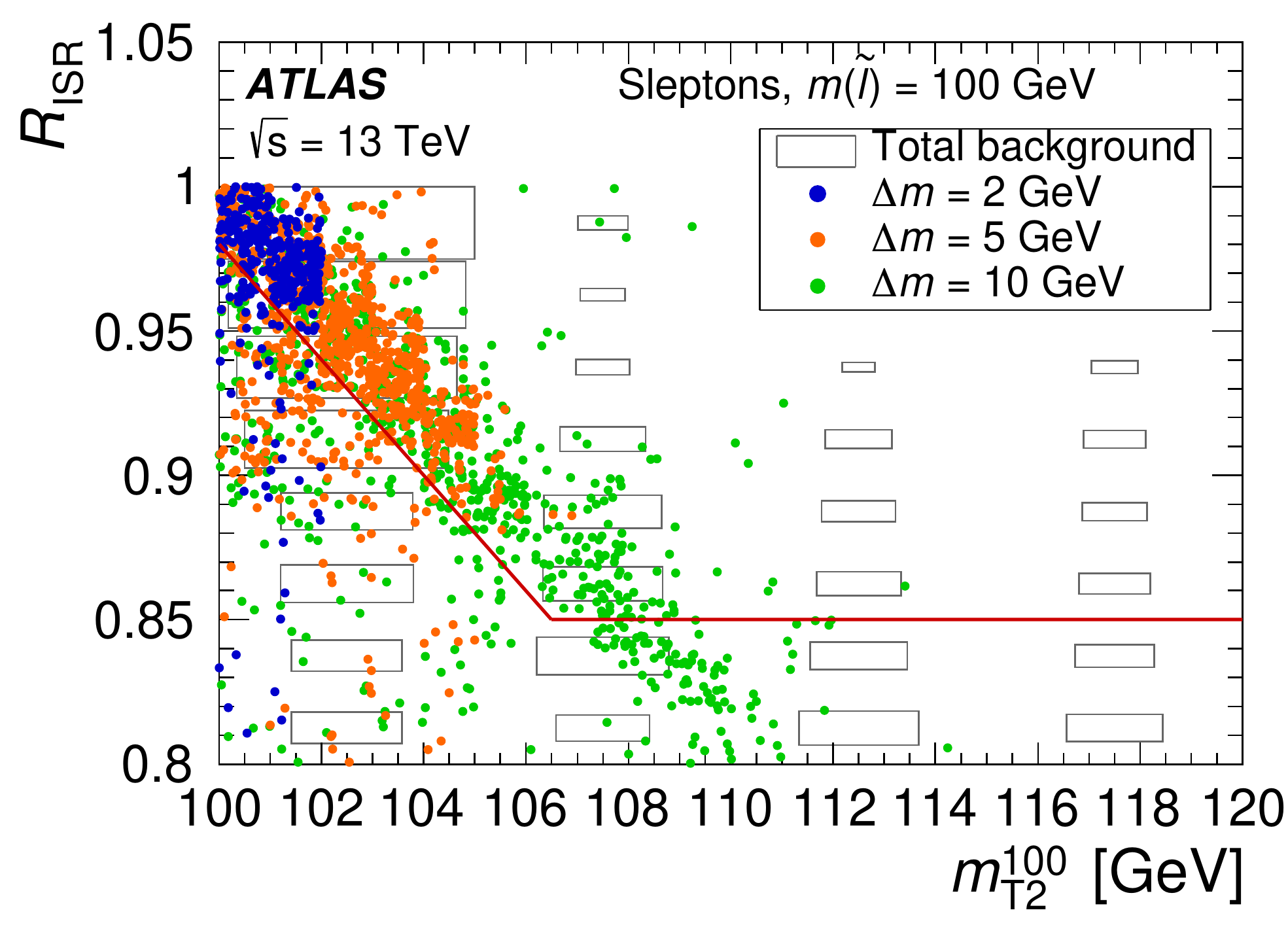}\\
\caption{Distributions of \RISR, the ratio of the \met{} to the transverse momentum of the hadronic ISR activity, for the electroweakino (left) and slepton (right) high-\met{} SRs. Distributions are shown after applying all signal selection criteria except those on \RISR.  The solid red line indicates the requirement applied in the signal region; events in the region below the red line are rejected. Representative benchmark signals for the higgsino (left) and slepton (right) simplified models are shown as circles. The gray rectangular boxes show the distribution of the total background prediction, which is primarily composed of top-like processes, diboson processes, and events with fake/nonprompt leptons.  Regions at larger \mll{} and \mtt{} are not populated by the representative signals shown here, but are useful probes of less-compressed signal models.}
\label{fig:signalreg:RISR}
\end{figure}
 
The \metoverhtlep{} variable, where \htlep{} is the scalar sum of the \pt\ of the two leptons, has been shown to be an
effective discriminant for SUSY signals~\cite{Aaboud:2017leg}.
The two low-\met\ electroweakino SRs are made orthogonal by requiring $\metoverhtlep>10$ for SR--E--med,
where \htlep{} is typically smaller for the SUSY signal, and $\metoverhtlep<10$ for SR--E--low, where \htlep{}
increases due to the larger mass splitting.
 
The \oneleponetrack{} channel targets SUSY signals with especially low values of \dm, which produce decay products with very low momentum.
The signal region SR--E--\oneleponetrack{} therefore requires that the identified lepton has $\pt<10~\GeV$ and that the track
has $\pt<5~\GeV$.
The lepton is also required to be within 1.0 radians of the \ptmiss{} in $\phi$, to reduce backgrounds with tracks associated with
nonprompt leptons or hadrons.
Finally, the SR--E--\oneleponetrack{} region requires $\metoverhtlep>30$, where in this case \htlep{} is the scalar sum of
lepton and track \pT, again exploiting the low values of \htlep{} expected for signal models with small mass splittings.
 
After all selection criteria are applied, the higgsino model with $m(\chitz) = 110~\GeV$ and $m(\chioz) = 100~\GeV$ has an
acceptance times efficiency of $1.1\times10^{-4}$ in the union of all SR--E regions.

Signal regions designed for sensitivity to electroweakinos produced through VBF are defined in Table~\ref{table:signalreg:vbfcuts}
and denoted SR--VBF.
VBF production is commonly characterized by the presence of two energetic jets with large dijet invariant mass and large
separation in pseudorapidity.
Two regions are constructed, distinguished by the pseudorapidity gap between the two leading jets: events with $2<\vbfdeta<4$ are
tested in SR--VBF--low, while events with $\vbfdeta>4$ are tested in SR--VBF--high.
The \met{} is required to be greater than 150~\GeV, which increases the acceptance relative to an $\met>200~\GeV$ requirement
while not introducing significant additional backgrounds.
Additional requirements on \ptltwo, \mtlone, and \metoverhtlep{} similarly reduce backgrounds for small losses in signal efficiency.
The \RVBF{} variable is constructed similarly to \RISR, with the vector sum of the two leading VBF jets in \RVBF{} taking
the place of the ISR system in \RISR.
Additionally, in the case that an energetic jet is well separated from the two leading VBF jets, this jet is added to the decay tree.
This forms an effective third-jet veto by altering the decay hemisphere, spoiling the back-to-back configuration in QCD-initiated
events, while in signal events the central hadronic activity is expected to be suppressed.
The \RVBF{} variable is also sensitive to the mass splitting,
so sliding requirements on \RVBF{} are used in both VBF SRs. The acceptance times efficiency of higgsinos with $m(\chitz) = 100~\GeV$ and $m(\chioz) = 95~\GeV$ produced through VBF in the SR--VBF is $2.9\times10^{-4}$.
 
\begin{table}[tbp]
\centering
\caption{Requirements applied to all events entering into signal regions used for searches for electroweakinos produced through VBF. The $2\ell$ preselection requirements from Table~\ref{table:signalreg:preselectioncuts} are implied.}
\begin{tabular}{l l l}
\toprule
\toprule
Variable       &\multicolumn{2}{l}{VBF SR Requirements}                           \\
\midrule
\mll~[\GeV]    & \multicolumn{2}{l}{$< 40$}          \\
Number of jets & \multicolumn{2}{l}{$\geq 2$}          \\
\ptjtwo~[\GeV] &\multicolumn{2}{l}{$>40$}                                     \\
\met~[\GeV]    &\multicolumn{2}{l}{$>150$}                                    \\
\met/\htlep    &\multicolumn{2}{l}{$>2$}                                      \\
\ptltwo~[\GeV] &\multicolumn{2}{l}{$>\min(10, 2 + \mll/3)$}                   \\
\mtlone~[\GeV] &\multicolumn{2}{l}{$<60$}                                     \\
\RVBF          &\multicolumn{2}{l}{$[\max(0.6,0.92-\mll/60),~1.0]$}           \\
\vbfetasign    &\multicolumn{2}{l}{$<0$}                                      \\
\mjj~[\GeV]	   &\multicolumn{2}{l}{$>400$}                                    \\
\vbfdeta	   &\multicolumn{2}{l}{$>2$}                                    \\
\midrule
 
&SR--VBF--low                                             & SR--VBF--high\\
\midrule
\vbfdeta	   &$<4$ 				                        &$>4$       \\
 
\bottomrule
\bottomrule
\end{tabular}
\label{table:signalreg:vbfcuts}
\end{table}

The SRs designed to provide sensitivity for slepton production, denoted SR--S, are defined in Table~\ref{table:signalreg:sleptoncuts}.  The slepton search exploits the relationship between the mass splitting and the lepton and \met{} kinematics via the stransverse mass (\mtt) variable~\cite{Lester:1999tx, Barr:2003rg}.  The stransverse mass is defined as:
\newcommand{\ptvec}{\mathbf{p}_\mathrm{T}}
\newcommand{\qtvec}{\mathbf{q}_\mathrm{T}}
\begin{linenomath*}
\begin{equation*}
\mtt^{m_\chi} \left(\ptvec^{\ell_1},\ptvec^{\ell_2},\ptvec^\mathrm{miss}\right)
= \underset{\qtvec}{\min}
\left(\max\left[
\mt\left(\ptvec^{\ell_1},\qtvec,m_\chi\right),
\mt\left(\ptvec^{\ell_2},\ptmiss-\qtvec,m_\chi\right)
\right]\right),
\end{equation*}
\end{linenomath*}
where $m_\chi$ is the hypothesized mass of the invisible particles,
and the transverse momentum vector $\qtvec$ with magnitude $q_\text{T}$ is chosen to minimize the larger of the two transverse masses, defined by
\begin{linenomath*}
\begin{equation*}
\mt\left(\ptvec,\qtvec,m_\chi\right)
= \sqrt{ m_\ell^2 + m_\chi^2
+ 2\left(\sqrt{\pt^2 + m_\ell^2} \sqrt{q_\mathrm{T}^2 + m_\chi^2} -
\ptvec \cdot \qtvec
\right)
}.
\end{equation*}
\end{linenomath*}
 
For signal events  with slepton mass $m(\slepton)$ and LSP mass $m(\chioz)$, the values of $\mtt^{m_\chi}$ are bounded
from above by $m(\slepton)$ when $m_\chi$ is equal to $m(\chioz)$
. The stransverse mass with $m_\chi = 100$~\GeV, denoted $\mtth$, is used in this paper.
The chosen value of 100~\GeV{} is based on the expected LSP masses of the signals studied.
The distribution of $\mtth$ does not vary significantly for the signals considered in which $m(\chioz) \neq 100~\GeV$.
 
The SR--S--low slepton region requires events with $150~\GeV{}<\met<200~\GeV$, while the SR--S--high region requires events with
$\met>200~\GeV$.
The SR--S--low region contributes most significantly for signals with $\dm\gtrsim 10~\GeV$, where the leptons satisfy
the \pt{} thresholds without needing a significant additional boost from ISR jets.
Both regions are constructed with sliding requirements on \ptltwo{} following the strategy for the electroweakino regions above.
The requirements on \RISR{} are looser in the SR--S--low region, targeting less compressed scenarios.
The SR--S--high region uses a sliding requirement on \RISR{} to maintain sensitivity to the most compressed scenarios
while reducing backgrounds for events with larger \mtth.
After all selection criteria are applied, the slepton model with $m(\slepton) = 100~\GeV$ and $m(\chioz) = 90~\GeV$ has an
acceptance times efficiency of $2.5\times10^{-3}$ when considering both SR--S regions.
Acceptances and efficiencies for left- and right-handed sleptons are consistent with each other for all slepton scenarios under study.
\begin{table}[tbp]
\centering
\caption{Requirements applied to all events entering into signal regions used for slepton searches. The $2\ell$ preselection requirements from Table~\ref{table:signalreg:preselectioncuts} are implied.}
\begin{tabular}{l l l}
\toprule
\toprule
&\multicolumn{2}{c}{Slepton SR Requirements}                                                         \\
\cline{2-3}\\[-1.0em]
Variable              &SR--S--low                                   &SR--S--high                                   \\
\midrule
\met~[\GeV]           &$[150, 200]$                                 &$>200$                                 \\
\mtth~[\GeV]           &$<140$                                       &$<140$  \\
 
\ptltwo~[\GeV]        &$>\min(15, 7.5 + 0.75\times(\mtt-100))$      &$>\min(20, 2.5+ 2.5\times(\mtt-100))$  \\
\RISR                 &$[0.8,~1.0]$                                 &$[\max(0.85,0.98-0.02\times(\mtt-100)),~1.0]$   \\
\bottomrule
\bottomrule
\end{tabular}
\label{table:signalreg:sleptoncuts}
\end{table}
 
After all selection requirements are applied, the SR--E and the SR--VBF regions are binned in \mll{}, with bin boundaries
at $\mll=1, 2, 3, 5, 10, 20, 30, 40$, and $60~\GeV{}$ for the $2\ell$ channels, and
at $m_{\ell\text{track}}=0.5, 1, 1.5, 2, 3, 4$, and $5~\GeV{}$ for the \oneleponetrack{} channel.
Events in the SR--E--med region with $\mll>30~\GeV{}$ have minimal sensitivity to the electroweakino signals studied and
are not considered. Similarly, events in the SR--E--\oneleponetrack{} with $m_{\ell\text{track}}>5~\GeV{}$ are discarded.
The slepton SR--S regions are instead binned in \mtth, with bin boundaries at $\mtth=100, 100.5, 101, 102, 105, 110, 120, 130$,
and $140~\GeV{}$.
Events with $\mtth$ above $140~\GeV$ have minimal sensitivity to compressed sleptons and are not considered in any of the regions.
Events with $\mll$ above $60~\GeV$ are rejected in preselection for all channels.
 
The binned \mll{} and \mtth{} distributions are used in two different types of statistical tests.
The first test is a search for excesses with minimal model dependence, in which any given fit considers a single inclusive SR.
An inclusive electroweakino SR is constructed by merging all SR--E--high, SR--E--med, SR--E--low, and SR--E--\oneleponetrack{} bins
below a $\mll$ bin boundary listed above, with each $2\ell$ electroweakino bin boundary corresponding to an inclusive SR.
Similarly, the inclusive slepton regions are constructed by merging all SR--S--high and SR--S--low bins below the \mtth{} bin
boundaries.
The inclusive VBF SRs are also constructed by merging the SR--VBF--low and SR--VBF--high bins below the $\mll$ boundaries.
Additional inclusive VBF SRs are defined using events in SR--VBF--high only.
 
The second type of test is referred to as an exclusion fit, which considers all relevant bins separately in the likelihood.
Dielectron and dimuon events in the $2\ell$ electroweakino SRs and in the slepton SRs are also fitted separately in the exclusion fits.
 

\section{Background estimation}
\label{sec:backgrounds}
The sources of SM background in regions with two leptons can be subdivided into two categories: reducible backgrounds from
events where at least one of the candidate leptons is FNP, and irreducible backgrounds from events that contain two prompt leptons.

Since MC simulation is not expected to model processes with FNP leptons accurately, a data-driven method, referred to as the
Fake Factor method~\cite{STDM-2011-24,ATL-PHYS-PUB-2010-005}, is employed to estimate these backgrounds. 
The yields obtained from this procedure are cross-checked in validation regions (VRs), which are not used to constrain the fit and are orthogonal in selection to the CRs and SRs.

The dominant sources of irreducible background are $\ttbar/tW$, $WW/WZ$, and $\Ztt$.
These backgrounds are estimated using MC simulations normalized to data in dedicated CRs.
Events originating from the production of a Drell--Yan lepton pair, triboson, Higgs boson or top quarks in association with gauge bosons
constitute a small fraction of the total background.
Their contributions in the regions with two leptons are estimated using the MC samples listed in Table~\ref{table:samples:BGs}.
Additional VRs
are used to validate the extrapolation of background in the fitting procedure within the same kinematic regime as the SRs.
 
The definitions of the CRs and VRs used in the electroweakino, VBF and slepton searches are summarized in
Tables~\ref{table:crs:vrs:ewkinocuts}, \ref{table:crs:vrs:vbfcuts} and~\ref{table:crs:vrs:sleptoncuts}, respectively. The VRSS regions are further described in Section~\ref{sec:2lfakes}, in the context of the FNP background estimation, while the remaining CRs and VRs are explained in Section~\ref{sec:2lcrs}.
 
 
\begin{table}[tbp]
\centering
\caption{Definition of control (``CR'' prefix) and validation (``VR'' prefix) regions used for background estimation in the electroweakino search, presented relative to the definitions of the corresponding signal regions SR--E--high, SR--E--med and SR--E--low. The $2\ell$ preselection criteria from Table~\ref{table:signalreg:preselectioncuts} and selection criteria from Table~\ref{table:signalreg:ewkinocuts} are implied, unless specified otherwise. }
\resizebox{\linewidth}{!}{
\renewcommand{\arraystretch}{1.1}
\begin{tabular}{llll}
\toprule
\toprule
Region                                & SR orthogonality                                   & Lepton Flavor
& Additional requirements \\
\midrule
CRtop--E--high             & \multirow{2}{*}{$\nbtagtwenty \geq 1$}    & \multirow{2}{*}{$ee+\mu\mu+e\mu+\mu e$}
& $\RISR \in [0.7,1.0]$, $\mtlone$ removed \\
CRtop--E--low   &                                                    &
& $\met/\HT^\text{lep}$ and $\mtlone$ removed \\
\midrule
CRtau--E--high             & \multirow{3}{*}{$m_{\tau\tau} \in [60, 120]~\GeV$} & \multirow{3}{*}{$ee+\mu\mu+e\mu+\mu e$}
& $\RISR \in [0.7,1.0]$, $\mtlone$ removed \\
CRtau--E--low   &                                                    &
& $\RISR \in [0.6,1.0]$, $\mtlone$ removed \\
VRtau--E--med    &                                                    &
& -- \\
\midrule
CRVV--E--high             &  $\RISR \in [0.7,0.85]$                               & \multirow{2}{*}{$ee+\mu\mu+e\mu+\mu e$}
& $\mtlone$ removed \\
CRVV--E--low   &  $\RISR \in [0.6,0.8]$                                &
& $\mtlone>30~\GeV$, $N_\mathrm{jets} \in [1,2]$, $\met/\HT^\text{lep}$ removed \\
\midrule
VRSS--E--high              & \multirow{3}{*}{Same sign $\ell^\pm \ell^\pm$}     & \multirow{3}{*}{$ee+\mu e,~\mu\mu+e\mu$}
& $\RISR \in [0.7,1.0]$, $\mtlone$ and $\pt^{\ell_2}$ removed \\
VRSS--E--low    &                                                    &
& $\met/\HT^\text{lep}$, $\mtlone$ and $\pt^{\ell_2}$ removed \\
VRSS--E--med     &                                                    &
& -- \\
\midrule
VRDF--E--high               & \multirow{3}{*}{$e \mu + \mu e$}                   & \multirow{3}{*}{$e\mu+\mu e$}
& -- \\
VRDF--E--low     &                                                    &
& -- \\
VRDF--E--med      &                                                    &
& -- \\
\bottomrule
\bottomrule
\end{tabular}}
\label{table:crs:vrs:ewkinocuts}
\end{table}

 
\begin{table}[tbp]
\centering
\caption{Definition of control (``CR'' prefix) and validation (``VR'' prefix) regions used for background estimation in the search for electroweakinos produced through VBF, presented relative to the definitions of the corresponding signal regions SR--VBF--high and SR--VBF--low. The $2\ell$ preselection criteria from Table~\ref{table:signalreg:preselectioncuts} and selection criteria from Table~\ref{table:signalreg:vbfcuts} are implied, unless specified otherwise.}
\resizebox{\linewidth}{!}{
\renewcommand{\arraystretch}{1.1}
\begin{tabular}{llll}
\toprule
\toprule
Region       & SR orthogonality                  & Lepton Flavor           & Additional requirements \\
\midrule
CRtop--VBF   & $\nbtagtwenty \geq 1$    & $ee+\mu\mu+e\mu+\mu e$  & $\RVBF$ and $\mtlone$ removed \\
\midrule
CRtau--VBF   & $m_{\tau\tau} \in [60, 120]~\GeV$ & $ee+\mu\mu+e\mu+\mu e$  & $\met/\HT^\text{lep} \in [2,10]$, $\RVBF$ and $\mtlone$ removed \\
\midrule
VRSS--VBF    & Same sign $\ell^\pm \ell^\pm$     & $ee+\mu e,~\mu\mu+e\mu$ & $\RVBF$, $\mtlone$  and $\pt^{\ell_2}$ removed \\
\midrule
VRDF--VBF--low    & $e \mu + \mu e$                   & $e\mu+\mu e$            & -- \\
VRDF--VBF--high    & $e \mu + \mu e$                   & $e\mu+\mu e$            & -- \\
\bottomrule
\bottomrule
\end{tabular}
}
\label{table:crs:vrs:vbfcuts}
\end{table}
 
\begin{table}[tbp]
\centering
\caption{Definition of control (``CR'' prefix) and validation (``VR'' prefix) regions used for background estimation in the slepton search, presented relative to the definitions of the corresponding signal regions SR--S--high and SR--S--low. The $2\ell$ preselection criteria from Table~\ref{table:signalreg:preselectioncuts} and selection criteria from Table~\ref{table:signalreg:sleptoncuts} are implied, unless specified otherwise. }
\renewcommand{\arraystretch}{1.1}
\begin{tabular}{llll}
\toprule
\toprule
Region                                 & SR orthogonality                                   & Lepton Flavor
& Additional requirements \\
\midrule
CRtop--S--high             & \multirow{2}{*}{$\nbtagtwenty \geq 1$}    & \multirow{2}{*}{$ee+\mu\mu+e\mu+\mu e$}
& $\RISR \in [0.7,1.0]$ \\
CRtop--S--low             &                                                  &
& -- \\
\midrule
CRtau--S--high             & \multirow{2}{*}{$m_{\tau\tau} \in [60, 120]~\GeV$} & \multirow{2}{*}{$ee+\mu\mu+e\mu+\mu e$}
& $\RISR \in [0.7,1.0]$ \\
CRtau--S--low              &                                                    &
& $\RISR \in [0.6,1.0]$\\
\midrule
CRVV--S--high              & $\RISR \in [0.7,0.85]$                               & \multirow{2}{*}{$ee+\mu\mu+e\mu+\mu e$}
& -- \\
CRVV--S--low               & $\RISR \in [0.6,0.8]$                                &
& $\mtlone>30$, $N_\mathrm{jets} \in [1,2]$ \\
\midrule
VRSS--S--high              & \multirow{2}{*}{Same sign $\ell^\pm \ell^\pm$}     & \multirow{2}{*}{$ee+\mu e,~\mu\mu+e\mu$}
& $\RISR \in [0.7,1.0]$, $\pt^{\ell_2}$ removed \\
VRSS--S--low              &                                                    &
& $\pt^{\ell_2}$ removed \\
\midrule
VRDF--S--high              & \multirow{2}{*}{$e \mu + \mu e$}                   & \multirow{2}{*}{$e\mu+\mu e$}
& -- \\
VRDF--S--low               &                                                    &
& -- \\
\bottomrule
\bottomrule
\end{tabular}
\label{table:crs:vrs:sleptoncuts}
\end{table}

 
The dominant source of background in the \oneleponetrack{} channel is combinatorial, from events containing one prompt lepton
and one random track, and is collectively estimated using data, as described in Section~\ref{sec:1ltbkg}.
 
\subsection{Reducible background in regions with two leptons}
\label{sec:2lfakes}
The FNP lepton background arises from jets misidentified as leptons, photon conversions, or semileptonic decays of heavy-flavor hadrons.
Studies based on simulated samples indicate that the last is the dominant component in the SRs with two leptons.
The contamination of the SRs by FNP lepton background is large at low values of \mll\ and \mtth, and decreases at the upper end of the distributions.

In the Fake Factor method, a two-lepton control sample is defined in data using leptons with modified signal lepton requirements.
At least one of the leptons, labeled as anti-ID, is required to fail one or more of the requirements applied to signal leptons, but is required to satisfy less restrictive requirements.
The other lepton can either meet all signal lepton requirements, in which case it is labeled as ID, or satisfy the anti-ID requirements.
This sample is enriched in FNP lepton backgrounds and is therefore referred to as the FNP control sample.
The contributions from processes with two prompt leptons in the FNP control sample are subtracted using simulated samples.
MC studies indicate that the leptons in the FNP control sample arise from processes similar to those for FNP leptons
passing the SR selections.
The FNP lepton background prediction in a given region is obtained by applying all selection requirements of that region to
the FNP control sample and scaling each event by a weight assigned to each anti-ID lepton, referred to as the fake factor.  Events in the FNP control sample containing a single anti-ID lepton have positive fake factors.  Events with two anti-ID leptons receive a weight corresponding to the product of the weights for the two anti-ID leptons, and enter with opposite sign to correct for events with two FNP leptons.
 
The fake factor is measured in a data sample collected with prescaled low-\pt\ single-lepton triggers. This sample is dominated
by multijet events with FNP leptons and is referred to as the measurement sample.
A selection of $\mtlone<40$~\GeV{} is applied to reduce the contributions from processes with prompt leptons in the measurement sample.
The contributions from these processes are subtracted using MC simulation, with negligible impact on the measured fake factors.
 
To enrich the sample in FNP leptons similar to those contaminating the SRs, the leading-jet \pt\ is required to be greater than 100~\GeV.
The fake factors are calculated as the ratio of ID to anti-ID leptons in the measurement sample, measured in bins of lepton \pt,
separately for electrons and muons.
The fake factors are also found to have a dependence on the number of $b$-tagged jets in the events.
Different fake factors are therefore computed in events with and without $b$-tagged jets.
 
The yields predicted by the Fake Factor method are cross-checked in dedicated VRs enriched in FNP lepton backgrounds, labeled VRSS.
As summarized in Tables~\ref{table:crs:vrs:ewkinocuts}, \ref{table:crs:vrs:vbfcuts} and~\ref{table:crs:vrs:sleptoncuts},
a dedicated VRSS is constructed for each SR by selecting events with two leptons with the same electric charge.
The kinematic requirements applied to each VRSS are mostly the same as the ones used in the corresponding SR, ensuring the FNP
lepton processes are similar in the two regions.
To guarantee high purity in FNP lepton background, the selection criteria designed to suppress these processes in the SRs,
such as the sliding cut on the \pt\ threshold of the subleading lepton, are loosened or removed in each VRSS.
The contribution of FNP background in the VRSS regions is typically above 91\%, with the remaining backgrounds originating
from $VV$ processes with two prompt leptons of the same electric charge.
The signal contamination is at most 14\%.
 
\subsection{Irreducible background in regions with two leptons}
\label{sec:2lcrs}
 
Several CRs are defined for the electroweakino, VBF and slepton searches, and are used to normalize the MC simulations
of $\ttbar/tW$ and $\Ztt$ background processes to the data in a simultaneous fit also including the SRs, as described in
Section~\ref{sec:result}.
In searches for electroweakinos and sleptons recoiling against ISR, CRs are also constructed to normalize the $WW/WZ$ background.
The event rates in the SRs are predicted by extrapolating from the CRs using the simulated MC distributions.
This extrapolation is validated using events in dedicated VRs.
 
The CRs are designed to be statistically disjoint from the SRs, to be enriched in a particular background process, to have minimal
contamination from the signals considered, and to exhibit kinematic properties similar to the SRs.
The CRs labeled as CRtop are defined by selecting events with at least one $b$-tagged jet.
The CRtop regions have purities ranging from 83\% to 94\% in processes with top quarks, and are used to constrain the
normalization of the $\ttbar$ and $tW$ processes with dilepton final states.
The CRtau regions, which are enriched in the $\Ztt$ process with purities of at least 75\%, are constructed by selecting events
satisfying $\mtautau \in [60,120]~\GeV$. 
Finally, the $\RISR$ selection used to define the SRs is modified to construct CRs enriched in $WW$ and $WZ$ processes, denoted CRVV.
In these CRs, $41\%$--$45\%$ of the events are $VV$ events.
 
The $\ttbar/tW$, $WW/WZ$ and $\Ztt$ processes containing two prompt leptons all yield same-flavor lepton pairs ($ee$ and $\mu\mu$)
at the same rate as for different-flavor pairs ($e\mu$ and $\mu e$, where the first lepton is the leading lepton).
This feature is used to enhance the statistical constraining power of the CRs, by selecting events with all possible flavor
assignments ($ee$, $\mu\mu$, $e\mu$, and $\mu e$).
It is also used to define additional VRs, denoted VRDF. One VRDF is defined for each $2\ell$ SR by requiring two different-flavor
leptons ($e\mu$ and $\mu e$), but otherwise keeping the same kinematic selections as the corresponding SR.
The relative fractions of each background process are similar in the SR and the corresponding VRDF.
The signal contamination in the VRDF regions is at most 16\%, originating from $\chiop\chiom$ or $\chitz\chiopm$ higgsino events
decaying fully leptonically.
 
In the search for electroweakinos recoiling against ISR, six single-bin CRs are defined as summarized in
Table~\ref{table:crs:vrs:ewkinocuts}.
Three CRs, labeled CR--E--high, employ a $\met>200~\GeV$ selection and are used to constrain the normalization
of $\ttbar/tW$, $WW/WZ$ and $\Ztt$ backgrounds in SR--E--high.
To minimize the impact of the mismodeling of the trigger efficiency in the simulation, three additional CRs, labeled CR--E--low, are
defined by selecting events with $\MET \in [120,200]~\GeV$.
These CRs are used to normalize the same background processes in SR--E--low.
Events with FNP leptons entering the CRs are suppressed using the same sliding cut on \ptltwo\ as the corresponding SRs.

The dominant source of irreducible background in the SR--E--med region is the \Ztt\ process.
It is difficult to construct a dedicated CR with enough events to constrain the normalization of the \Ztt\ background
in the SR--E--med region. The CRtau--E--low region is therefore used for this purpose.
The extrapolation from CRtau--E--low to SR--E--med is tested in an additional VR, labeled VRtau--E--med, defined by
selecting events with $\mtautau \in [60,120]~\GeV$, but otherwise applying the same kinematic selections as in the SR--E--med region,
as summarized in Table~\ref{table:crs:vrs:ewkinocuts}.
 
Two control regions are defined for the VBF search, as summarized in Table~\ref{table:crs:vrs:vbfcuts}.
The CRtop--VBF and the CRtau--VBF regions are designed with a $\vbfdeta > 2$ requirement, and are used to constrain the
normalizations of the $\ttbar/tW$ and $\Ztt$ processes in both the SR--VBF--low and the SR--VBF--high regions.
The number of events in the VBF CRs is increased by removing the \mtlone\ and \RVBF\ selections used in the SRs.
 
Six CRs are used to normalize the $\ttbar/tW$, $WW/WZ$ and $\Ztt$ background processes entering the SR--S--low and
SR--S--high regions, as summarized in Table~\ref{table:crs:vrs:sleptoncuts}.
The CRs used in the search for sleptons are designed similarly to the CRs used in the search for electroweakinos.
One notable difference is the sliding cut on the \ptltwo\ threshold, which is chosen to match the requirements used
in the slepton SRs and therefore depends on \mtth.
 
\subsection{Background in the \oneleponetrack{} signal region}
\label{sec:1ltbkg}
The background in the SR--E--\oneleponetrack{} region is suppressed by requiring that the selected track be associated with a
reconstructed lepton candidate.
Simulation studies show that this background is dominated by events with one prompt lepton and one track from hadrons or
nonprompt leptons.
The MC samples used to model SM processes with two prompt leptons contribute negligibly in the \oneleponetrack{} SR.
 
The amount of background in the \oneleponetrack{} channel is estimated using a data-driven procedure.
A control sample is defined in data with events that satisfy the same selection criteria as the SR--E--\oneleponetrack{} region.
Instead of selecting OS events with one lepton and one track, the lepton and the track in the control sample are required to have
the same electric charge (SS).
The contamination of the SS control sample by signal is negligible.
The data in the SS sample are directly used as the estimate of the background in SR--E--\oneleponetrack{}.
The background estimate assumes that the background events are produced with equal rates for OS and SS events.
This is expected to be the case because the track is randomly selected and its electric charge is not correlated with the
charge of the prompt lepton.
 
The assumption that OS and SS background events are produced with equal rates in the \oneleponetrack{} signal region is tested in simulation using $W+$jets events.
The ratio of OS to SS $W+$jets events was found to be compatible with one, with a statistical uncertainty of 12\% determined
by the size of the MC sample. A VR, denoted VR--\oneleponetrack{}, is constructed to test the assumption using data.
The VR--\oneleponetrack{} is designed using the same kinematic selections as the \oneleponetrack{} SR,
except that $\dphilepmet > 1.5$ is required to ensure that the samples are disjoint.
The upper bound on $\Delta R_{\ell\mathrm{track}}$ used in SR--E--\oneleponetrack{} is removed to reduce the signal contamination,
and the $\met/\htlep$ requirement is loosened to $\met/\htlep>15$ to increase the number of events in the VR.
The kinematic distributions of the SS and OS data events in the VR--\oneleponetrack{} are compared and found to agree.

\section{Systematic uncertainties}
Systematic uncertainties are evaluated for all background processes and signal samples.
As the predictions for the main SM background processes modeled via MC simulation are normalized to data in dedicated
control regions, the systematic uncertainties only affect the extrapolation to the signal regions in these cases.
 
Figure~\ref{fig:syst:summary}} illustrates the dominant classes of uncertainties in the expected background yields
in the electroweakino, VBF and slepton SRs.
The main sources of experimental uncertainty affect the FNP background predictions obtained with the Fake Factor method.
These systematic uncertainties stem from the size of the FNP control samples, as well as from the size of the measurement
sample used to compute the fake factors.
The uncertainties associated with the subtraction of processes involving prompt leptons in the FNP control samples and
in the measurement sample are estimated from simulation and found to be negligible.
Uncertainties are also assigned to cover the differences in the event and lepton kinematics between the measurement
region and the signal regions.
Moreover, additional uncertainties are computed as the differences between the FNP background predictions and
observed data in the VRSS regions.
 
Other sources of significant experimental systematic uncertainties are the jet energy scale (JES) and resolution (JER).
The jet uncertainties are derived as a function of \pt\ and $\eta$ of the jet, as well as of the pileup conditions and
the jet flavor composition of the selected jet sample.
They are determined using a combination of simulated samples and studies of data, such as measurements of the jet \pt\ balance
in dijet, $Z$+jet and $\gamma$+jet events~\cite{PERF-2016-04}.
The systematic uncertainties related to the modeling of \met\ in the simulation are estimated by propagating the
uncertainties in the energy and momentum scale of each of the objects entering the calculation, as well as the uncertainties
in the soft-term resolution and scale~\cite{ATLAS-CONF-2018-023}.
 
The reconstruction, identification and isolation efficiencies for low-$\pt$ leptons, as well as the momentum resolution and scale,
are measured and calibrated following methods similar to those employed for higher-$\pt$
electrons~\cite{EGAM-2018-01} and muons~\cite{PERF-2015-10}.
The associated systematic uncertainties are in general found to be small.
 
The MC samples simulating the dominant background processes, $\ttbar/tW$, $\Ztt$ and $VV$, are also affected by different
sources of theoretical modeling uncertainty.
The uncertainties related to the choice of QCD renormalization and factorization scales are assessed by varying the
corresponding generator parameters up and down by a factor of two around their nominal values.
Uncertainties in the resummation scale and the matching scale between matrix elements and parton showers for
the $\Ztt$ samples are evaluated by varying up and down by a factor of two the corresponding parameters in \textsc{Sherpa}.
The uncertainties associated with the choice of PDF set, NNPDF~\cite{Ball:2012cx,Ball:2014uwa}, and uncertainty in the
strong coupling constant, $\alpha_\mathrm{s}$, are also considered.
 
As discussed in Section~\ref{sec:backgrounds}, the background predictions in the \oneleponetrack{} SR, selecting OS lepton--track pairs,
are extracted from a SS data control sample.
Two different types of systematic uncertainty are associated with the OS--SS extrapolation.
For $\mlt<2$~\GeV{}, low-mass resonances can cause higher production rates for OS events than for SS events. A 30\% uncertainty is assigned, based on an exponential fit to the OS/SS ratio as a function of \met{} in the $\dphilepmet>1.5$ region. This OS/SS ratio was found to be constant and equal to 1 for $\met>200$~\GeV{}, indicating low-mass resonances do not contribute significantly to the OS sample in the SR--E--\oneleponetrack{} region. The uncertainty is computed as the value of the fitting function at $\met=200$~\GeV{}, where the deviation from unity is largest, summed linearly with the corresponding fit uncertainty.
The $\mlt>2$~\GeV{} region is instead mainly populated by $W+$jets events, in which the correlation between the lepton and the track charge may introduce differences between the SS and OS expectations. A 12\% uncertainty, extracted from $W+$jets simulated events, is assigned.

The $\mlt>2$~\GeV{} region is instead mainly populated by $W+$jets events, in which the correlation between the lepton
and the track charge may introduce differences between the SS and OS expectations.
A 12\% uncertainty, extracted from $W+$jets simulated events, is assigned.
 
\begin{figure}[tbp]
\centering
\includegraphics[width=0.69\columnwidth]{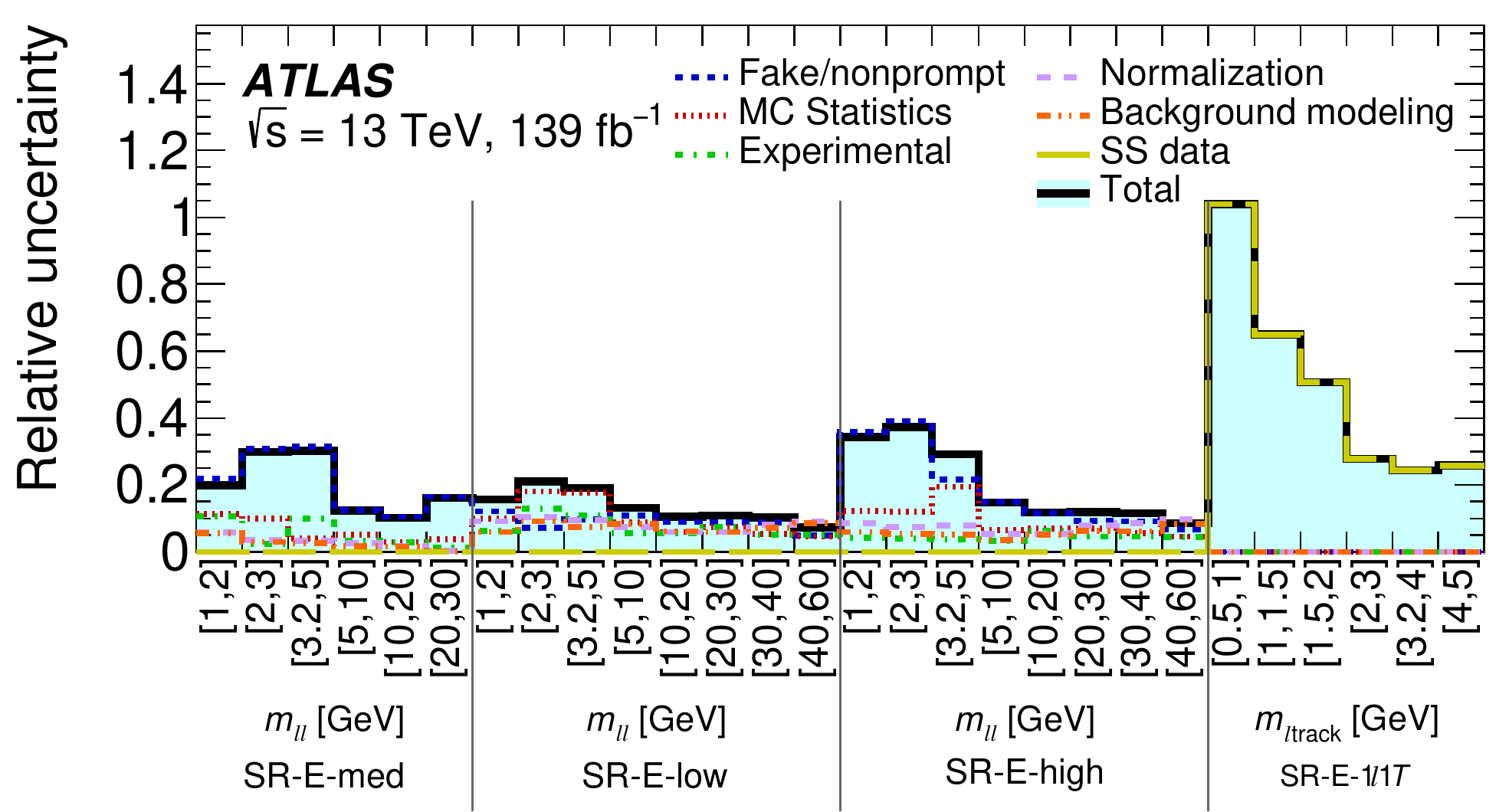}\\
\vspace{0.2cm}
\includegraphics[width=0.69\columnwidth]{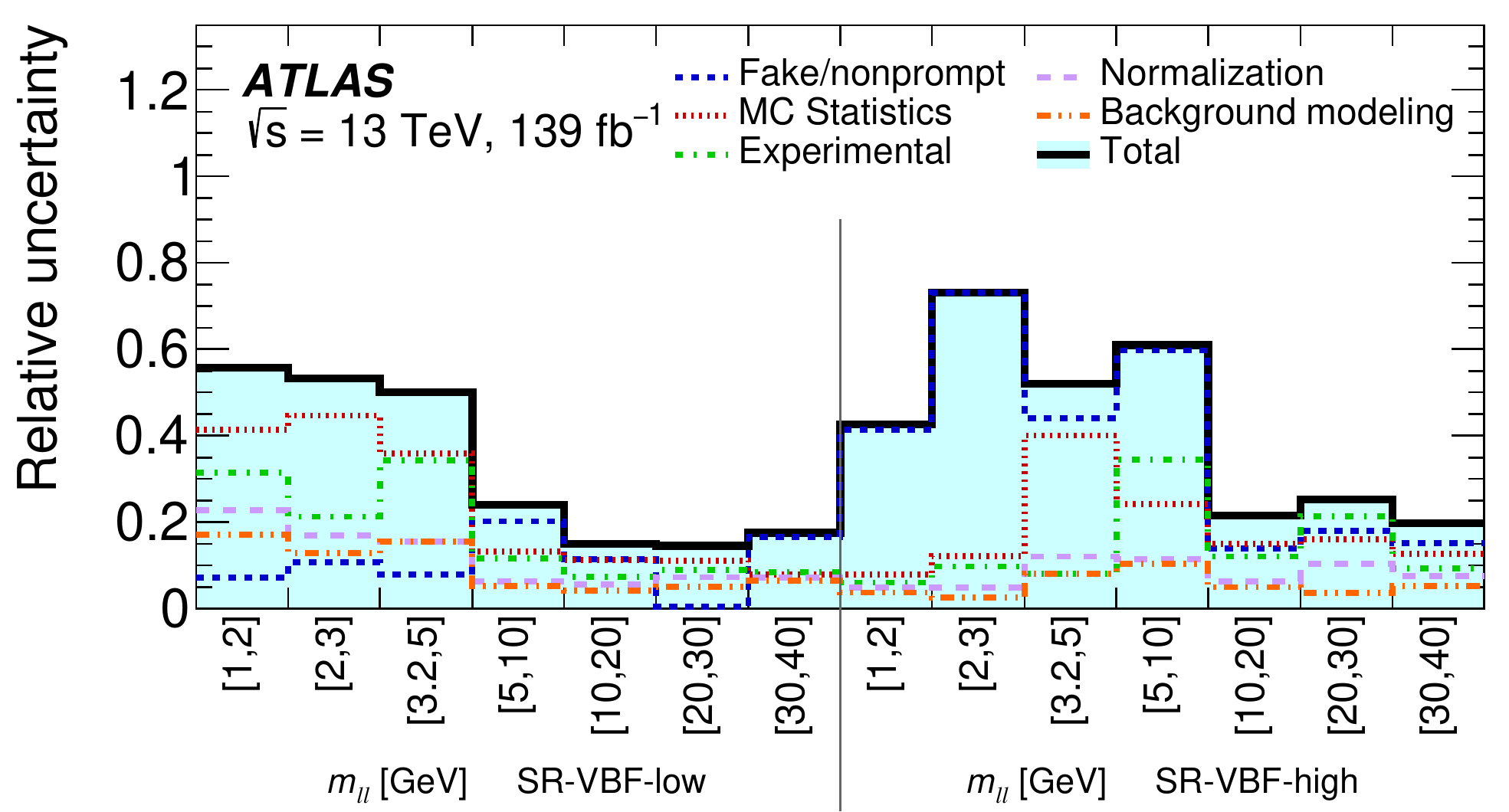}\\
\vspace{0.2cm}
\includegraphics[width=0.69\columnwidth]{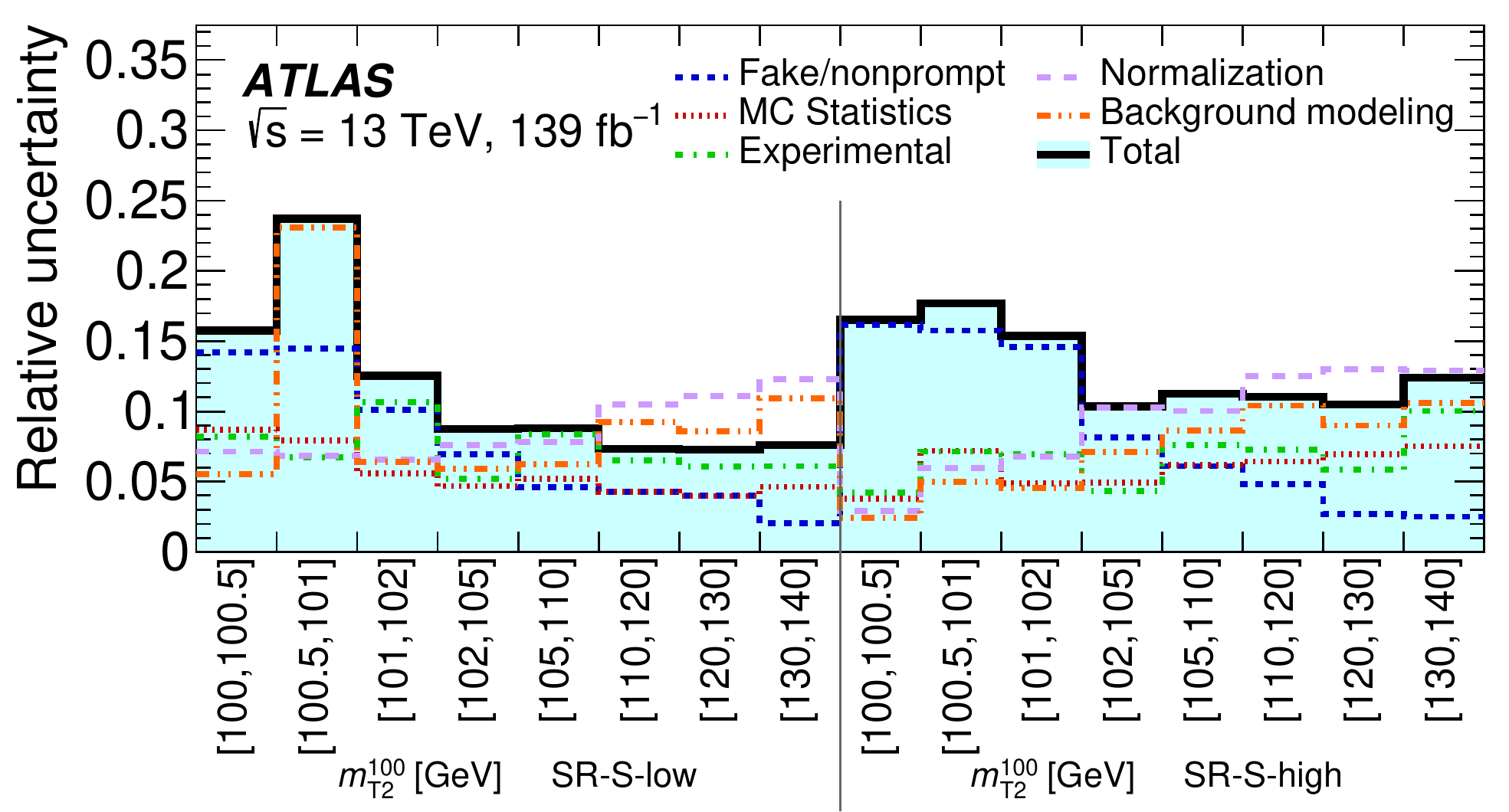}\\
\caption{The relative systematic uncertainties in the fitted SM background as obtained from \SRCRBGonlyfits{} for the electroweakino SRs (top),
VBF SRs (middle),
and slepton SRs (bottom). The uncertainty in the \textit{SS data} includes a statistical component due to the size of the SS data sample used to estimate the background in the SR-E-\oneleponetrack{} region, and a systematic component from the SS--OS extrapolation.
The \textit{MC Statistics} uncertainty originates from the limited size of the MC samples used to model the irreducible background contributions.
The \textit{Normalization} uncertainty arises from the use of CRs to normalize the contributions of $\ttbar/tW$, $\Ztt$ and $WW/WZ$ backgrounds, while \textit{Background modeling} includes the different sources of theoretical modeling uncertainties in the \mll{} or \mtth{} lineshapes for the irreducible backgrounds. All sources of uncertainty affecting the FNP background estimate are included under \textit{Fake/nonprompt}. The uncertainties arising from the reconstruction and selection of signal leptons, jets and $\met$ are included under the \textit{Experimental} category.
The individual uncertainties can be correlated and do not necessarily add up in quadrature to the total uncertainty.}
\label{fig:syst:summary}
\end{figure}

Uncertainties in the expected yields for non-VBF SUSY samples arising from generator modeling are determined \textit{in situ} by
comparing the yields from $Z\to\mu\mu$ events in data with those from $Z(\to\mu\mu)$+jets events generated using the same MG5\_aMC@NLO configuration
as the signal samples.
The muon four-momenta are added to the \met{} to emulate the \pt{} of the SUSY system in signal events, and uncertainties are
derived from observed differences in \met{} between data and simulation.
The largest modeling uncertainties are approximately 20\% for samples with the most compressed mass spectrum and in
high-\met{} channels, while low-\met{} channels and noncompressed signal points have uncertainties ranging from 1\% to 10\%.
Uncertainties in the signal acceptance due to PDF uncertainties are evaluated following the
PDF4LHC15 recommendations~\cite{Butterworth:2015oua} and amount to at most 15\% for large \chitz~or $\widetilde{\ell}$ masses.
Uncertainties in the shape of the $\mll$ or $\mtth$ signal distributions due to the sources above are found to be small,
and are neglected.
 
Uncertainties due to generator modeling in the acceptance of the VBF signal samples are evaluated by varying by a factor of
two the {MG5\_aMC@NLO} parameters corresponding to the renormalization, factorization and CKKW-L matching scales, as well as
the \textsc{Pythia~8} shower tune parameters and $\alpha_\mathrm{s}$.
The largest uncertainties arise from renormalization and factorization scale variations (13\%--22\%), with smaller
contributions from matching and $\alpha_\mathrm{s}$ variations (0.5\%--5\%).
 
Additional uncertainties are assigned to the predictions from signal simulation in the \oneleponetrack{} SR.
An uncertainty in the modeling of the rate for reconstructed tracks that do not match a generated charged particle is accounted for.
It is estimated by comparing the nonlinear component of the per-event track multiplicity as a function of pileup, in data and
simulation.
Furthermore, the calibration procedure applied to MC events to match the track impact parameter resolution in different data-taking
periods is also a source of systematic uncertainty.
Finally, uncertainties are assigned to the track--lepton matching efficiency and the track isolation efficiency, as derived
from the studies of events with a $J/\psi$ meson or $Z$ boson decaying into a lepton and a track, described in Section~\ref{sec:selection}.

\FloatBarrier
\section{Results}
\label{sec:result}
Data in the control regions, validation regions, and signal regions are compared with SM predictions using a profile
likelihood method~\cite{Cowan:2010js} implemented in the \textsc{HistFitter} package~\cite{Baak:2014wma}.
Most systematic uncertainties are treated as nuisance parameters with Gaussian constraints in the likelihood,
apart from those of statistical nature, for which Poisson constraints are used.
Experimental systematic uncertainties are correlated between signal and backgrounds for all regions.
 
\subsection{Control and validation regions}
 
A \CRBGonlyfit{} is constructed using only the control regions to constrain the fit parameters.
The data in the control regions CRtop, CRtau and CRVV are fit simultaneously in each search to constrain overall
normalization factors for the $\ttbar/Wt$, $\Ztt$, and $VV$ background predictions.
The resulting normalization parameters are presented in Table~\ref{table:results:normfactors}.
 

\begin{table}[h]
\centering
\caption{Normalization factors obtained from a \CRBGonlyfit{} defined for electroweakino, slepton and VBF searches. The uncertainties include statistical and systematic contributions combined.}
\renewcommand{\arraystretch}{1.1}
\begin{tabular}{llccc}
\toprule
\toprule
&&\multicolumn{3}{c}{Normalization Parameters}\\
Backgrounds &\met{} region   & electroweakino & slepton & VBF\\
\midrule
\multirow{2}{*}{$\ttbar/Wt$} & high & $1.08  \pm 0.20$    & $1.05  \pm 0.20$ & \multirow{2}{*}{$1.04 \pm  0.04$}\\
& low & $1.08  \pm 0.18 $   & $1.09  \pm 0.19$ &  \\
\midrule
\multirow{2}{*}{$\Ztt$} & high & $0.96  \pm 0.14$    & $0.80  \pm 0.17$ & \multirow{2}{*}{$0.97 \pm 0.13$} \\
& low & $1.02  \pm 0.15$    & $1.08  \pm 0.17$ &  \\
\midrule
\multirow{2}{*}{$VV$} & high &  $0.89  \pm 0.27$    & $0.85  \pm 0.28$ & \multirow{2}{*}{--}\\
& low &  $0.69  \pm 0.22$    & $0.71  \pm 0.23$ & \\
\bottomrule
\bottomrule
\end{tabular}
\label{table:results:normfactors}
\end{table}

\begin{figure}[tbp]
\centering
\includegraphics[width=0.60\columnwidth]{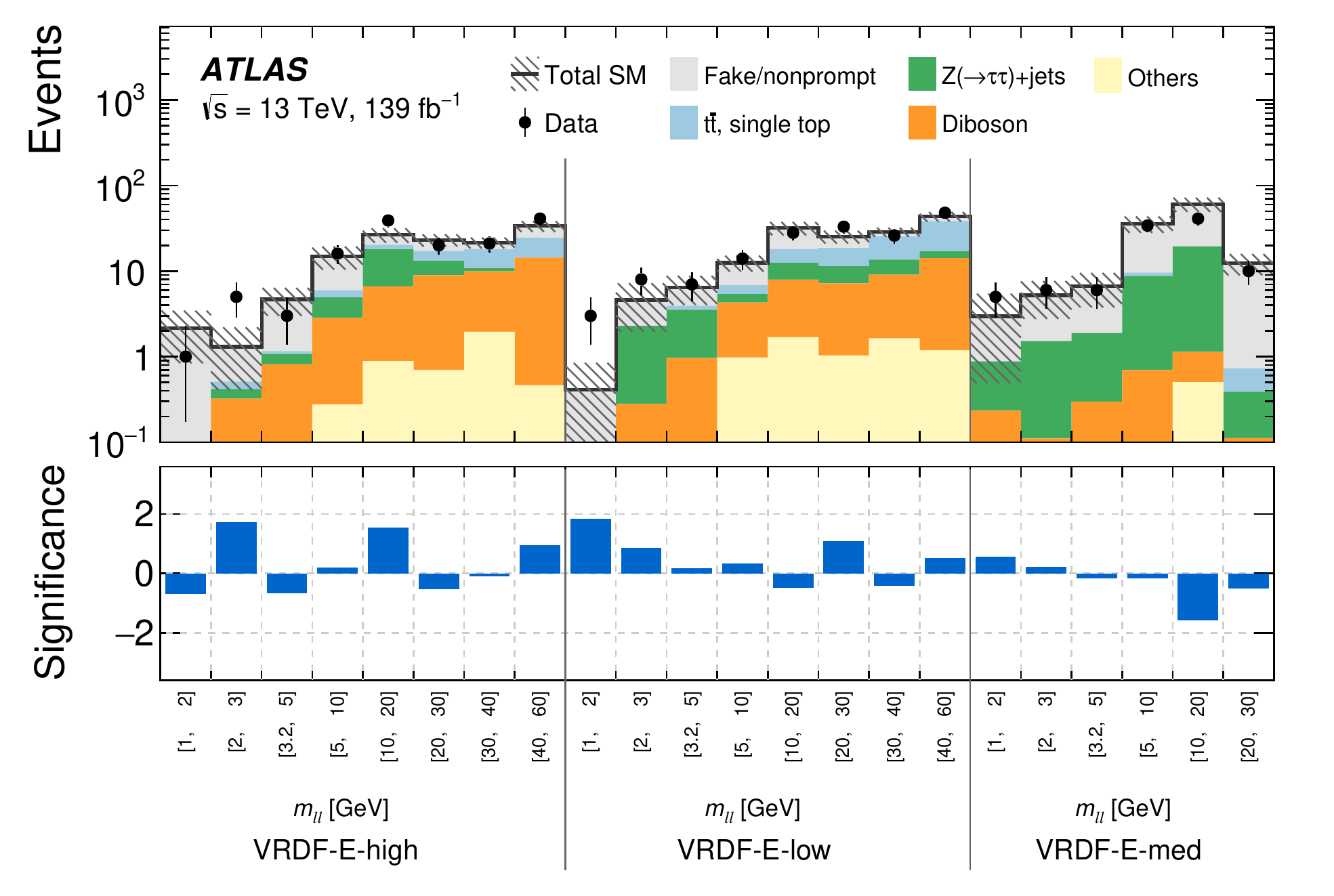}\\
\includegraphics[width=0.60\columnwidth]{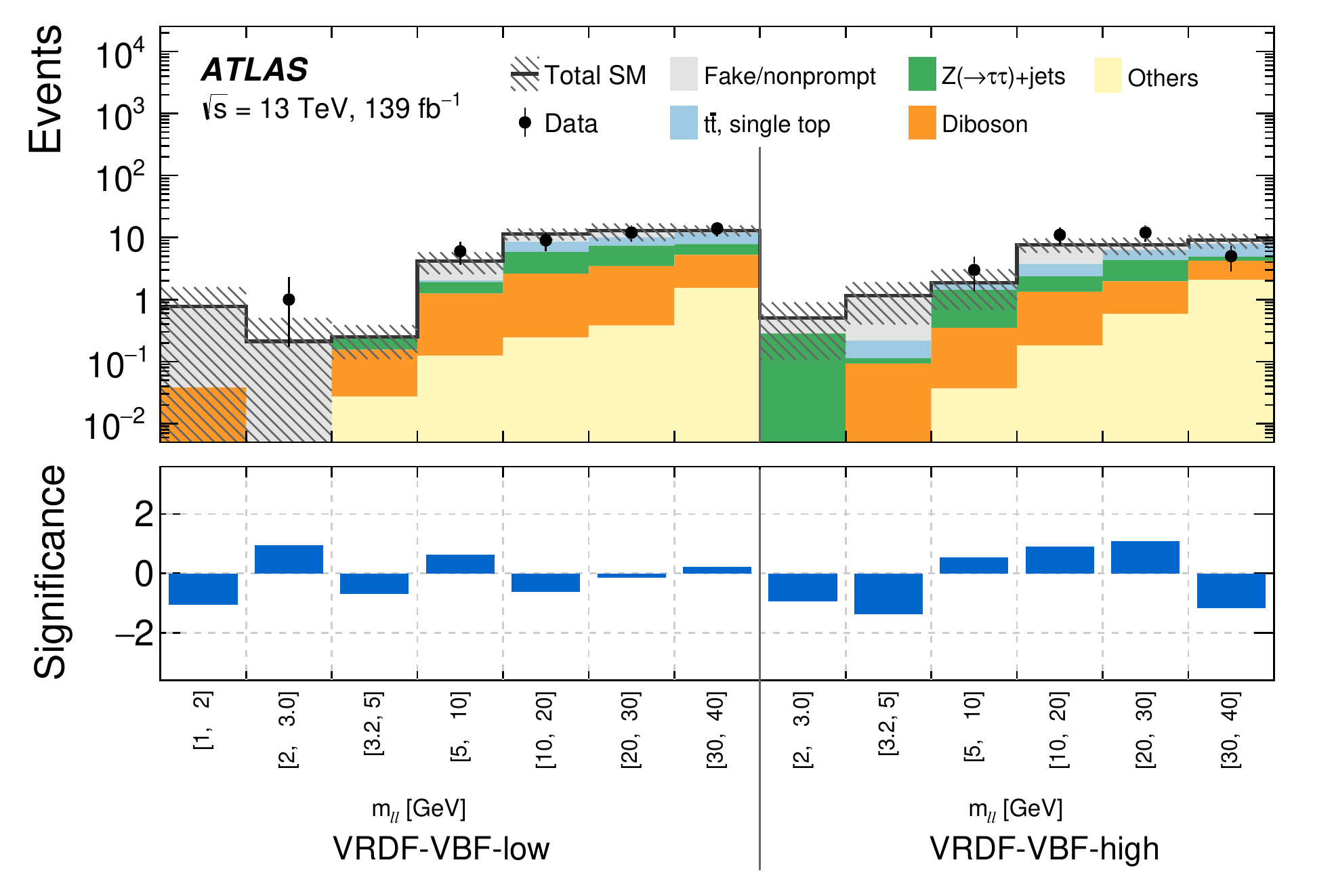}\\
\includegraphics[width=0.60\columnwidth]{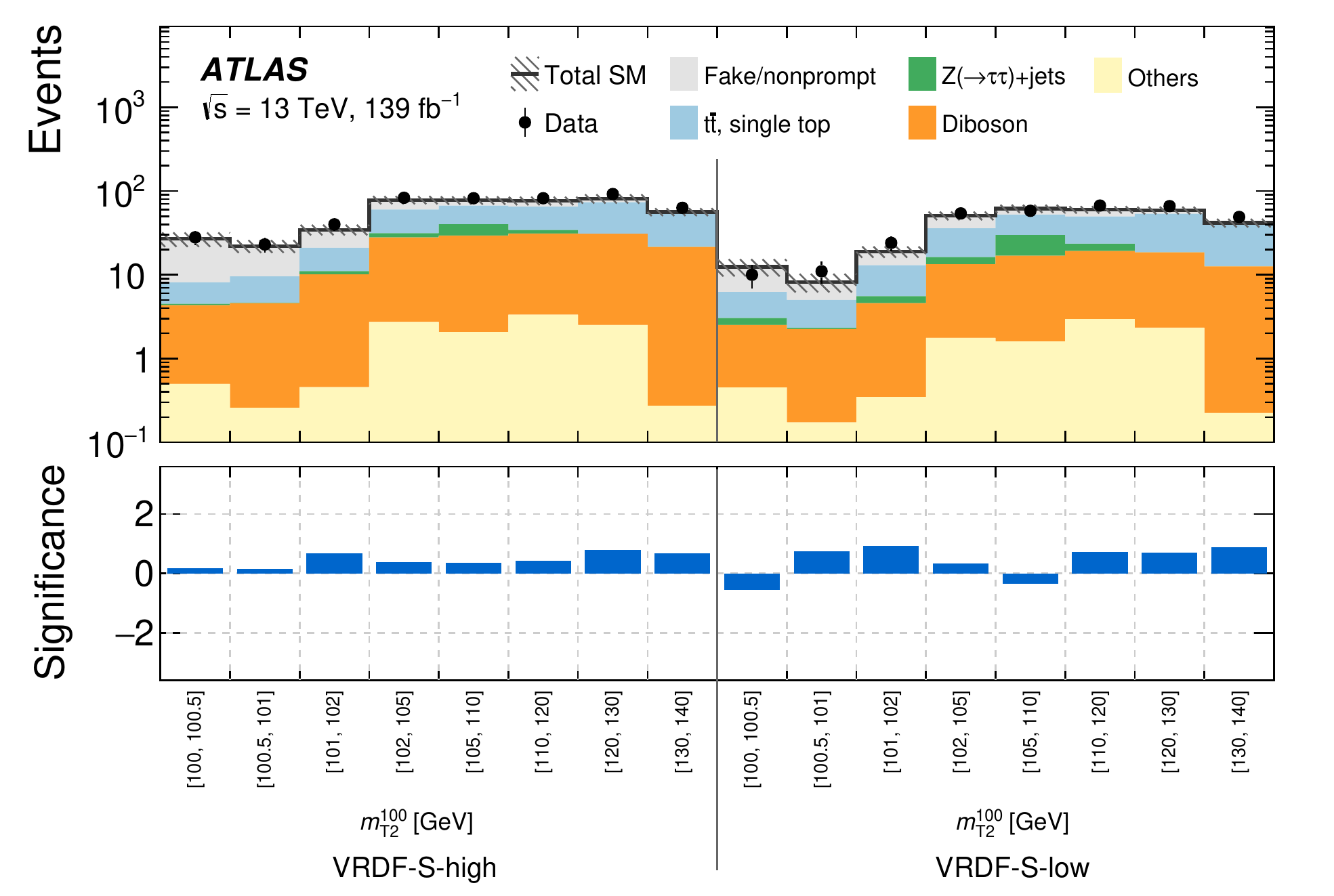}\\
\caption{Comparison of observed and expected event yields in the VRDF regions after a \CRBGonlyfit. The three VRDF--E regions are shown at the top, binned in $\mll$ as the corresponding electroweakino SRs.  The two VRDF--VBF regions are shown in the middle, also binned in $\mll$.   The bin $1~\GeV < \mll < 2~\GeV$  is omitted from the VRDF-VBF-high region because both the expected and observed event yields are zero.  Finally, the two VRDF--S regions are shown at the bottom, binned in $\mtth$ as the corresponding slepton SRs. Uncertainties in the background estimates include both the statistical and systematic uncertainties.
The bottom panel in all three plots shows the significance of the difference between the expected and observed yields, computed following the profile likelihood method of Ref.~\cite{2008NIMPA.595..480C} in the case where the observed yield exceeds the prediction, and using the same expression with an overall minus sign if the yield is below the prediction.}
\label{fig:results:CRVRpulls}
\end{figure}

\begin{figure}[tbp]
\centering
\includegraphics[width=0.49\columnwidth]{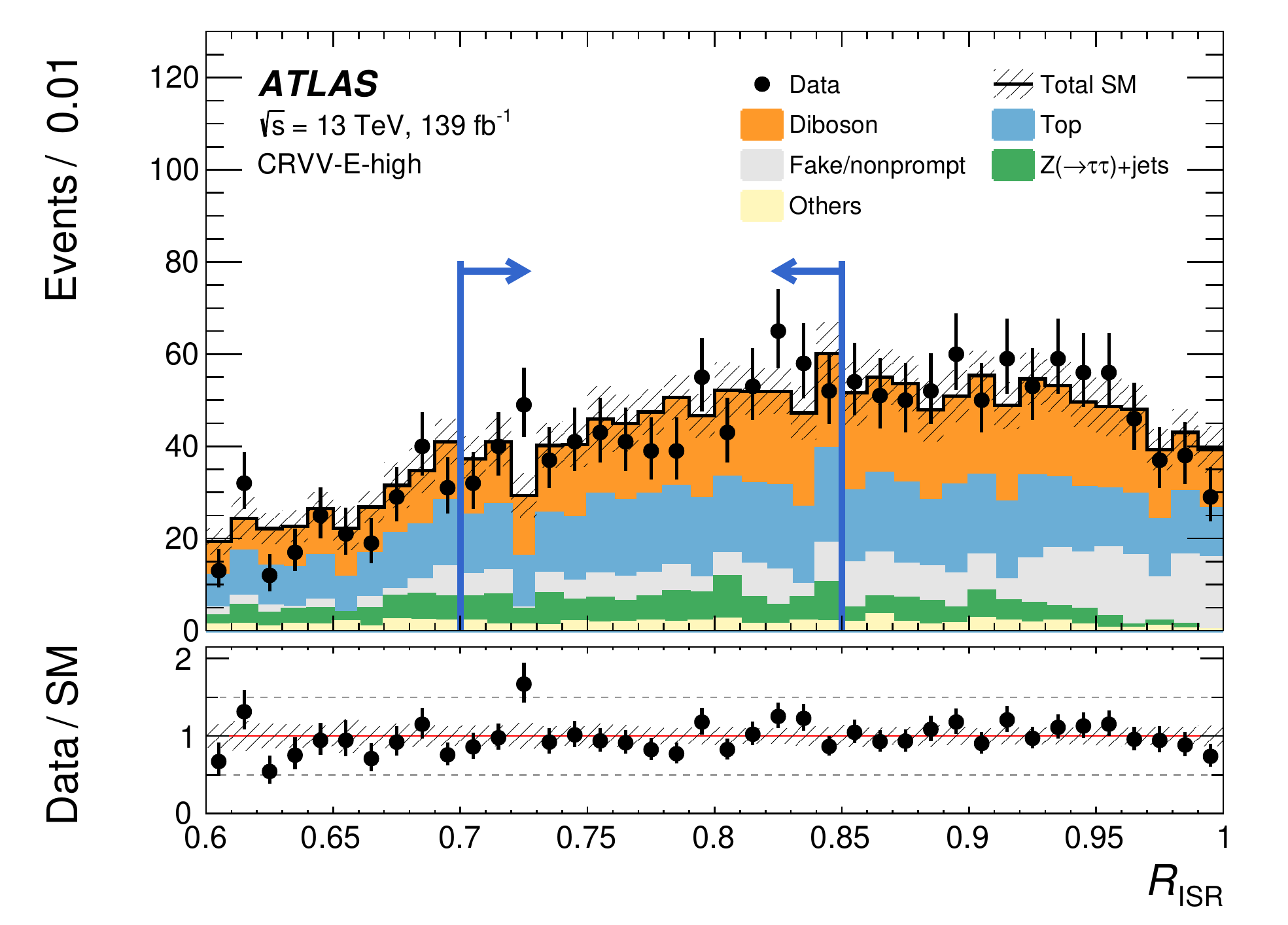}
\includegraphics[width=0.49\columnwidth]{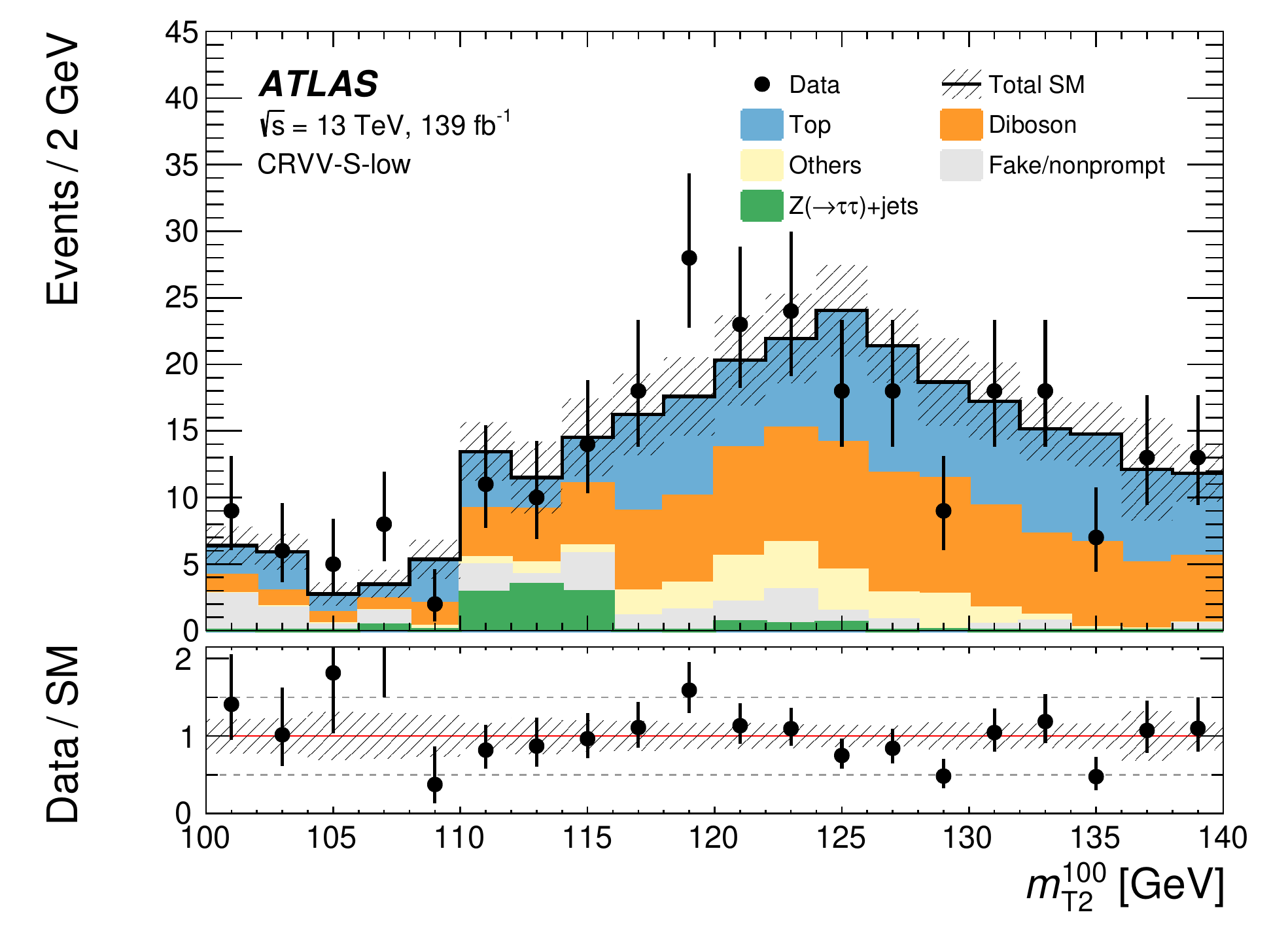}\\
\includegraphics[width=0.49\columnwidth]{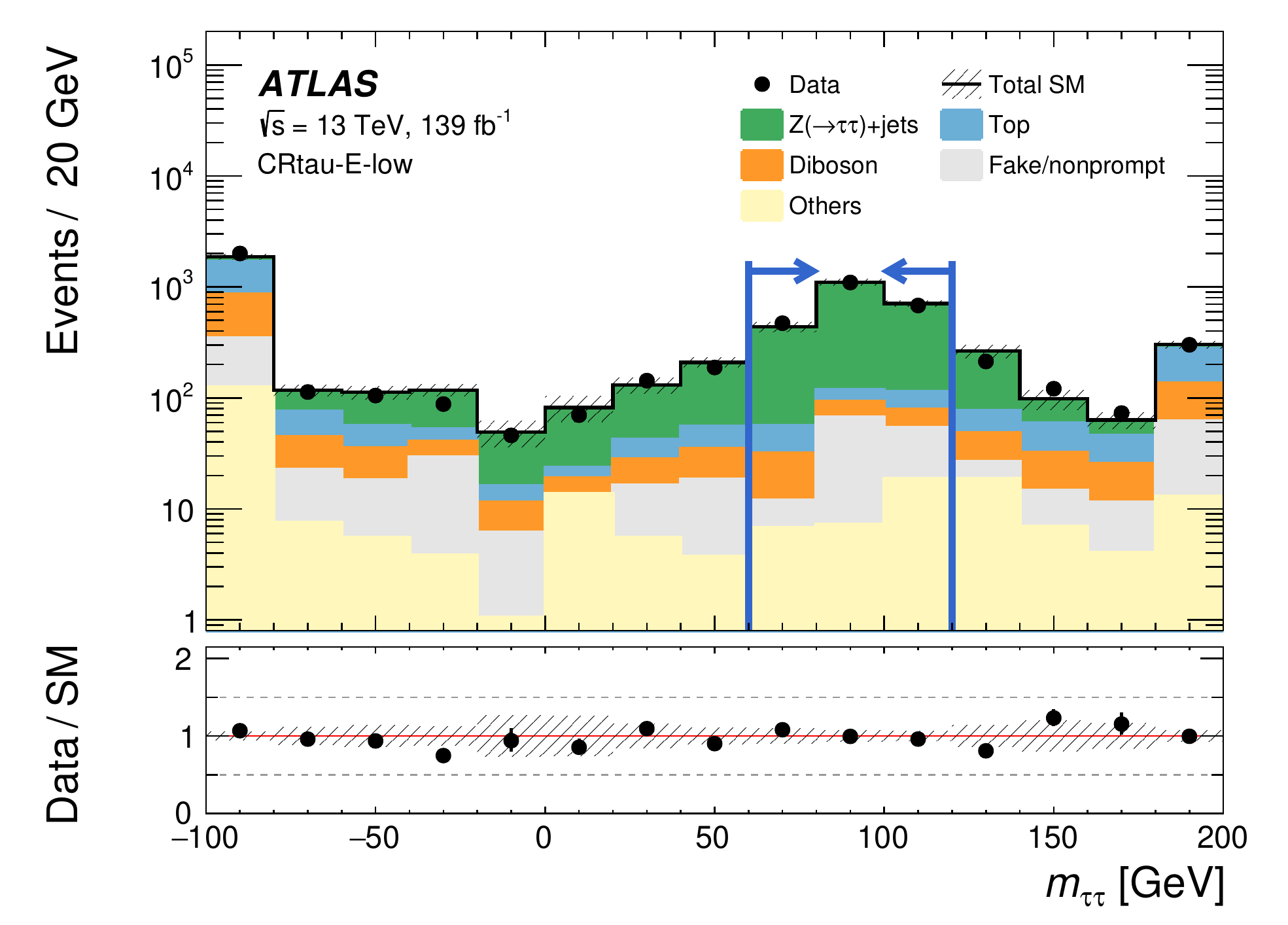}
\includegraphics[width=0.49\columnwidth]{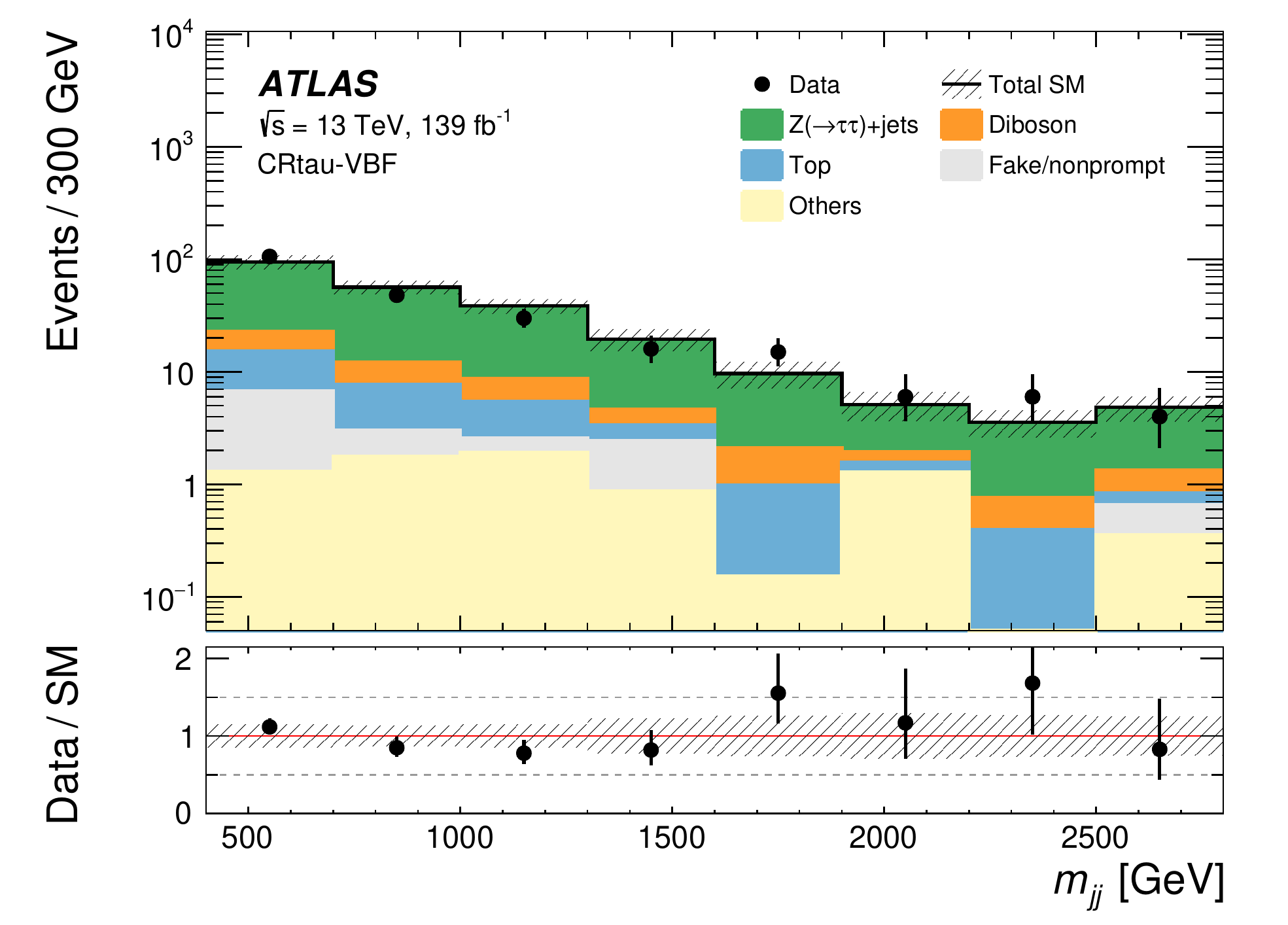}\\
\includegraphics[width=0.49\columnwidth]{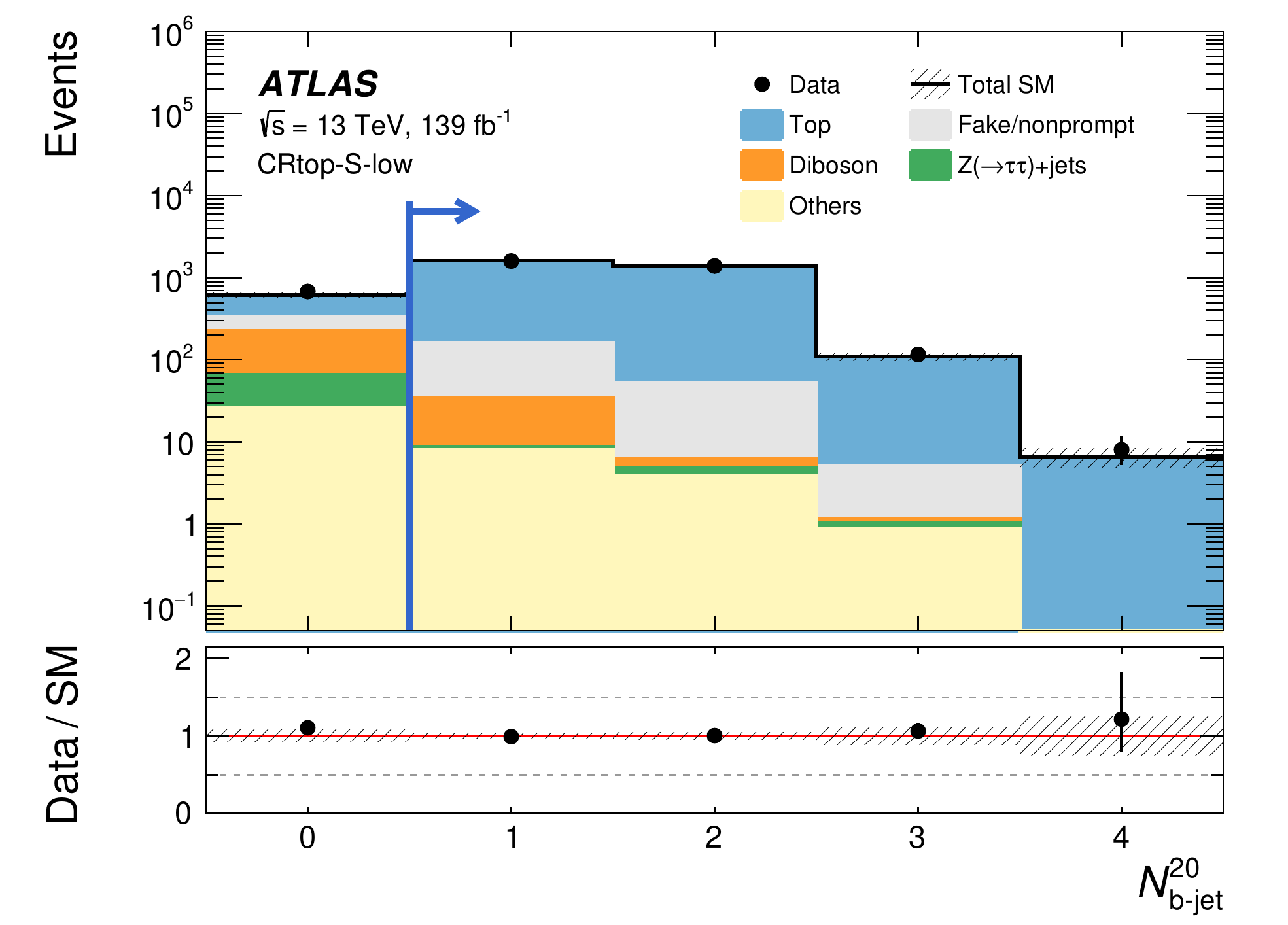}
\includegraphics[width=0.49\columnwidth]{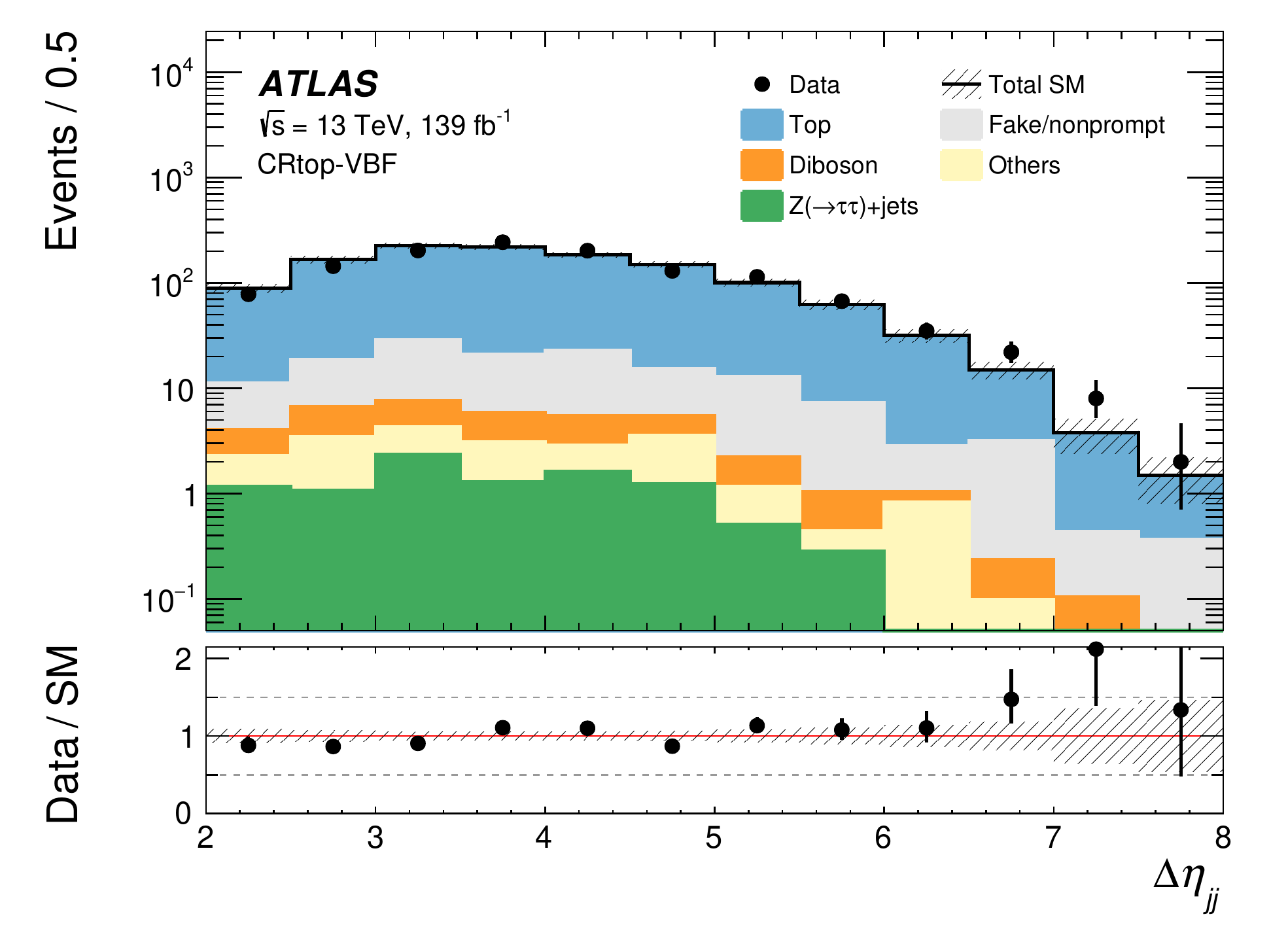}\\
\caption{Examples of kinematic distributions after the \CRBGonlyfit{} showing the data as well as the expected background in the control regions CRVV--E--high (top left), CRVV--S--low (top right), CRtau--E--low (middle left), CRtau--VBF (middle right), CRtop--S--low (bottom left) and CRtop--VBF (bottom right). The full event selection of the corresponding regions is applied, except for distributions showing blue arrows, where the requirement on the variable being plotted is removed and indicated by the arrows in the distributions instead. The first~(last) bin includes underflow~(overflow). The uncertainty bands plotted include all statistical and systematic uncertainties.}
\label{fig:results:CRkinematics}
\end{figure}

\begin{figure}[tbp]
\centering
\includegraphics[width=0.49\columnwidth]{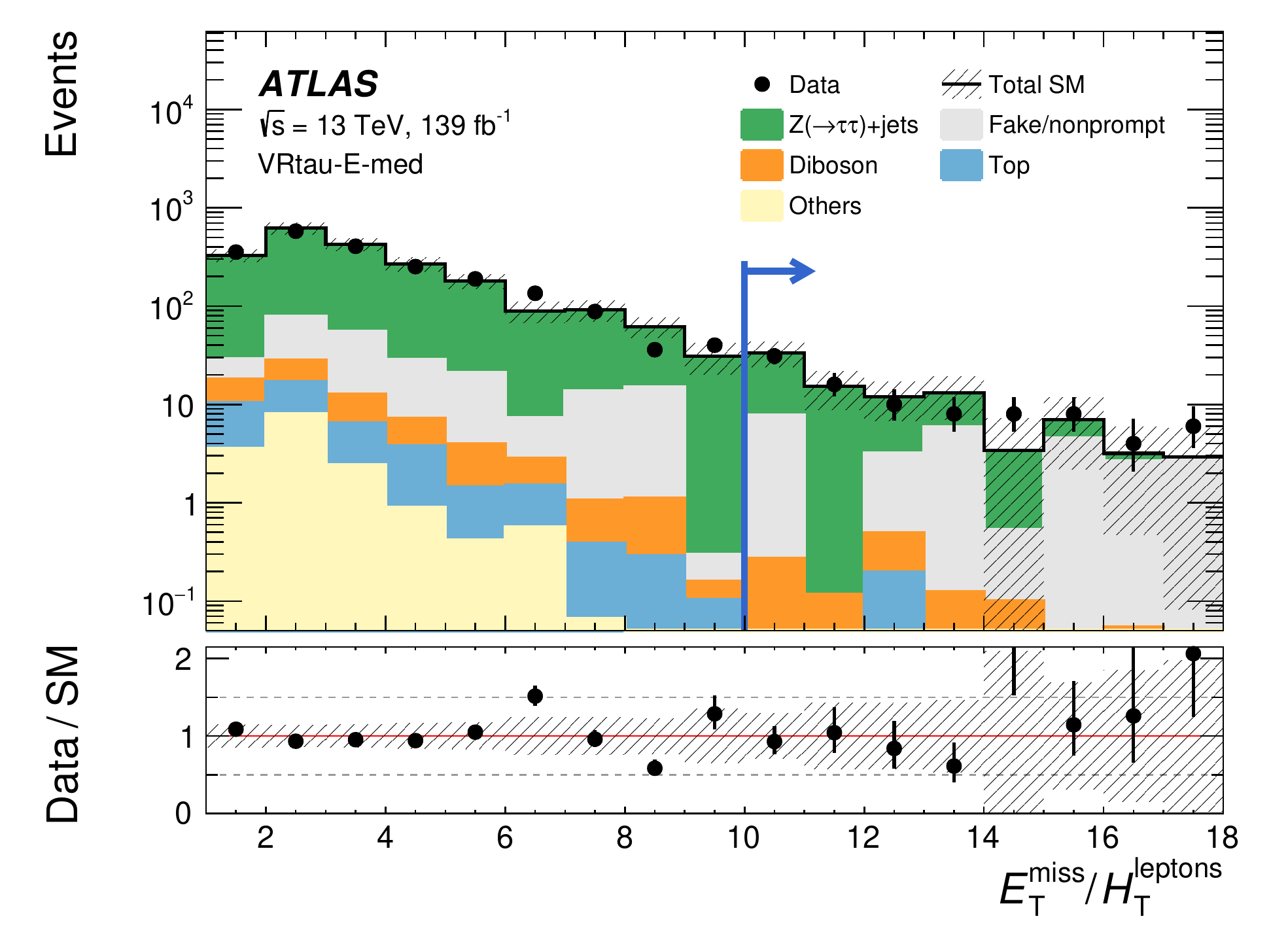}
\includegraphics[width=0.49\columnwidth]{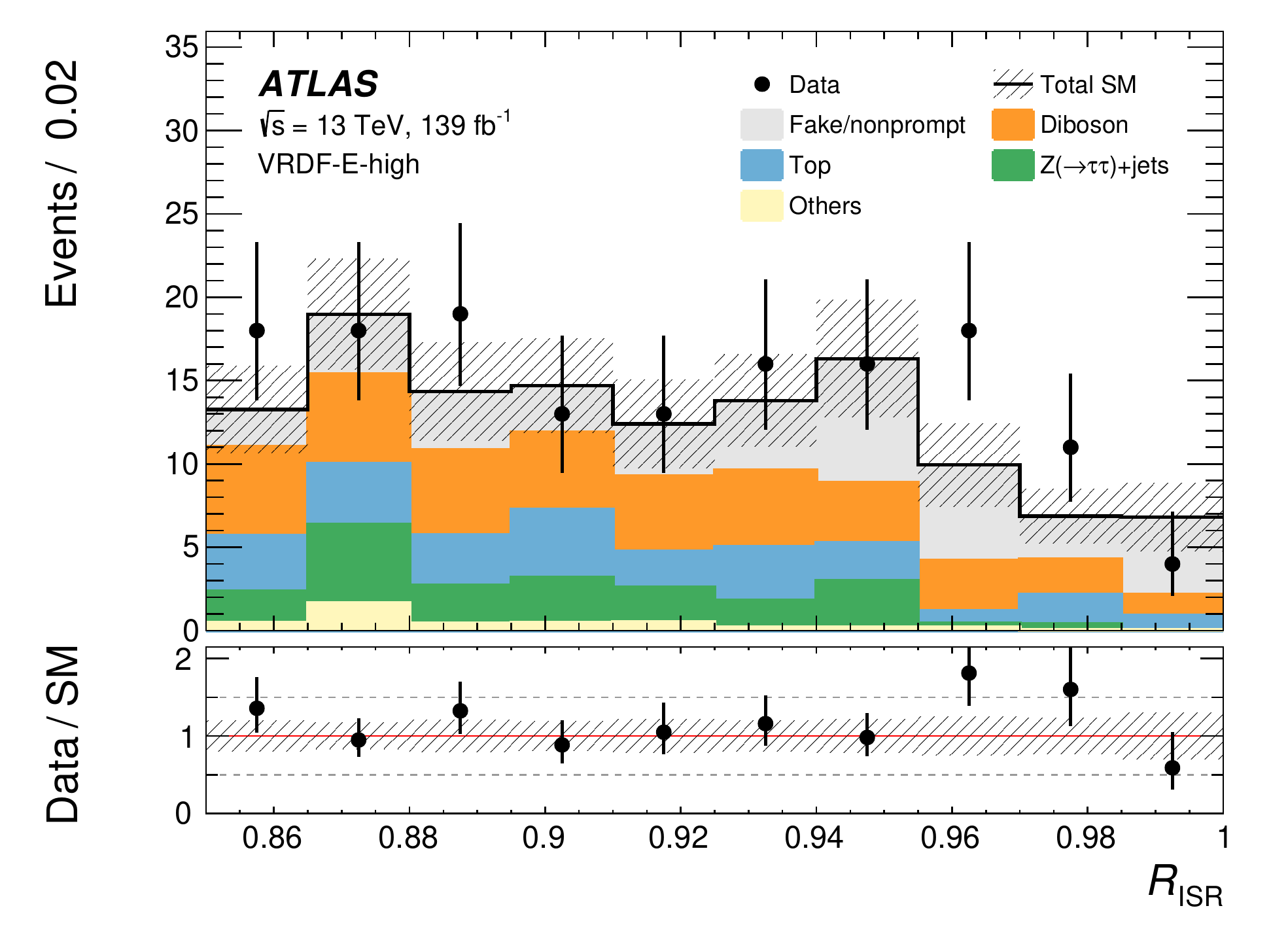}\
\includegraphics[width=0.49\columnwidth]{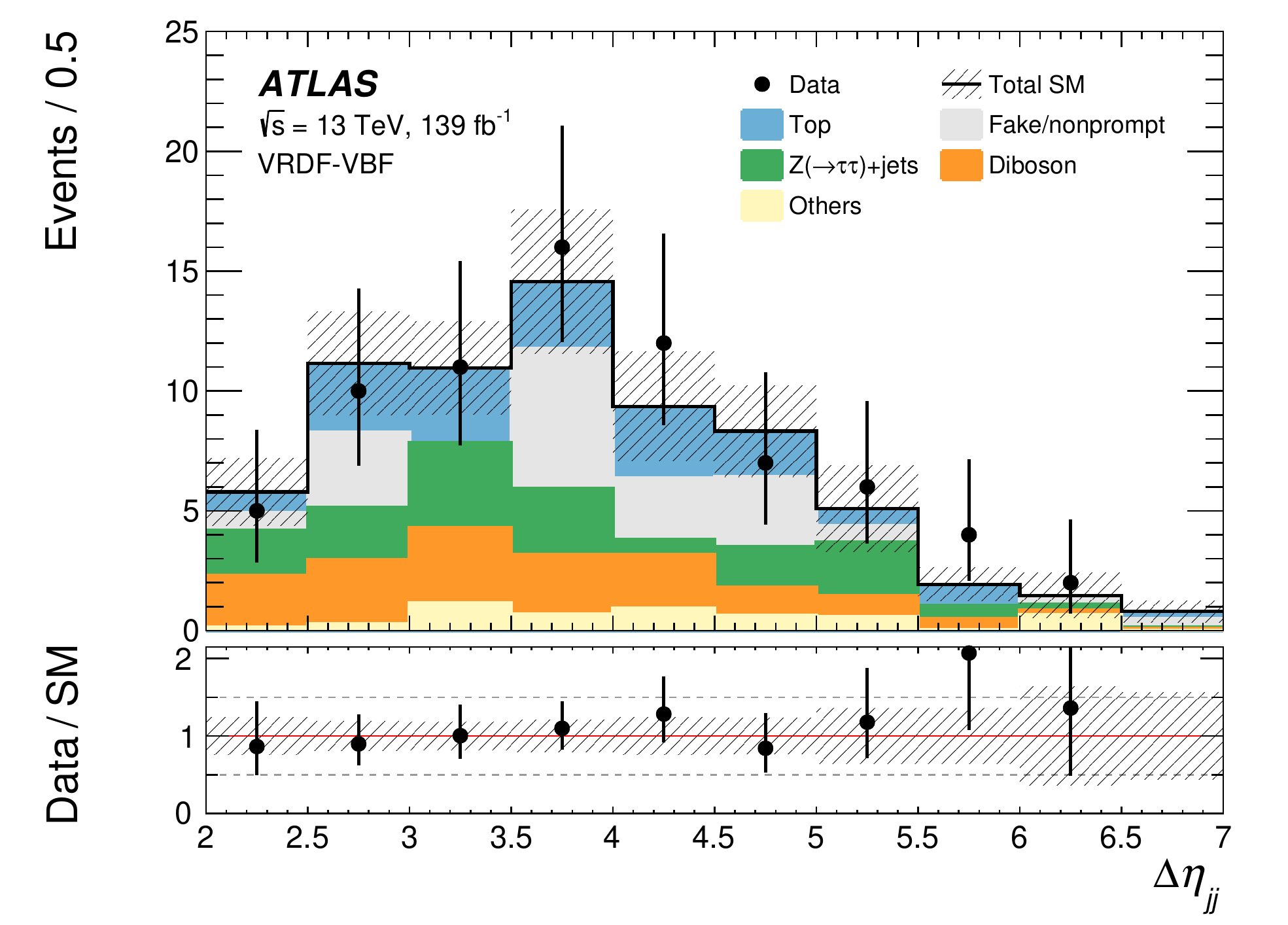}
\includegraphics[width=0.49\columnwidth]{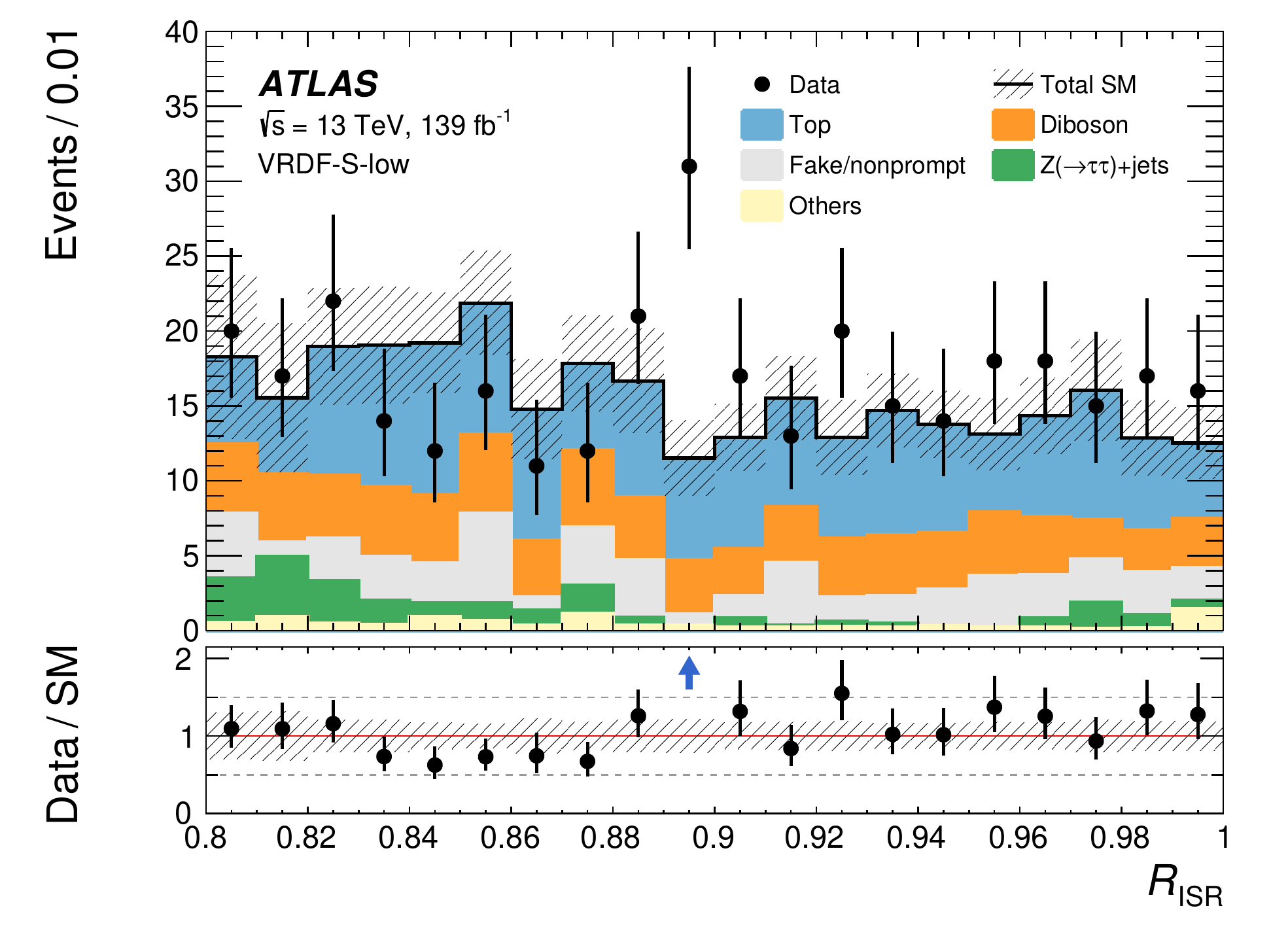}\\
\caption{Examples of kinematic distributions after the \CRBGonlyfit{} showing the data as well as the expected background in the validation regions VRtau--E--med (top left), VRDF--E--high (top right), VRDF--VBF, including both VRDF-VBF-high and VRDF-VBF-low (bottom left) and VRDF--S--low (bottom right). The full event selection of the corresponding regions is applied, except for distributions showing blue arrows, where the requirement on the variable being plotted is removed and indicated by the arrows in the distributions instead. The first~(last) bin includes underflow~(overflow). The uncertainty bands plotted include all statistical and systematic uncertainties.}
\label{fig:results:VRkinematics1}
\end{figure}

\begin{figure}[tbp]
\centering
\includegraphics[width=0.49\columnwidth]{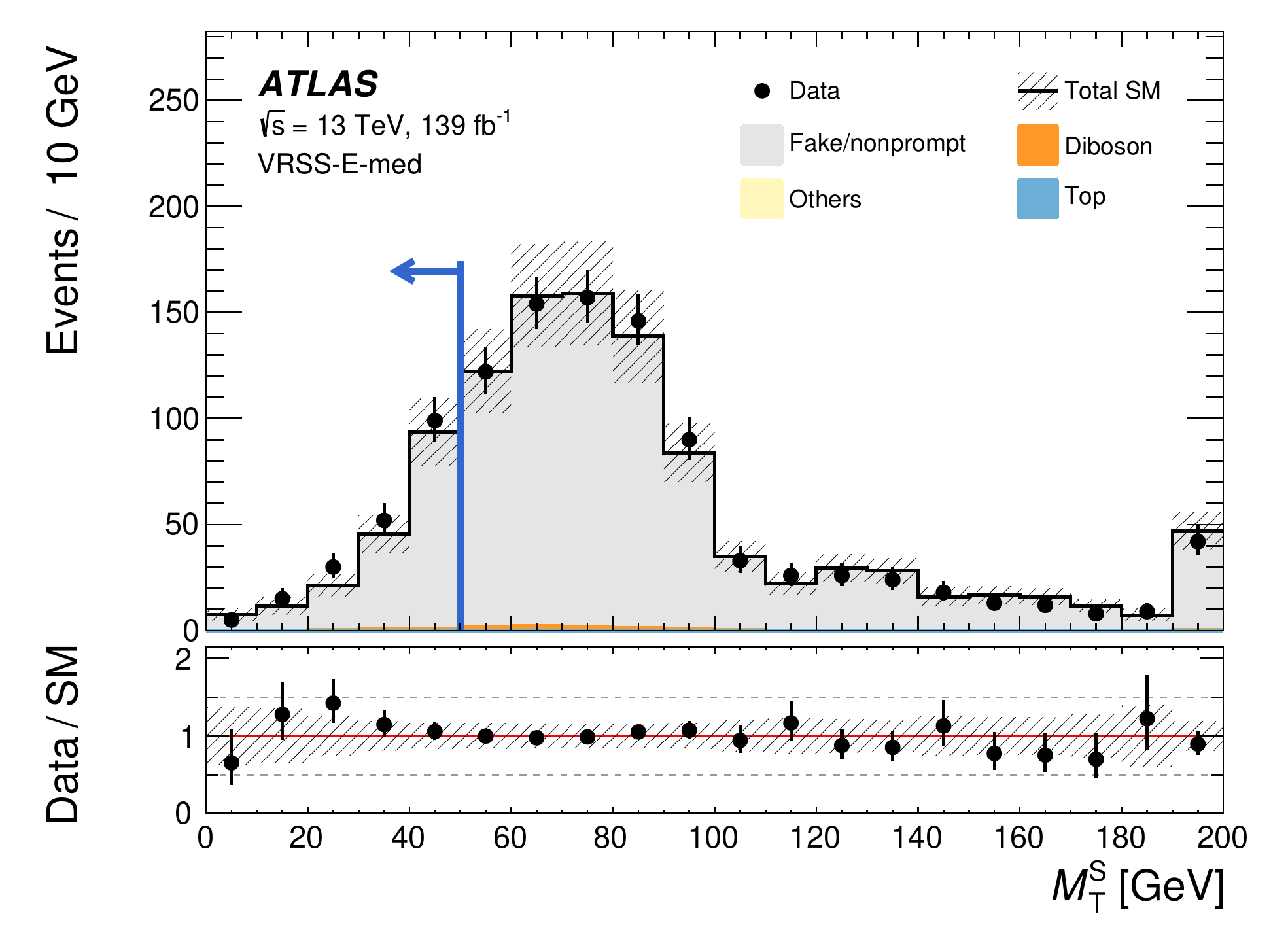}
\includegraphics[width=0.49\columnwidth]{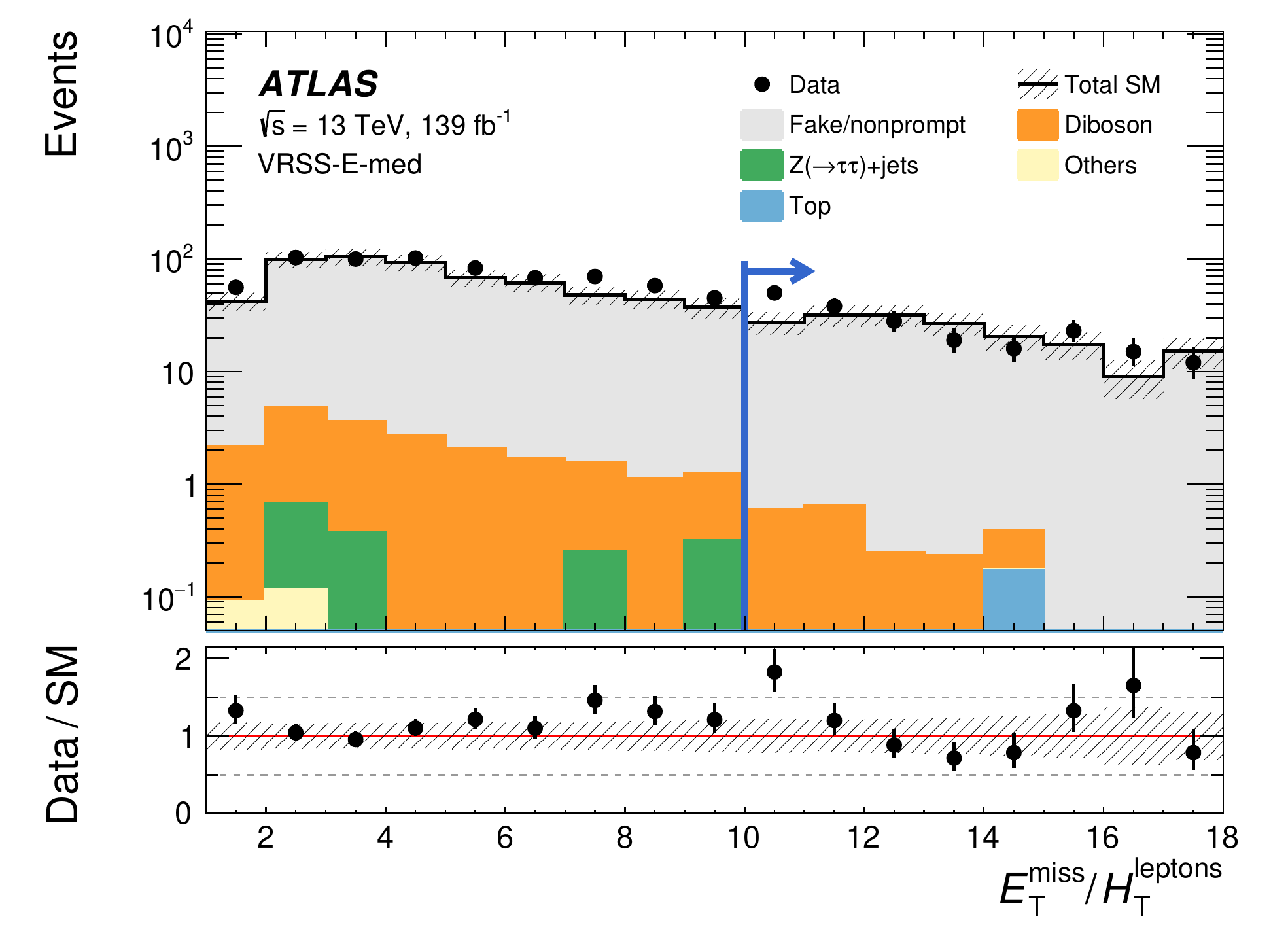}\\
\includegraphics[width=0.49\columnwidth]{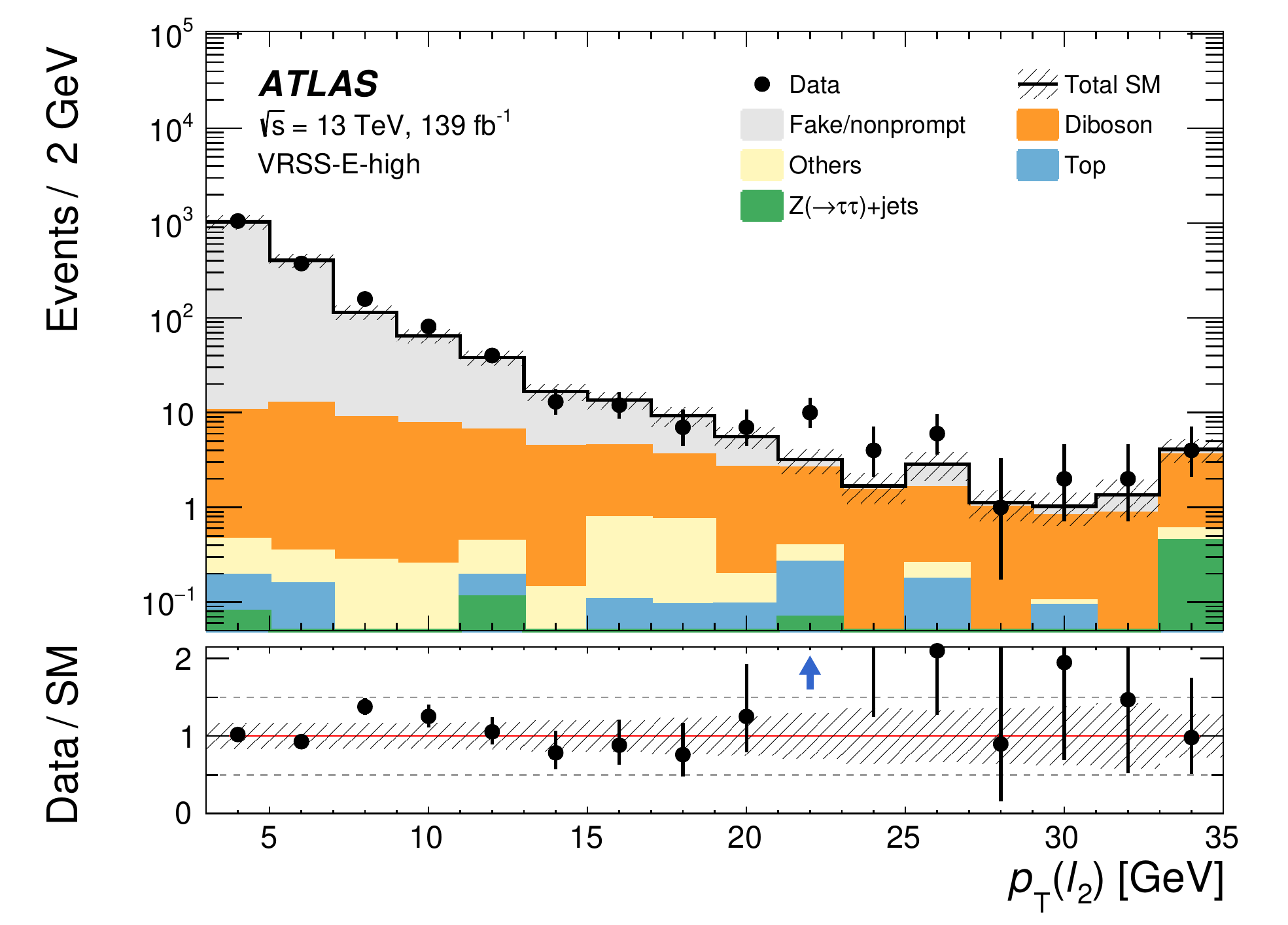}
\includegraphics[width=0.49\columnwidth]{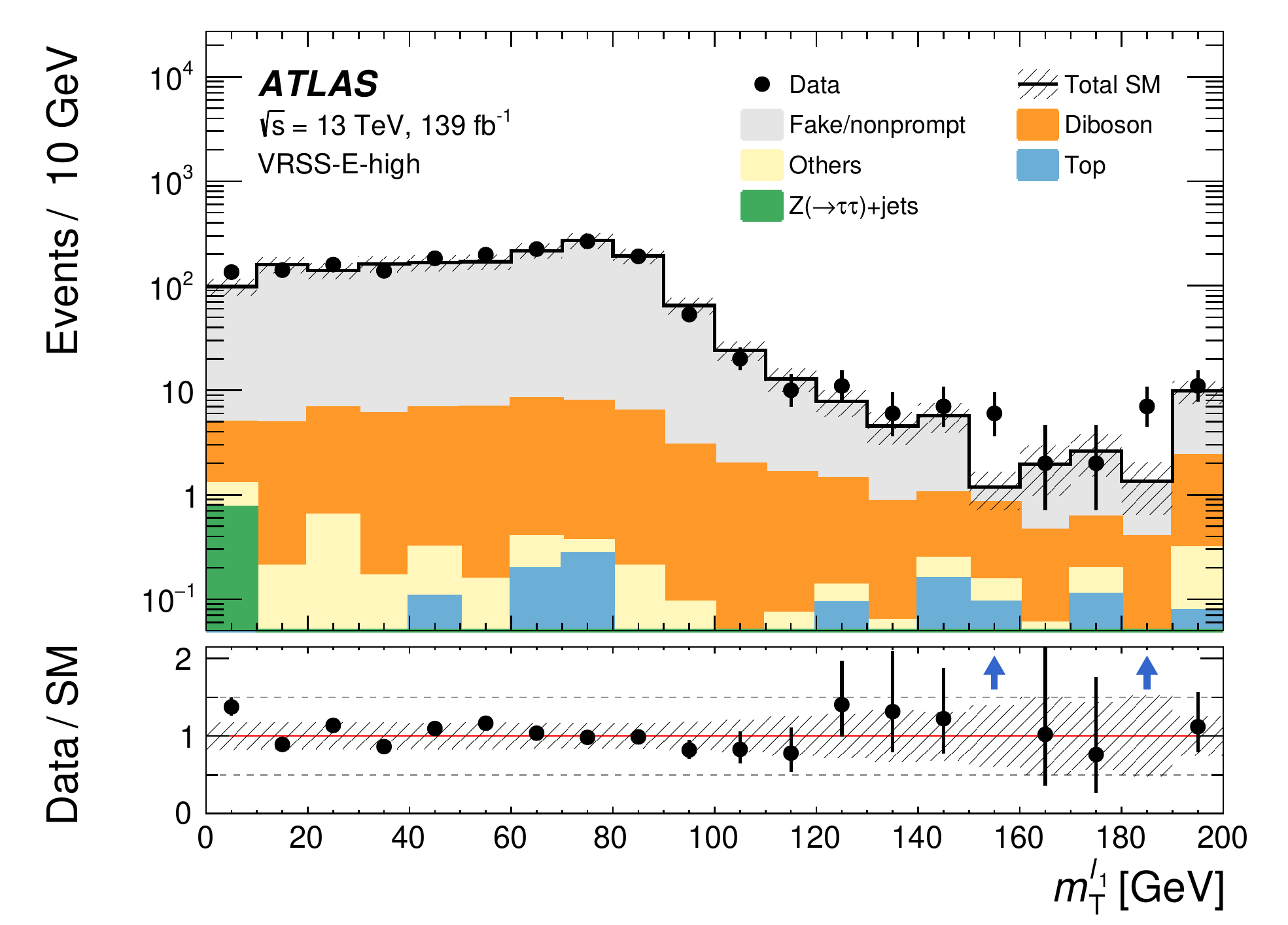}\\
\includegraphics[width=0.49\columnwidth]{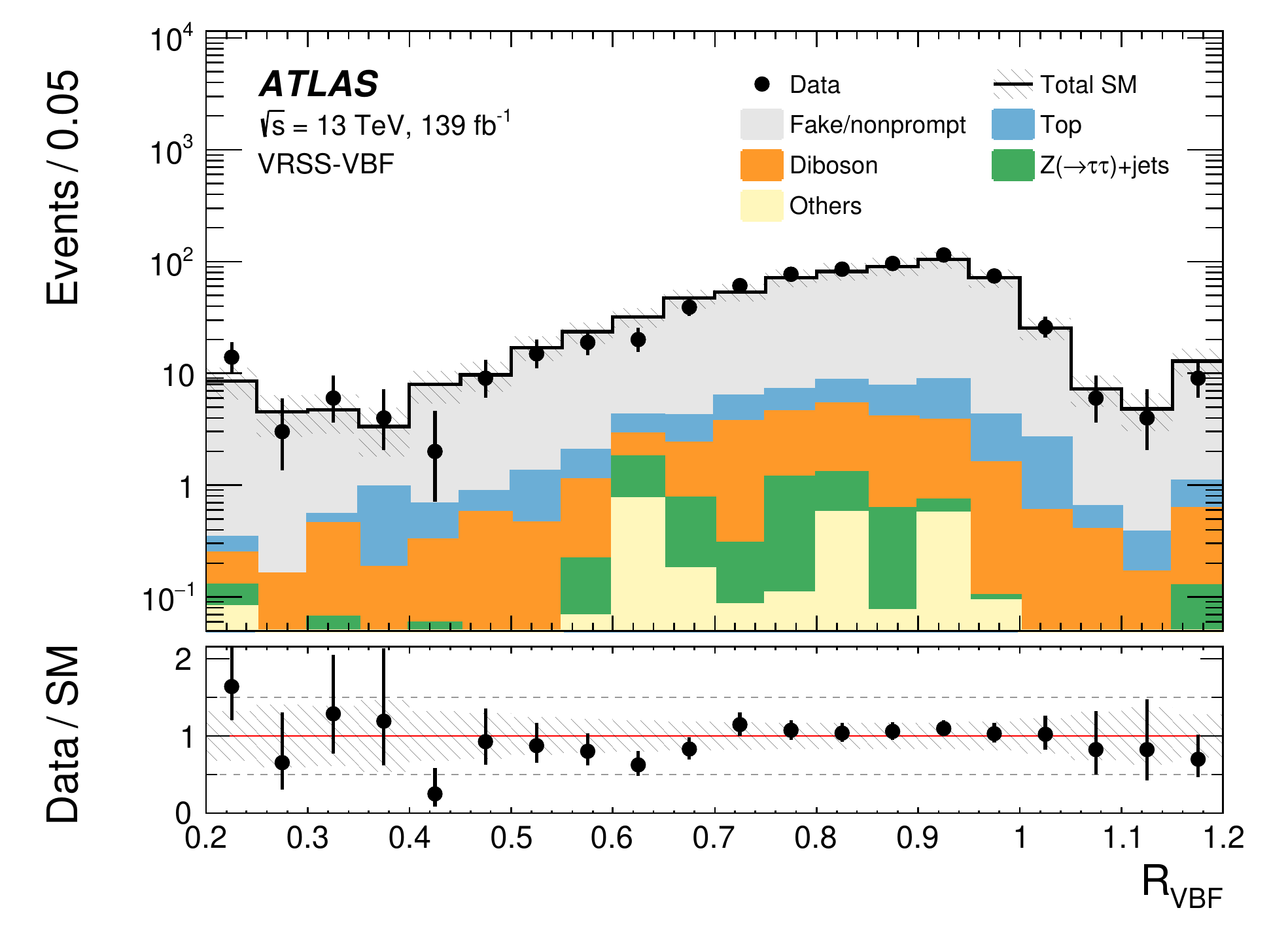}
\includegraphics[width=0.49\columnwidth]{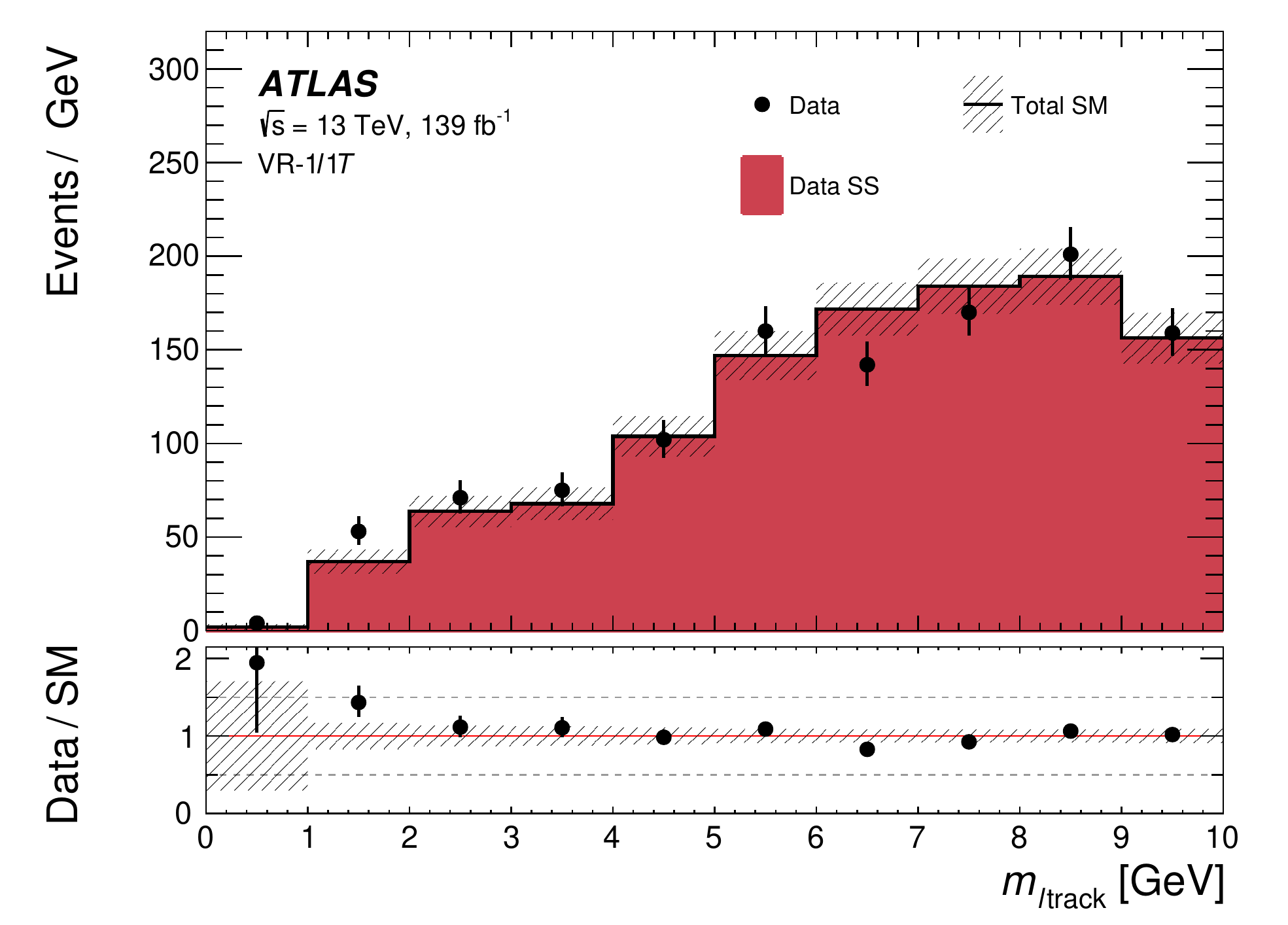}\\
\caption{Examples of kinematic distributions after the \CRBGonlyfit{} showing the data as well as the expected background in the validation regions VRSS--E--med (top), VRSS--E--high (middle), VRSS--VBF (bottom left), and VR--\oneleponetrack{} (bottom right). The full event selection of the corresponding regions is applied, except for distributions showing blue arrows, where the requirement on the variable being plotted is removed and indicated by the arrows in the distributions instead. The first~(last) bin includes underflow~(overflow). The uncertainty bands plotted include all statistical and systematic uncertainties.}
\label{fig:results:VRkinematics2}
\end{figure}

 
The background prediction as obtained from the \CRBGonlyfit{} is then compared with data in the validation regions to
verify the accuracy of the background modeling.
Figure~\ref{fig:results:CRVRpulls} shows a comparison of the data yields with background predictions in the VRDF regions,
binned in $\mll$ and $\mtth$ using the same intervals as defined for the corresponding SRs.
Good agreement is observed in all event selection categories, with deviations below $2\sigma$.
Examples of kinematic distributions in control and validation regions are presented in Figures~\ref{fig:results:CRkinematics} through~\ref{fig:results:VRkinematics2}, where good agreement between data and MC simulation is seen in both the shape and
normalization of the discriminating variables.
 
\FloatBarrier
 
\subsection{Inclusive signal regions}
 
The inclusive signal regions defined in Section~\ref{sec:signalreg} are used to test for excesses of events above the SM predictions.
Each fit only considers one single-bin inclusive signal region, and includes a signal model with an unconstrained
normalization parameter to estimate the contributions of any phenomena beyond those predicted by the Standard Model.
The signal region is fit simultaneously with the control regions, which are assumed to contain no signal, resulting in background
estimates constrained by the \CRBGonlyfit.
 
To quantify the probability under the background-only hypothesis to produce event yields greater than or equal to the
observed data, the discovery $p$-values are calculated for each inclusive signal region. The results for the electroweakino, VBF, and slepton regions are shown in Table~\ref{tab:Upperlimit_SRSF_allmet}. Several electroweakino regions have low $p$-values, with the lowest observed in the $\mll<20~\GeV$ bin corresponding to a
local significance of \MaxDiscoverySignif.
The CL$_\text{s}$ prescription~\cite{Read:2002hq} is used to perform a hypothesis test that sets upper limits at the 95\% confidence
level (CL) on the observed (expected) number of signal events $S_\text{obs (exp)}^{95}$ in each inclusive signal region.
Dividing $S_\text{obs}^{95}$ by the integrated luminosity defines the upper limits on the visible
cross-sections $\langle\epsilon\sigma\rangle_\text{obs}^{95}$.
 
\newenvironment{DIFnomarkup}{}{}
\begin{table}
\centering
\caption[Breakdown of upper limits.]{
Left to right: The first column indicates the inclusive signal region under study, defined as the union of the individual SRs defined in Section~\ref{sec:signalreg} and by upper bounds on \mll{} or \mtth{} in \GeV.  The \mll{} regions include events in both the 2$\ell$ and \oneleponetrack{} channels, while the \mtth{} regions only include 2$\ell$ events.  The next two columns present observed ($N_{\mathrm{obs}}$) and expected ($N_{\mathrm{exp}}$) event yields in the inclusive signal regions.  The latter are obtained by the \CRBGonlyfit, and the errors include both the statistical and systematic uncertainties. The next two columns show the observed 95\% CL upper limits on the visible cross-section
$\left(\langle\epsilon\sigma\rangle_{\mathrm{obs}}^{95}\right)$ and on the number of
signal events $\left(S_{\mathrm{obs}}^{95}\right)$.  The next column
$\left(S_{\mathrm{exp}}^{95}\right)$ shows the 95\% CL upper limit on the number of
signal events, given the expected number (and $\pm 1\sigma$
deviations from the expectation) of background events.
The last column indicates the discovery $p$-value ($p(s = 0)$).}
\renewcommand{\arraystretch}{1.1}
\setlength{\tabcolsep}{0.0pc}
\begin{tabular*}{\textwidth}{l@{\extracolsep{\fill}}lc r@{~$\pm$~\hspace{-4.85ex}}l cc r@{\hspace{-4.85ex}}l l}
\toprule
\toprule
& {{Signal Region}} & $N_{\mathrm{obs}}$ & \multicolumn{2}{c}{$N_{\mathrm{exp}}$} & $\langle\epsilon{\mathrm{\sigma}}\rangle_{\mathrm{obs}}^{95}$ [fb]  &  $S_{\mathrm{obs}}^{95}$  & \multicolumn{2}{c}{$S_{\mathrm{exp}}^{95}$} & $p(s=0)$ \\
\midrule
\vtlab{\large SR--E}{9}
&$\mll<1$    &   $0$   &   $1.0$ &$1.0 $     & $0.022$ &  $3.0$ & $ { 3.0 }$ &$^{ +1.3 }_{ -0.0 }$ & $0.50$   \\%
&$\mll<2$    &   $46$   &   $44$ &$6.8 $     & $0.15$ &  $21$ & $ { 19 }$ &$^{ +7 }_{ -5 }$ & $0.38$   \\%
&$\mll<3$    &   $90$   &   $77$ &$12 $     & $0.29$ &  $41$ & $ { 31 }$ &$^{ +11 }_{ -9 }$ & $0.18$   \\%
&$\mll<5$    &   $151$   &   $138$ &$18 $     & $0.38$ &  $52$ & $ { 43 }$ &$^{ +16 }_{ -11 }$ & $0.24$   \\%
&$\mll<10$    &   $244$   &   $200$ &$19 $     & $0.62$ &  $86$ & $ { 49 }$ &$^{ +26 }_{ -13 }$ & $0.034$   \\%
&$\mll<20$    &   $383$   &   $301$ &$23 $     & $0.95$ &  $132$ & $ { 61 }$ &$^{ +22 }_{ -16 }$ & $0.0034$   \\%
&$\mll<30$    &   $453$   &   $366$ &$27 $     & $1.04$ &  $144$ & $ { 70 }$ &$^{ +26 }_{ -20 }$ & $0.0065$   \\%
&$\mll<40$    &   $492$   &   $420$ &$30 $     & $0.96$ &  $134$ & $ { 74 }$ &$^{ +29 }_{ -20 }$ & $0.027$   \\%
&$\mll<60$    &   $583$   &   $520$ &$35 $     & $0.97$ &  $135$ & $ { 84 }$ &$^{ +32 }_{ -23 }$ & $0.063$   \\%
\midrule
\vtlab{\large SR--VBF}{7}
&$\mll<2$    &   $0$     & $2.8$ &$1.6$      &    $0.022$ & $3.0$ & $ { 3.9 }$&$^{ +1.6 }_{ -0.9 }$ & $0.50$  \\%
&$\mll<3$    &   $1$     & $3.1$ &$1.7$      &    $0.030$ & $3.6$ & $ { 4.4 }$&$^{ +2.0 }_{ -1.0 }$ & $0.50$  \\%
&$\mll<5$    &   $2$     & $3.3$ &$1.7$      &    $0.035$ & $4.8$ & $ { 5.2 }$&$^{ +2.1 }_{ -1.1 }$ & $0.50$  \\%
&$\mll<10$   &   $9$     & $8.4$ &$2.7$      &    $0.068$ & $9.5$ & $ { 8.8 }$&$^{ +3.2 }_{ -2.2 }$ & $0.43$  \\%
&$\mll<20$   &   $36$    & $32$ &$5$         &    $0.14$  & $20$  & $ { 16 }$&$^{ +6 }_{ -4 }$      & $0.27$ \\%
&$\mll<30$   &   $58$    & $52$ &$7$         &    $0.19$  & $26$  & $ { 21 }$&$^{ +8 }_{ -6 }$      & $0.28$ \\%
&$\mll<40$   &   $82$    & $74$ &$10$        &    $0.24$  & $33$  & $ { 27 }$&$^{ +10 }_{ -7 }$     & $0.27$ \\%
\midrule
\vtlab{\large SR--VBF--high}{7}
&$\mll<2$    &   $0$    &  $2.4$ &$1.1$      &    $0.022$ & $3.0$ & $ { 4.0 }$&$^{ +1.6 }_{ -0.9 }$ & $0.50$  \\%
&$\mll<3$    &   $1$    &  $3.0$ &$1.4$      &    $0.025$ & $3.5$ & $ { 4.6 }$&$^{ +1.8 }_{ -1.2 }$ & $0.50$  \\%
&$\mll<5$    &   $2$    &  $3.0$ &$1.4$      &    $0.034$ & $4.7$ & $ { 5.1 }$&$^{ +2.0 }_{ -1.3 }$ & $0.50$  \\%
&$\mll<10$   &   $3$    &  $3.8$ &$1.7$      &    $0.041$ & $5.6$ & $ { 5.8 }$&$^{ +2.1 }_{ -1.3 }$ & $0.50$  \\%
&$\mll<20$   &   $9$    &  $11.7$ &$2.8$     &    $0.055$ & $8$   & $ { 9 }$&$^{ +4 }_{ -2.3 }$     & $0.50$ \\%
&$\mll<30$   &   $17$   &  $20$ &$5$         &    $0.079$ & $11$  & $ { 13 }$&$^{ +5 }_{ -3.2 }$    & $0.50$ \\%
&$\mll<40$   &   $26$   &  $28$ &$6$         &    $0.10$  & $14$  & $ { 15 }$&$^{ +6 }_{ -4 }$      & $0.50$ \\%
\midrule
\vtlab{\large SR--S}{8}
&$\mtth<100.5$ &   $24$   &   $27$ &$4.8 $     & $0.09$ &  $13$ & $ { 14 }$ &$^{ +5 }_{ -4 }$ & $0.50$   \\%
&$\mtth<101$   &   $41$   &   $46$ &$6.5 $     & $0.11$ &  $16$ & $ { 18 }$ &$^{ +7 }_{ -5 }$ & $0.50$   \\%
&$\mtth<102$   &   $91$   &   $82$ &$10 $     & $0.25$ &  $35$ & $ { 28 }$ &$^{ +10 }_{ -8 }$ & $0.25$   \\%
&$\mtth<105$   &   $158$   &   $158$ &$17 $     & $0.30$ &  $41$ & $ { 41 }$ &$^{ +16 }_{ -11 }$ & $0.50$   \\%
&$\mtth<110$   &   $243$   &   $242$ &$21 $     & $0.38$ &  $52$ & $ { 52 }$ &$^{ +19 }_{ -14 }$ & $0.36$   \\%
&$\mtth<120$   &   $328$   &   $312$ &$24 $     & $0.51$ &  $71$ & $ { 60 }$ &$^{ +22 }_{ -17 }$ & $0.26$   \\%
&$\mtth<130$   &   $419$   &   $388$ &$28 $     & $0.66$ &  $92$ & $ { 68 }$ &$^{ +27 }_{ -18 }$ & $0.17$   \\%
&$\mtth<140$   &   $472$   &   $443$ &$31 $     & $0.69$ &  $95$ & $ { 74 }$ &$^{ +28 }_{ -21 }$ & $0.19$   \\%
\bottomrule
\bottomrule
\end{tabular*}
\label{tab:Upperlimit_SRSF_allmet}
\end{table}
 

\FloatBarrier
 
\subsection{Exclusive signal regions and model-dependent interpretations}
 
\begin{figure}[tbp]
\centering
\includegraphics[width=0.60\columnwidth]{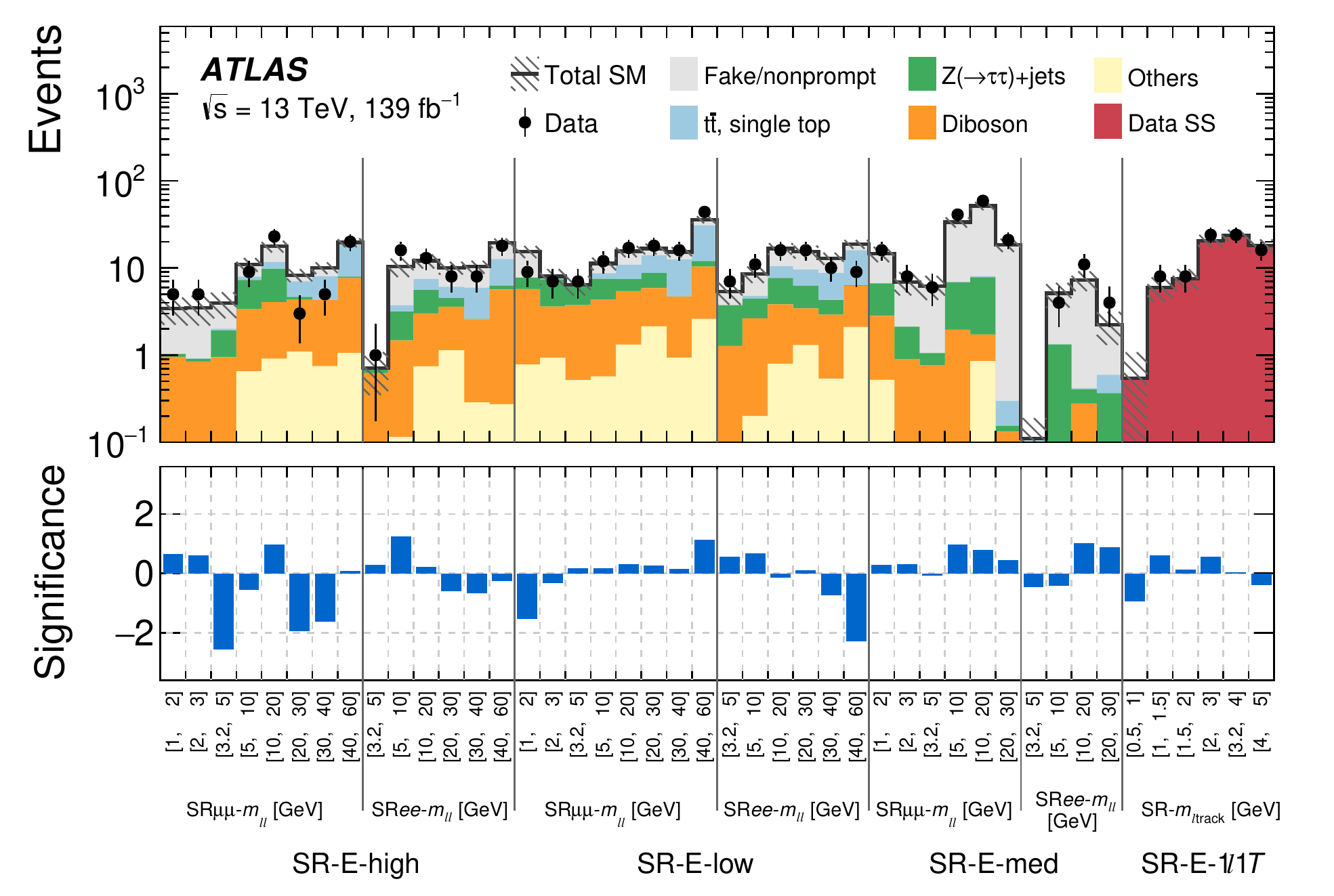}\\
\includegraphics[width=0.60\columnwidth]{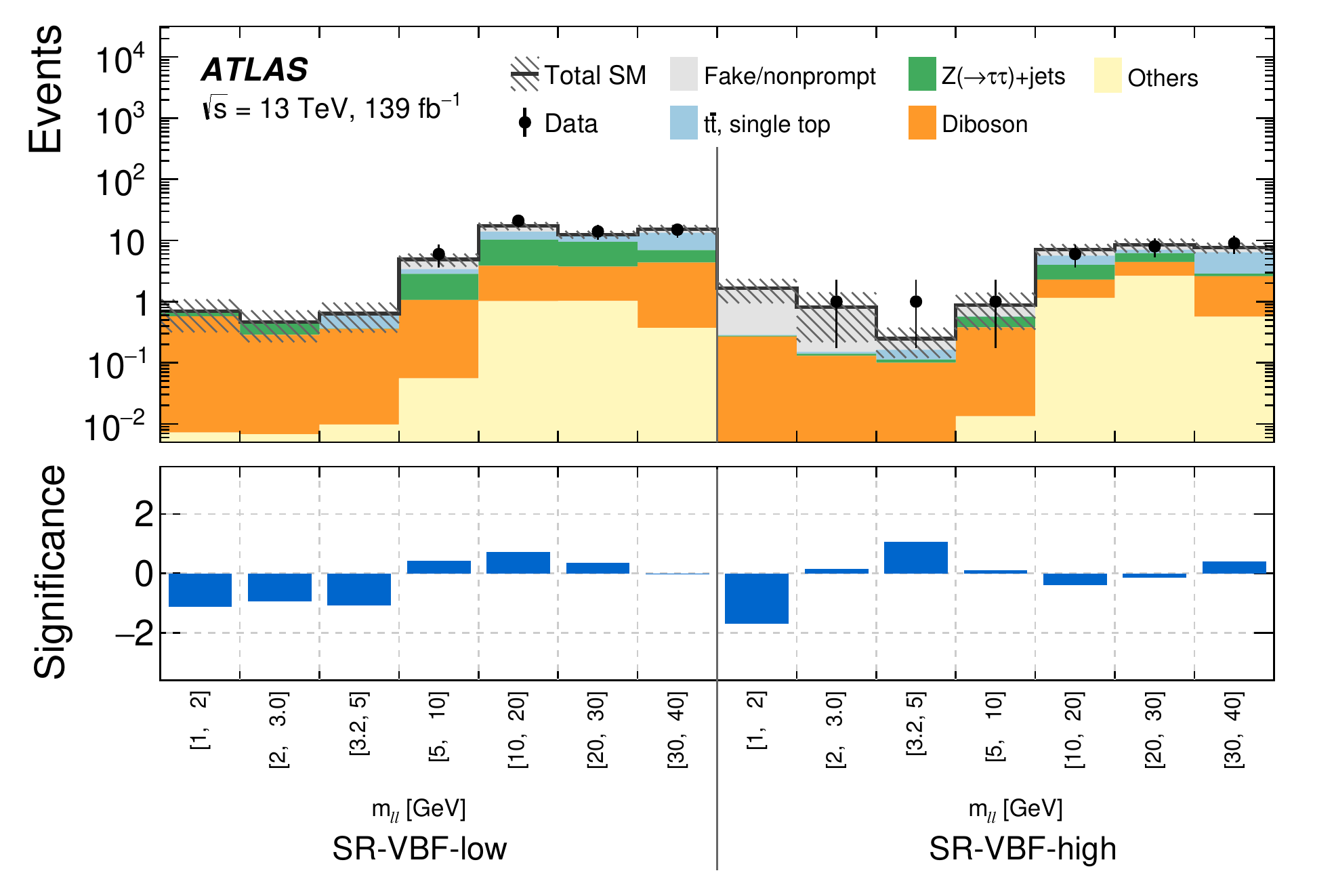}\\
\includegraphics[width=0.60\columnwidth]{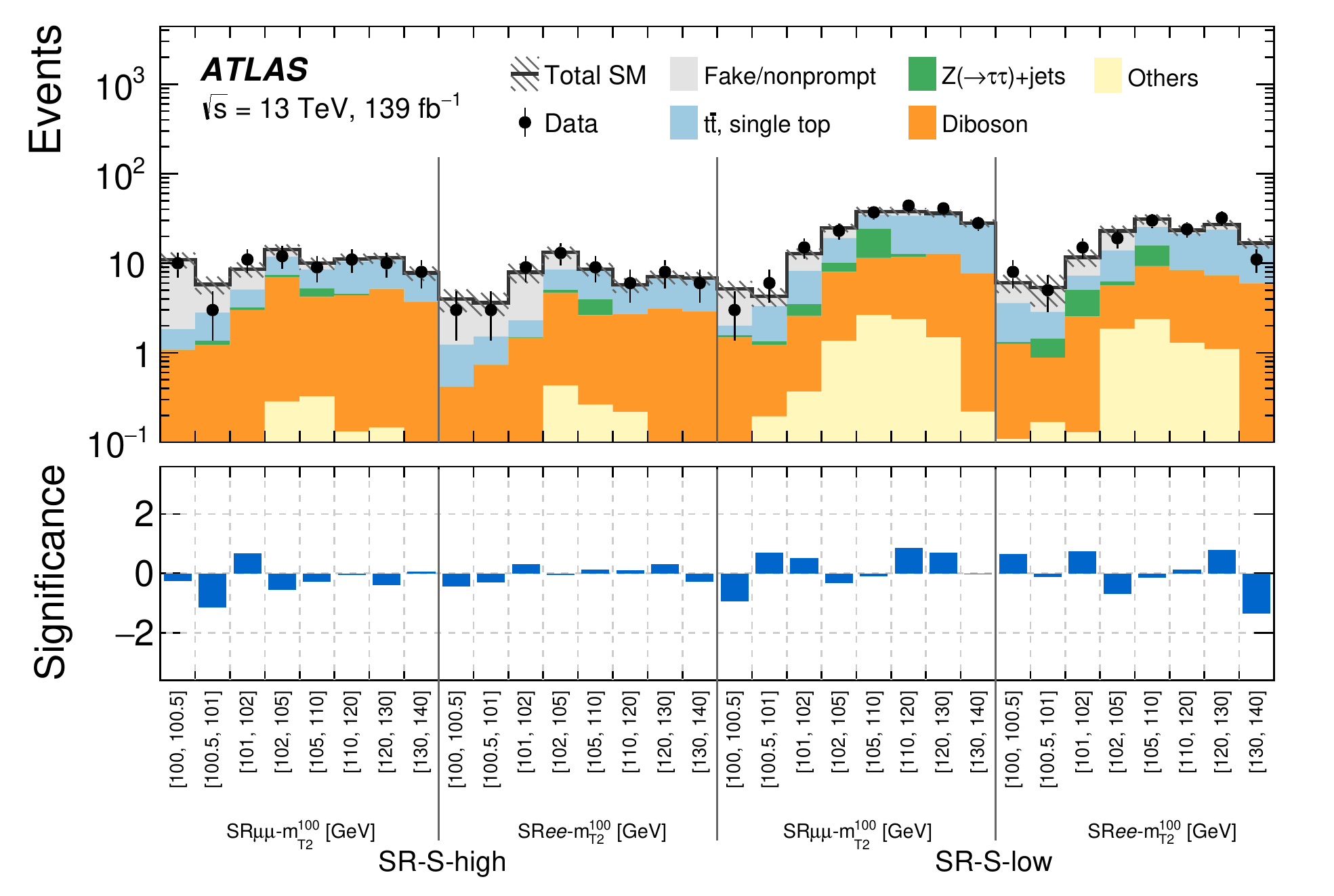}\\
\caption{Comparison of observed and expected event yields in the SRs after the \SRCRBGonlyfits. The SRs used in searches for electroweakinos recoiling against ISR are shown at the top, and the SRs used for the VBF electroweakino search are shown in the middle, all binned in $\mll$. The SRs used in searches for sleptons recoiling against ISR are shown at the bottom, binned in $\mtth$. Uncertainties in the background estimates include both the statistical and systematic uncertainties.  The bottom panel in all three plots shows the significance of the difference between the expected and observed yields, computed following the profile likelihood method of Ref.~\cite{2008NIMPA.595..480C} in the case where the observed yield exceeds the prediction, and using the same expression with an overall minus sign if the yield is below the prediction.}
\label{fig:results:mllmt2}
\end{figure}

\begin{figure}[tbp]
\centering
\includegraphics[width=0.45\columnwidth]{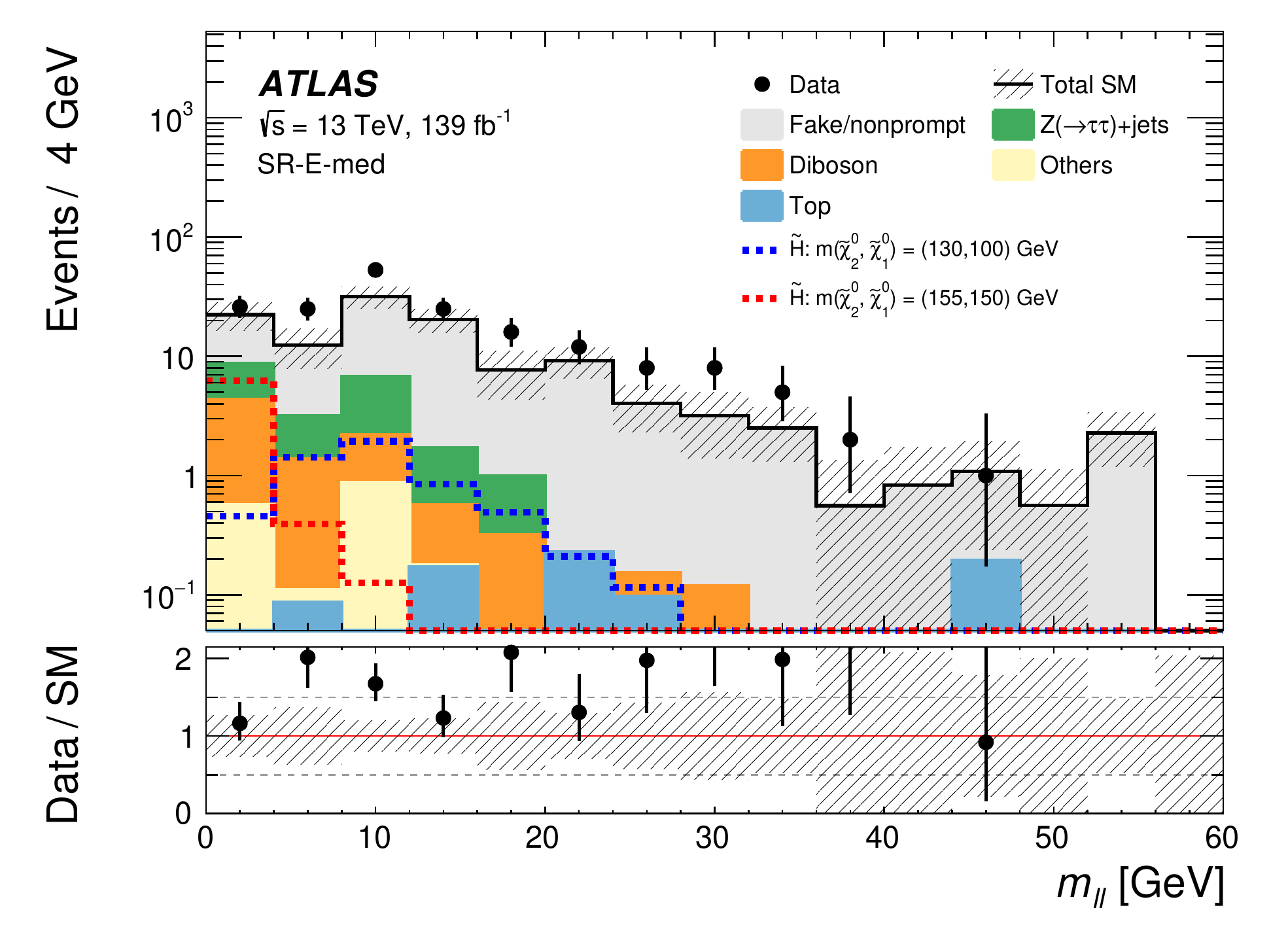}
\includegraphics[width=0.45\columnwidth]{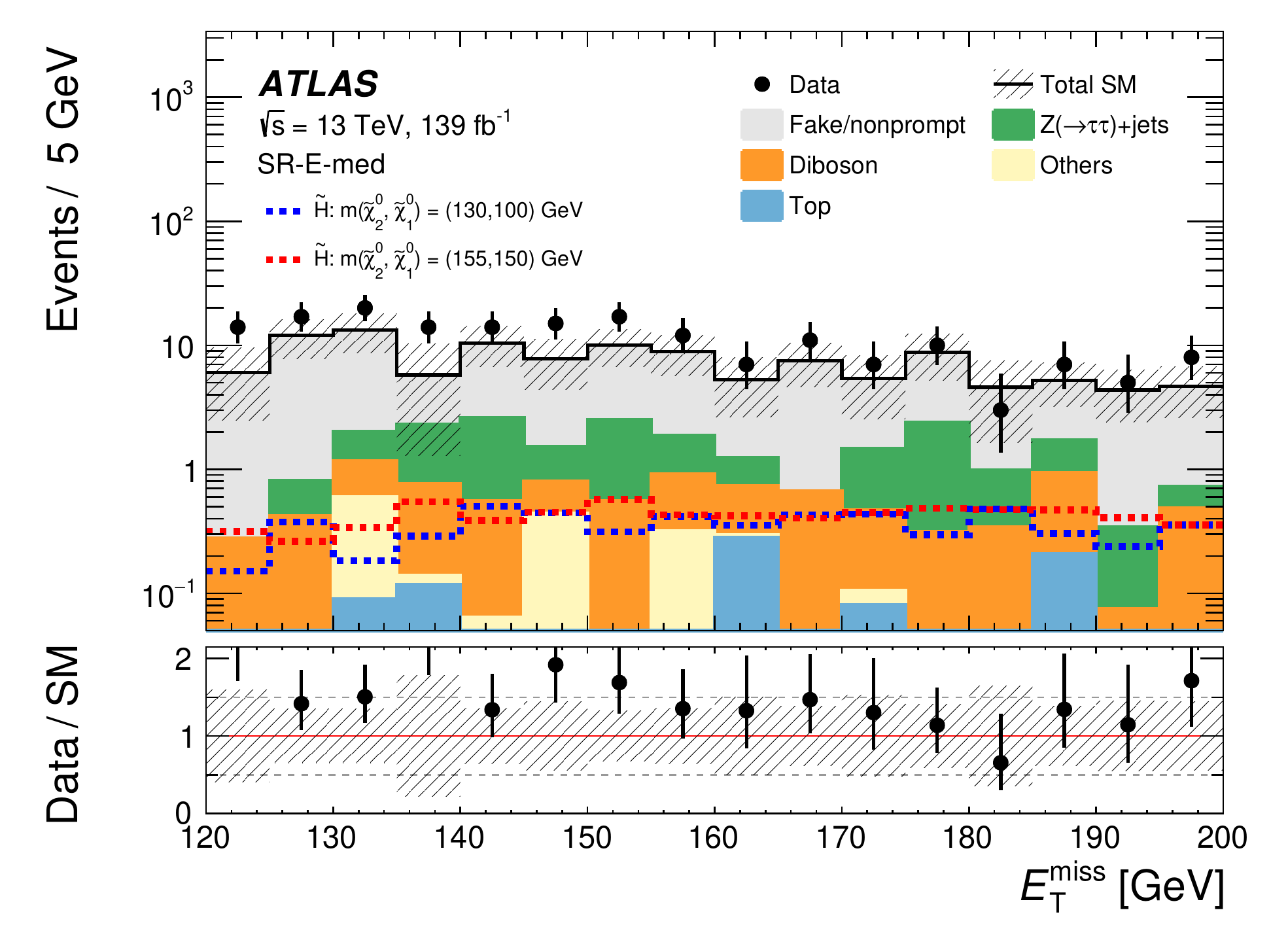}\\
\includegraphics[width=0.45\columnwidth]{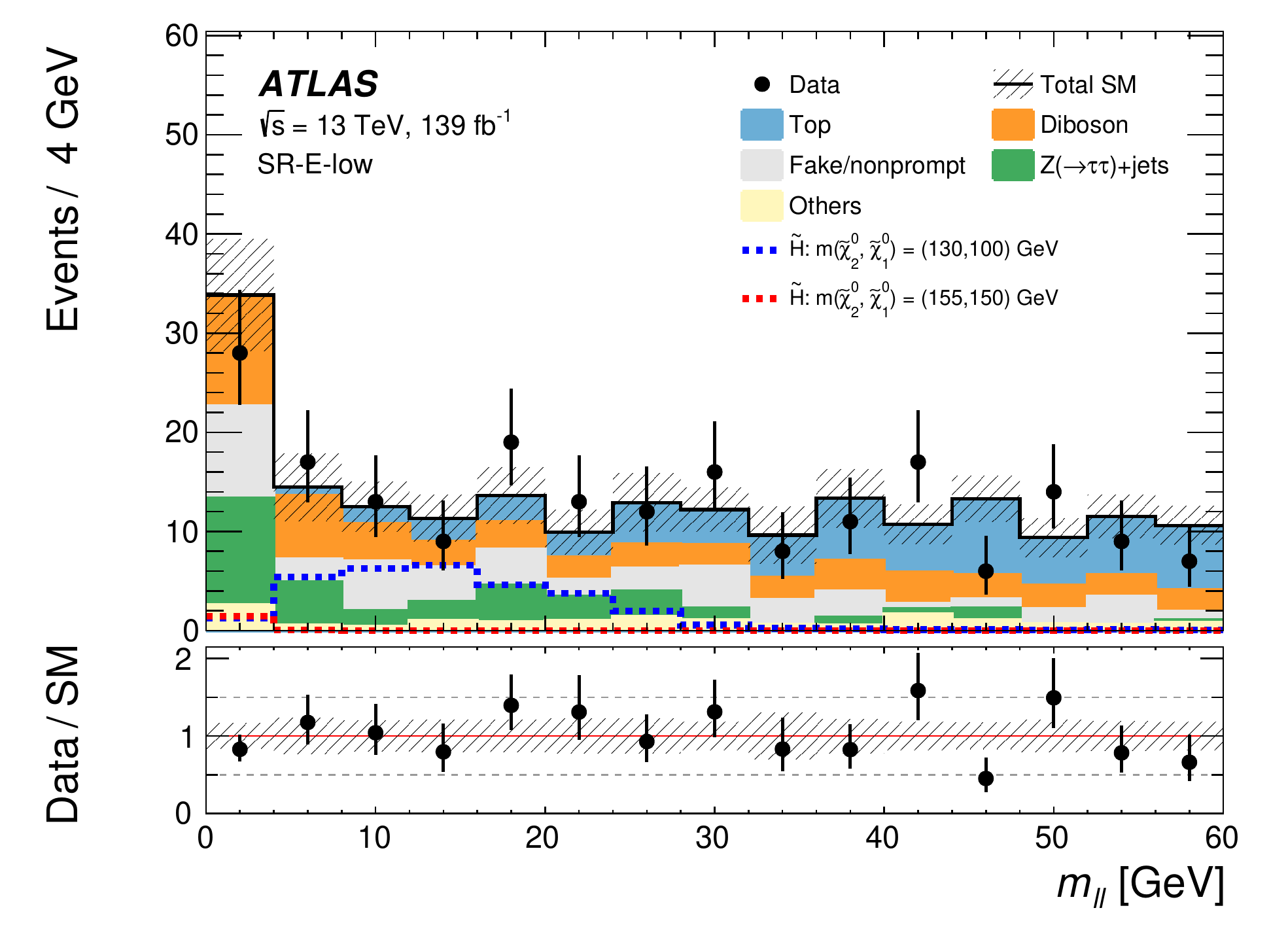}
\includegraphics[width=0.45\columnwidth]{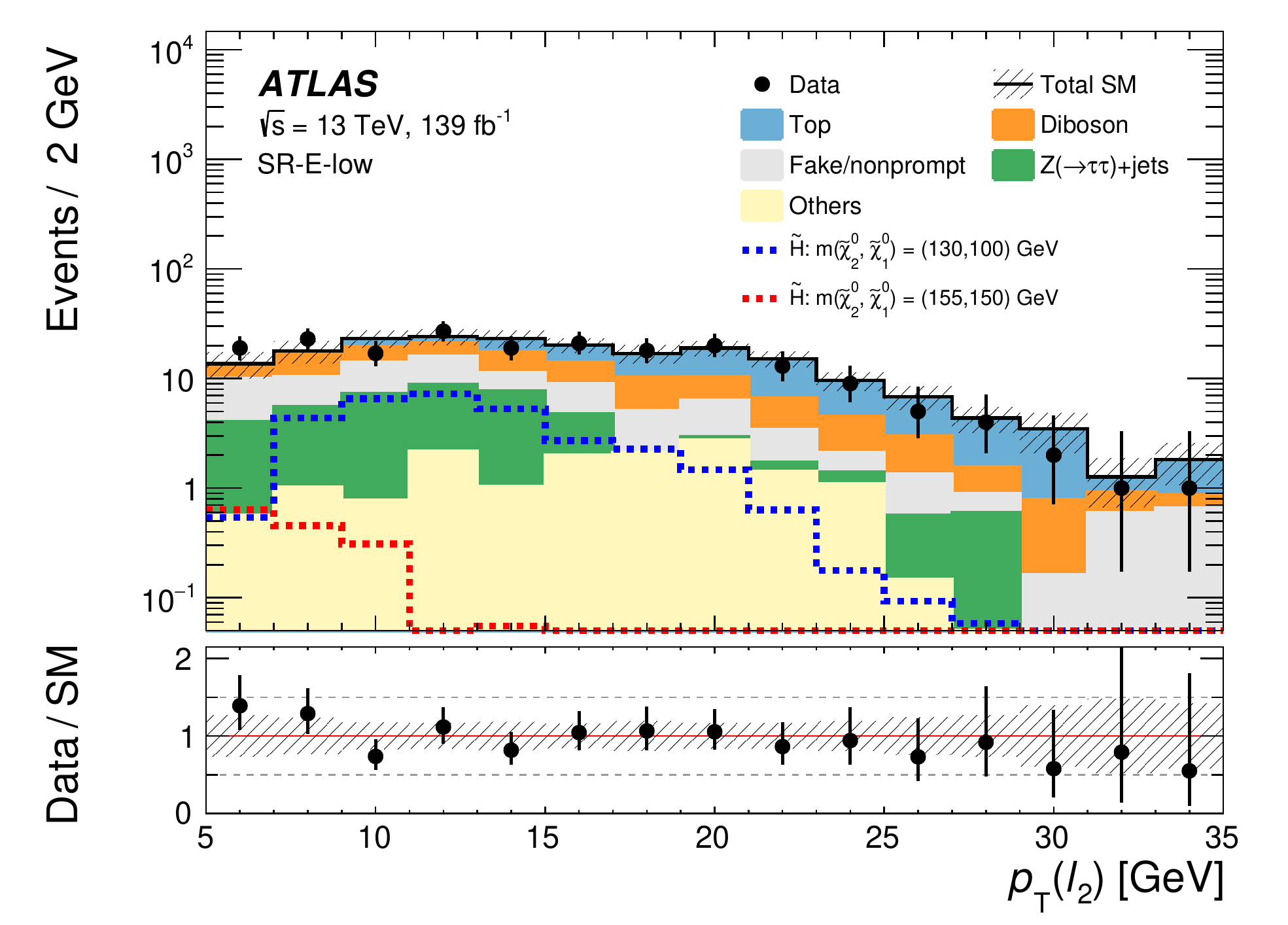}\\
\includegraphics[width=0.45\columnwidth]{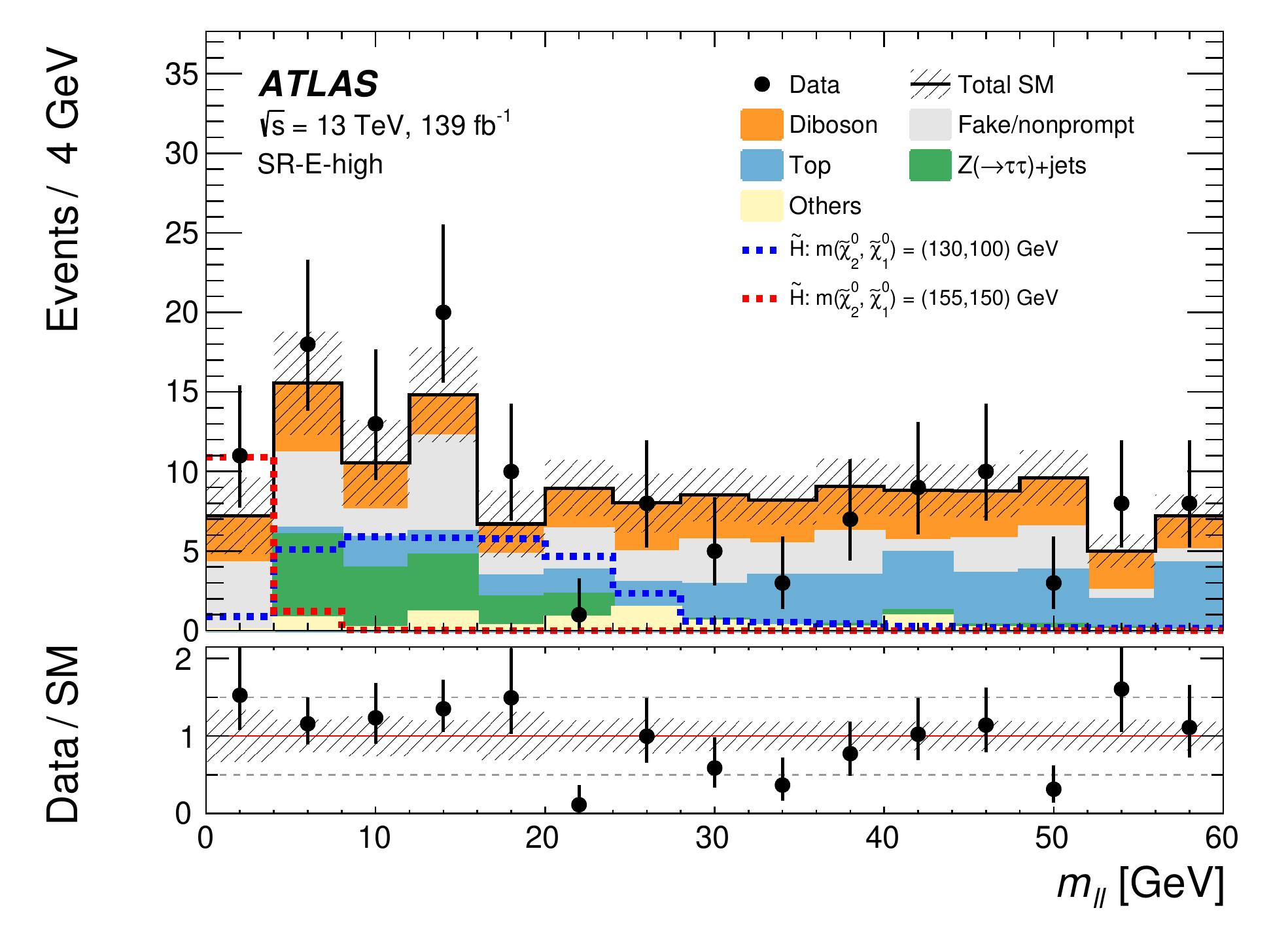}
\includegraphics[width=0.45\columnwidth]{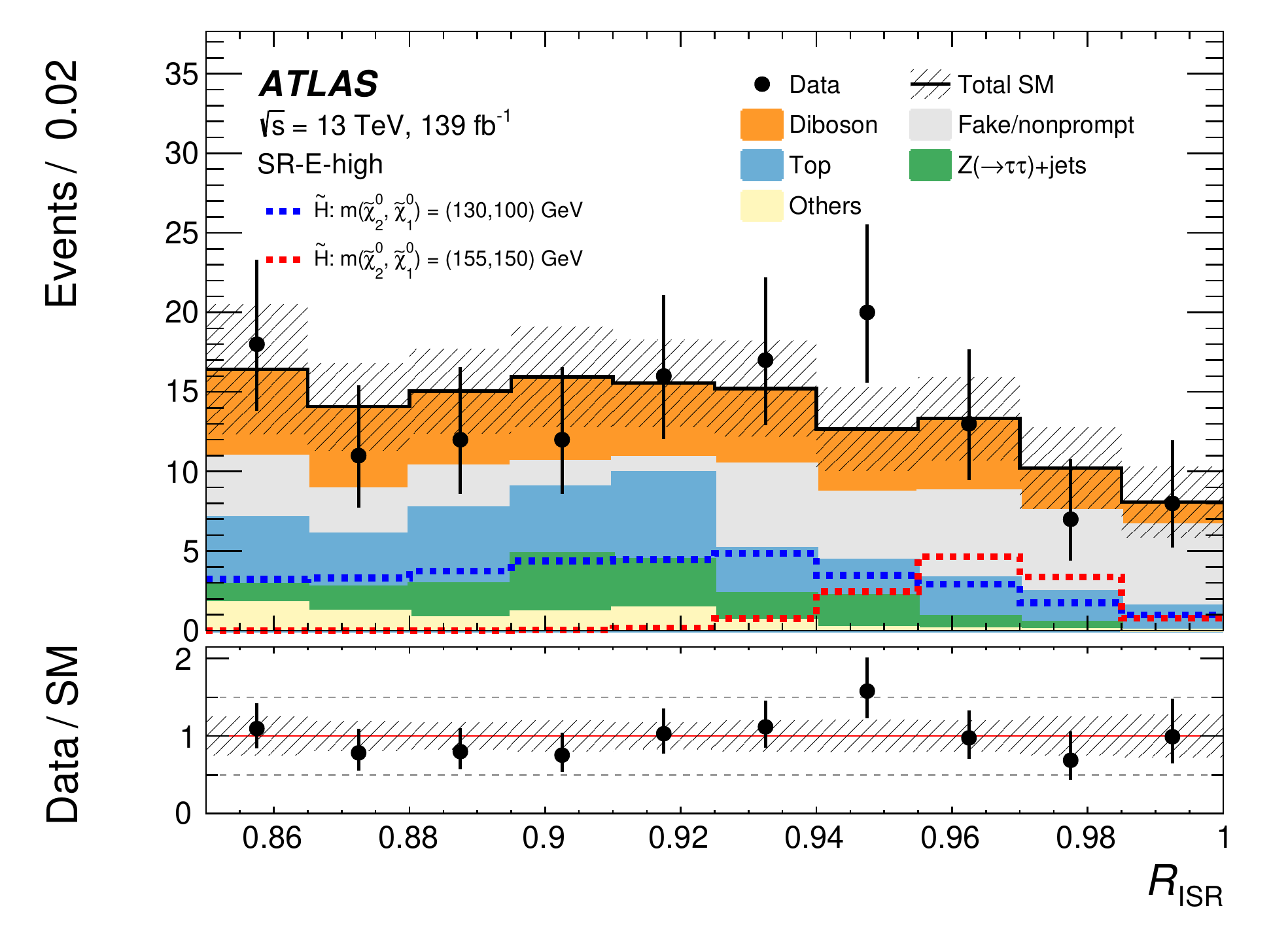}\\
\includegraphics[width=0.45\columnwidth]{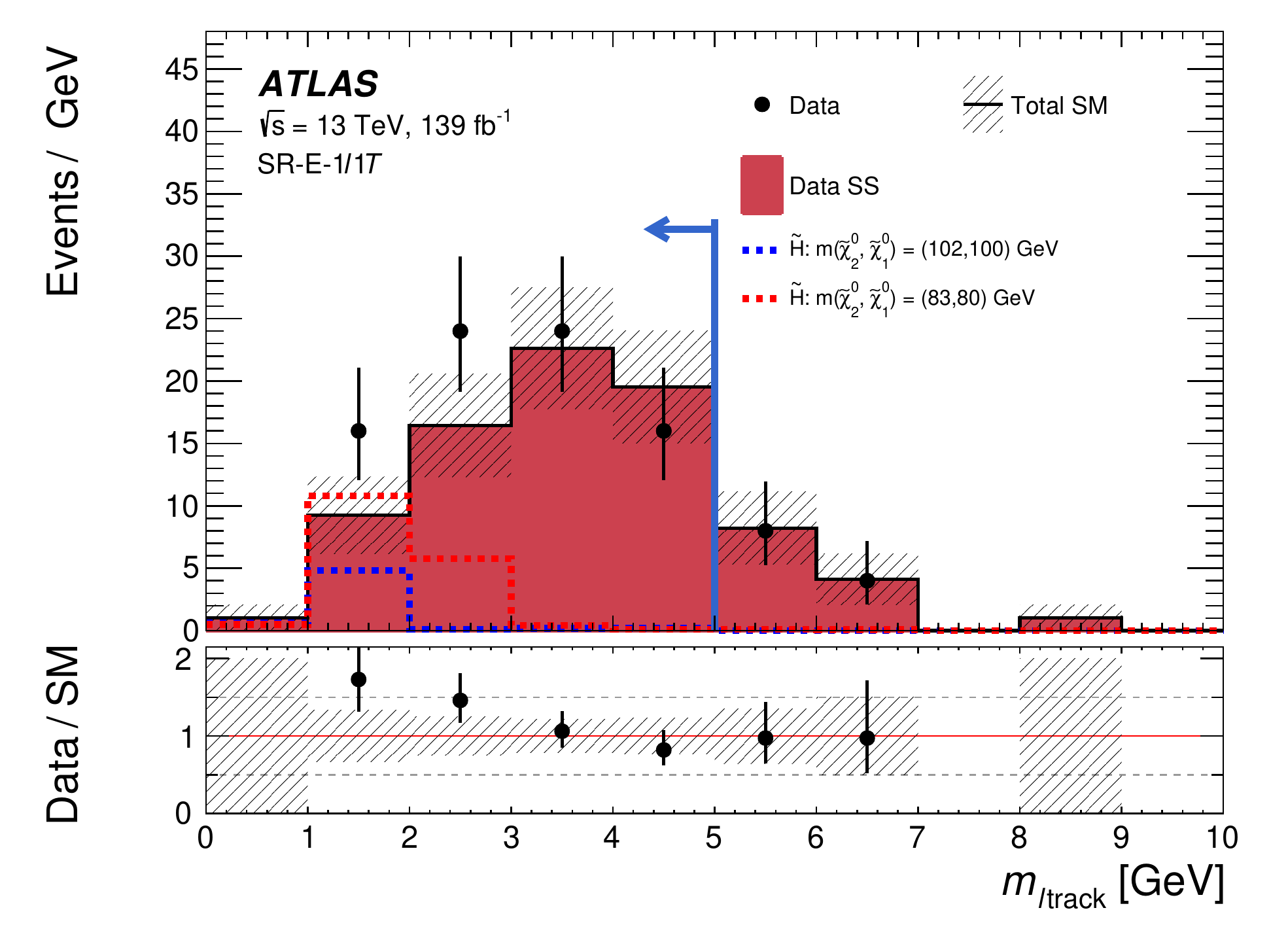}
\includegraphics[width=0.45\columnwidth]{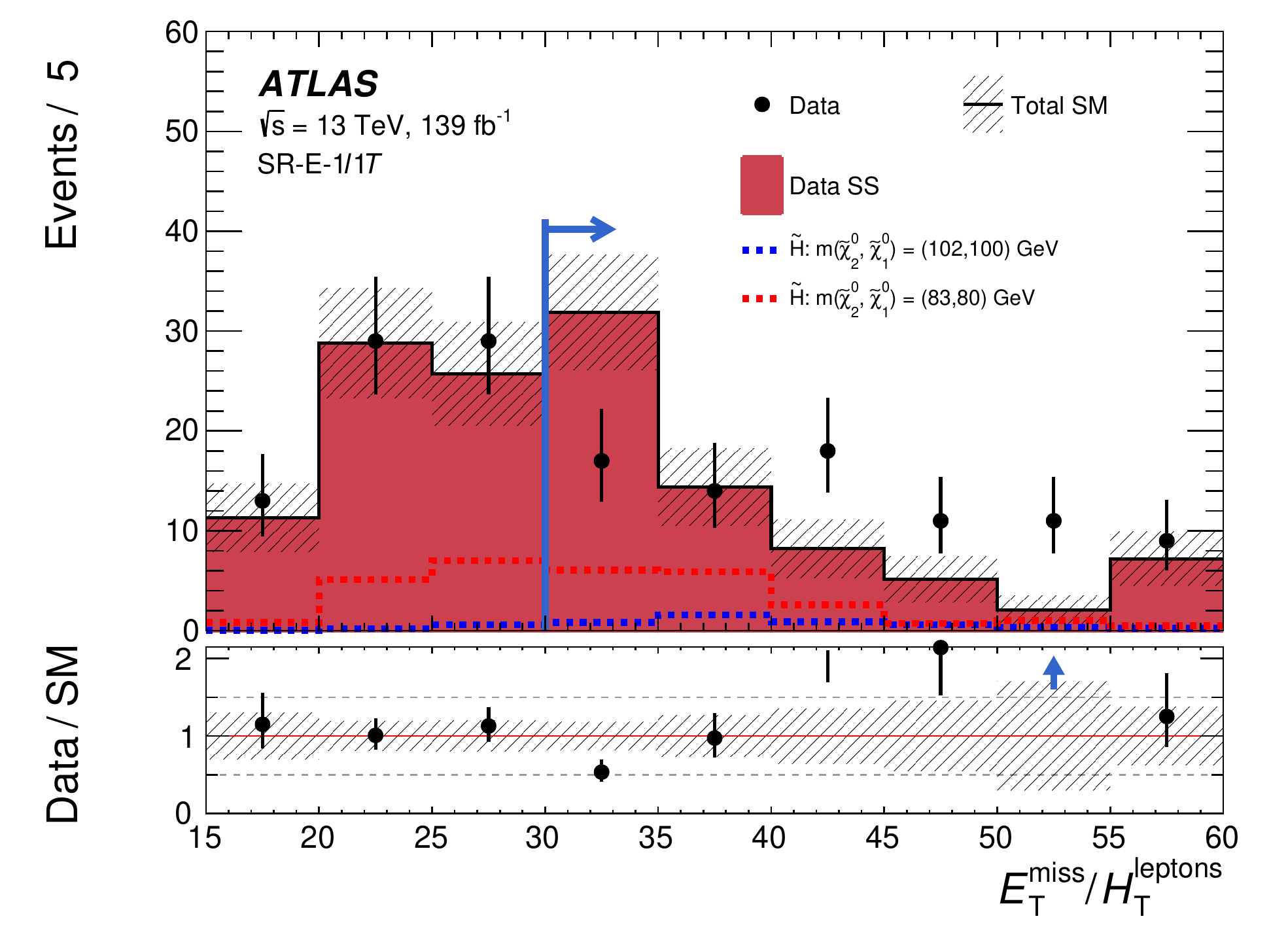}\\
\caption{Examples of kinematic distributions after the \CRBGonlyfit{} showing the data as well as the expected background in the signal regions sensitive to electroweakinos. The full event selection of the corresponding regions is applied, except for distributions showing blue arrows, where the requirement on the variable being plotted is removed and indicated by the arrows in the distributions instead. The first~(last) bin includes underflow~(overflow). The uncertainty bands plotted include all statistical and systematic uncertainties.}
\label{fig:results:ewkinoSRkinematics}
\end{figure}

\begin{figure}[tbp]
\centering
\includegraphics[width=0.48\linewidth]{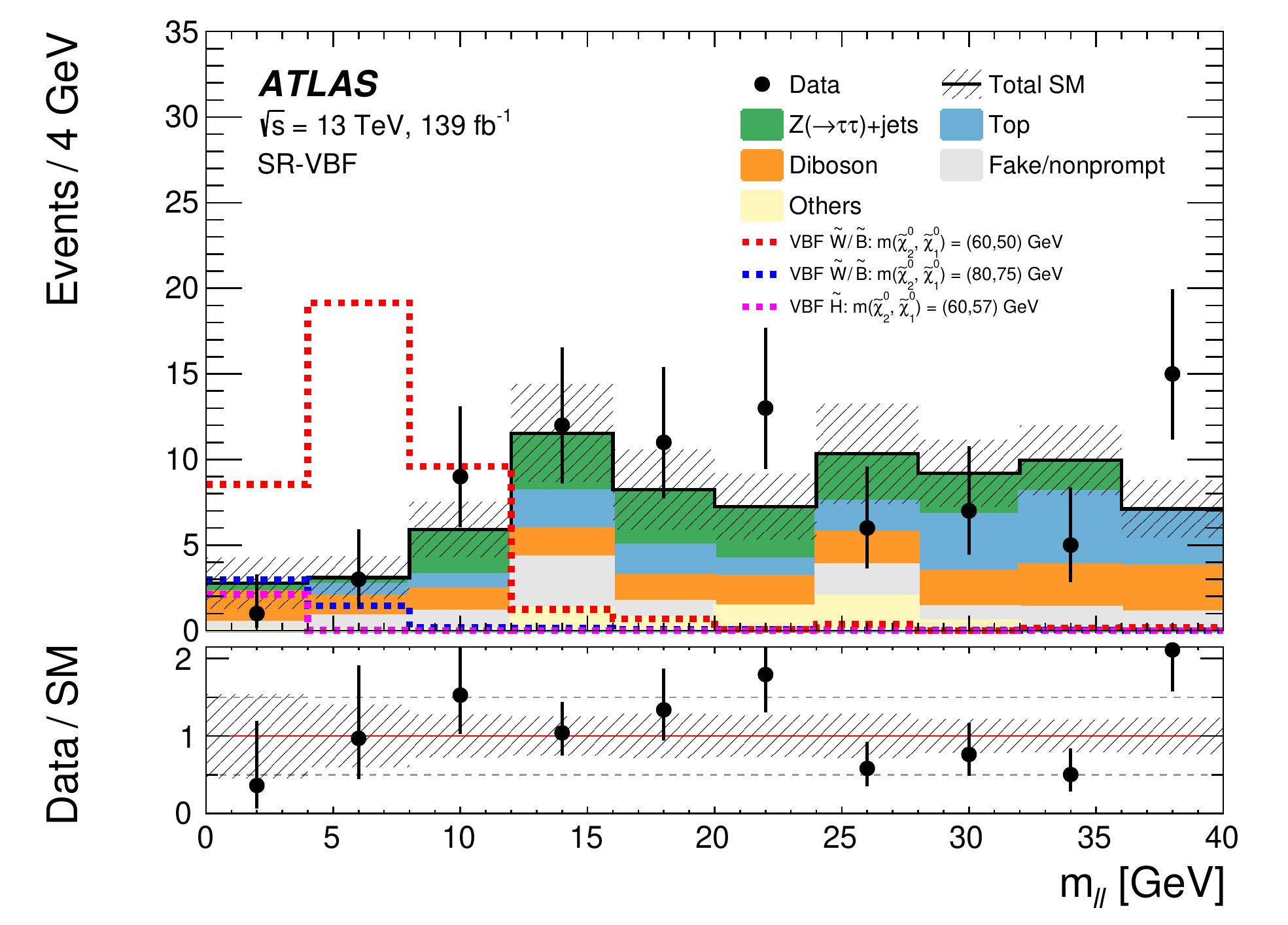}
\includegraphics[width=0.48\linewidth]{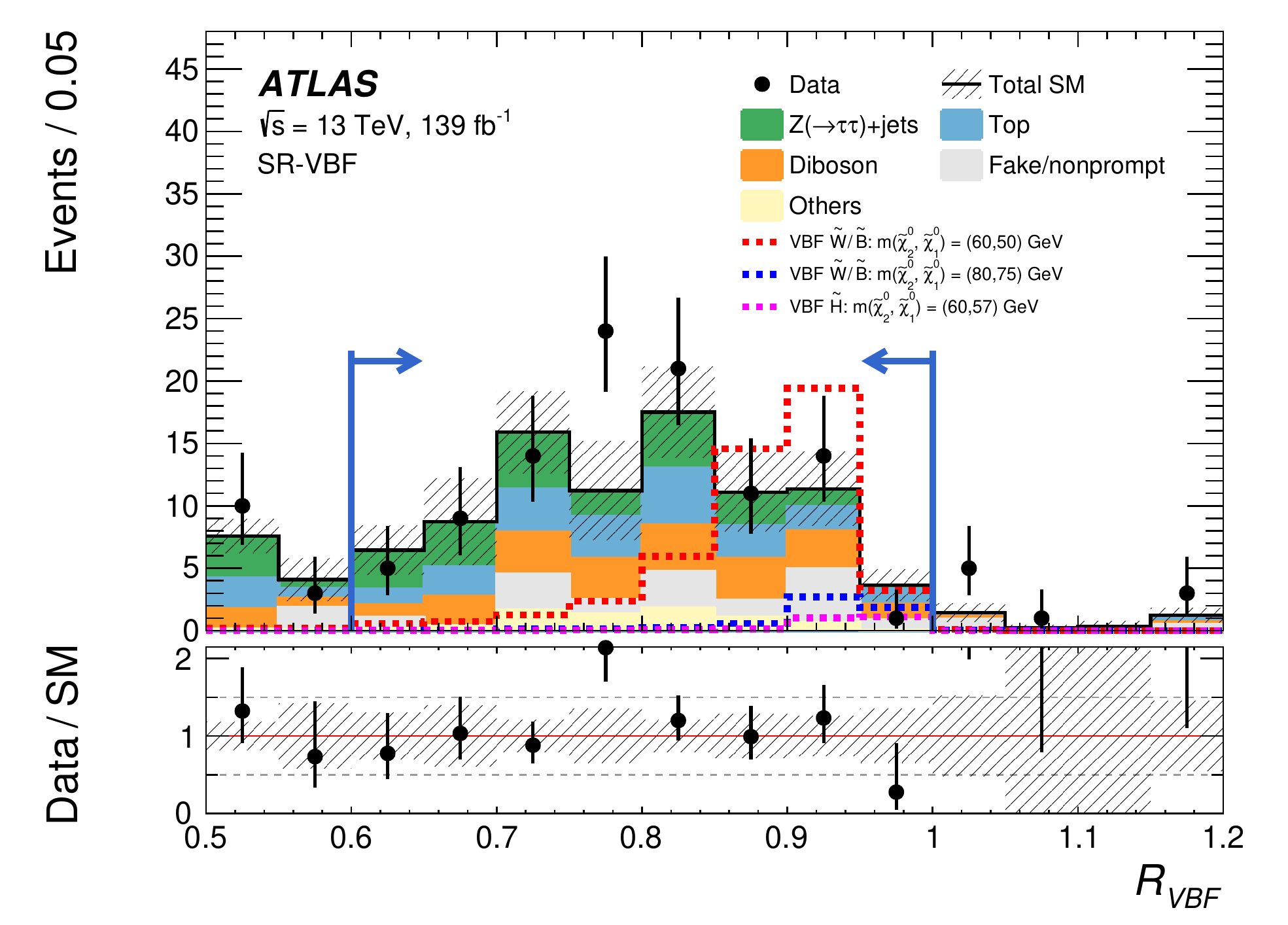}\\
\includegraphics[width=0.48\linewidth]{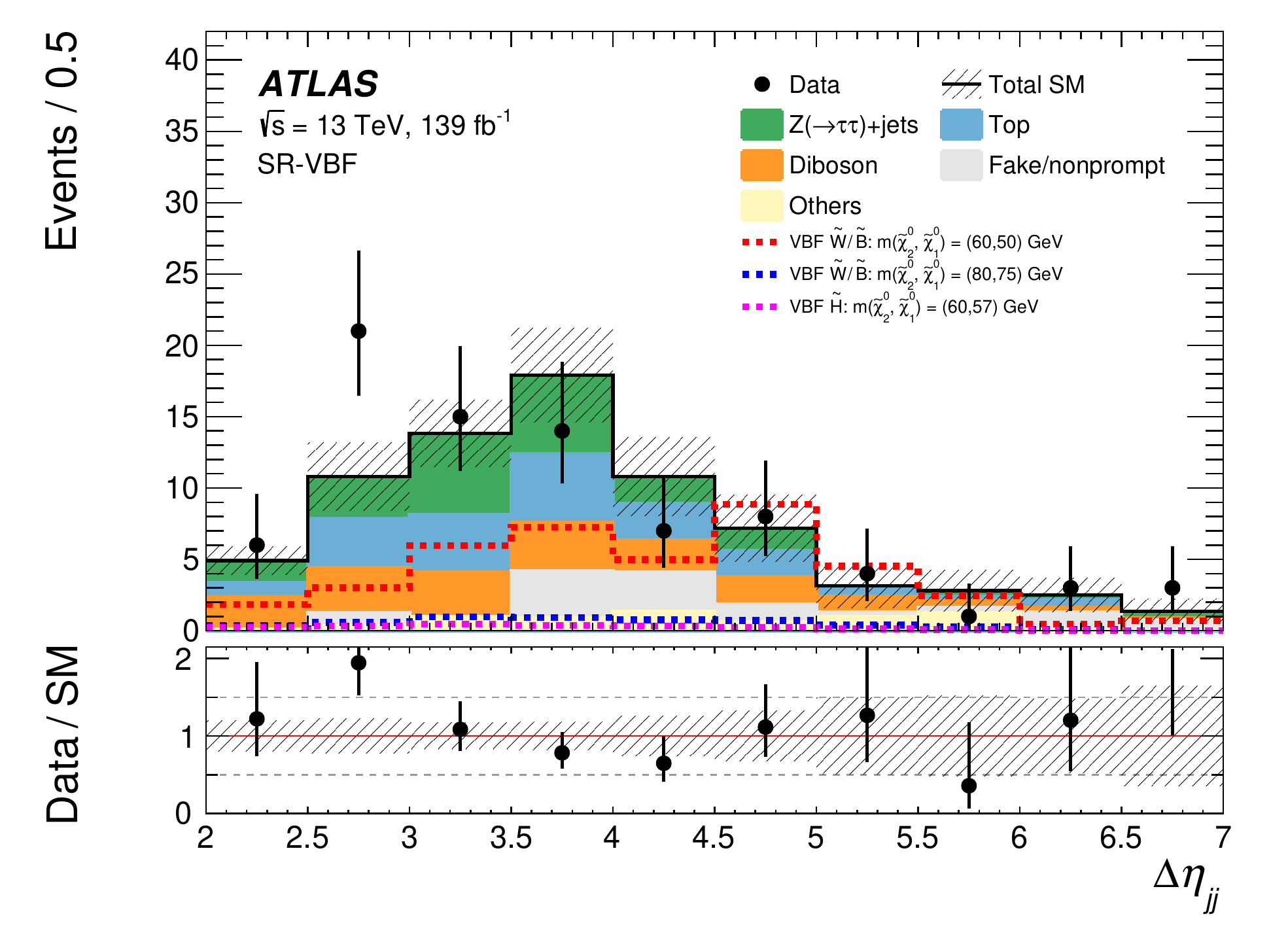}
\includegraphics[width=0.48\linewidth]{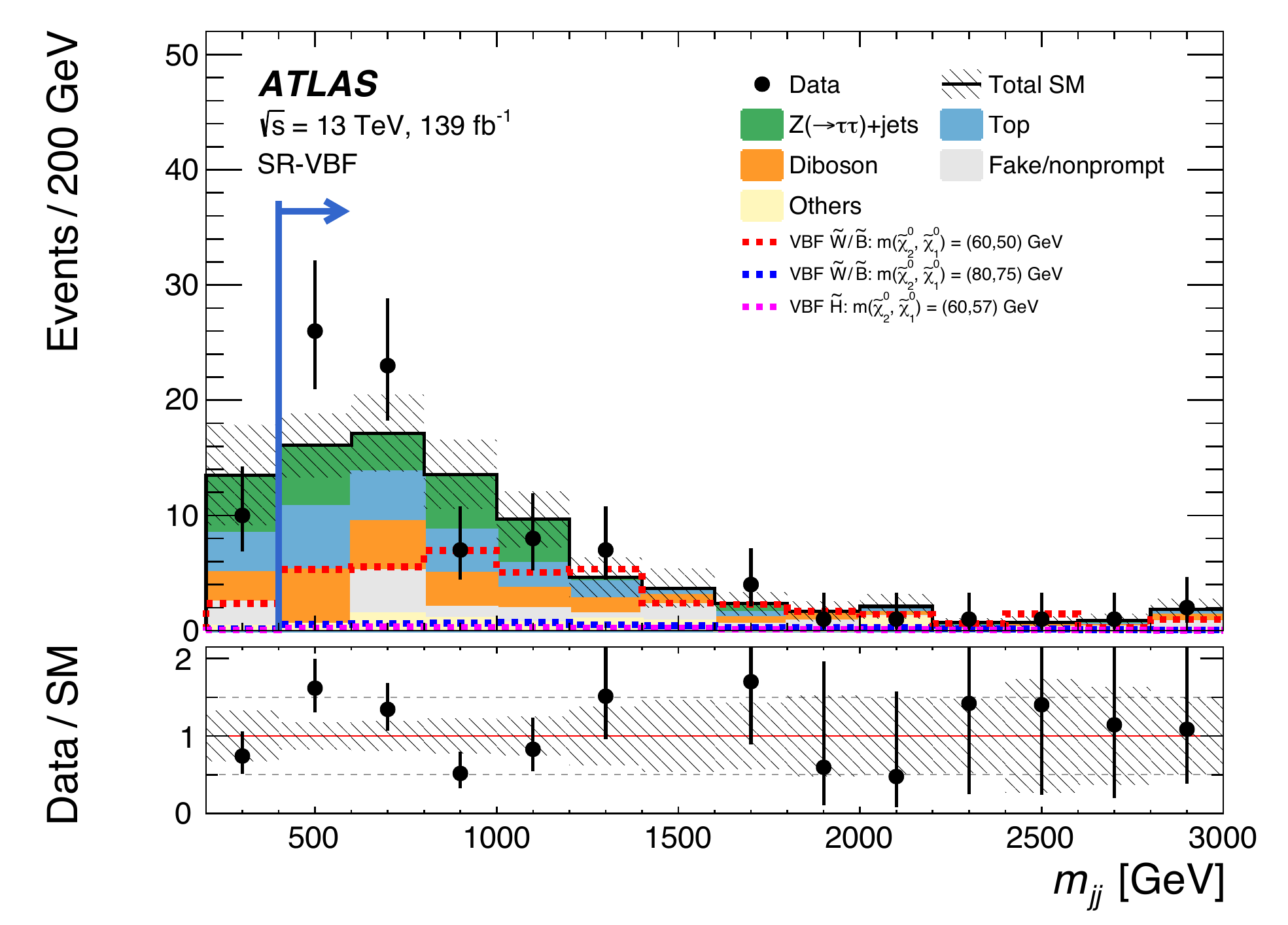}\\
\caption{Examples of kinematic distributions after the \CRBGonlyfit{} showing the data as well as the expected background in the signal regions sensitive to electroweakinos produced through VBF. The full event selection of the corresponding regions is applied, except for distributions showing blue arrows, where the requirement on the variable being plotted is removed and indicated by the arrows instead in the distributions. The first~(last) bin includes underflow~(overflow). The uncertainty bands plotted include all statistical and systematic uncertainties.}
\label{fig:results:vbfSRkinematics}
\end{figure}

\begin{figure}[tbp]
\centering
\includegraphics[width=0.45\columnwidth]{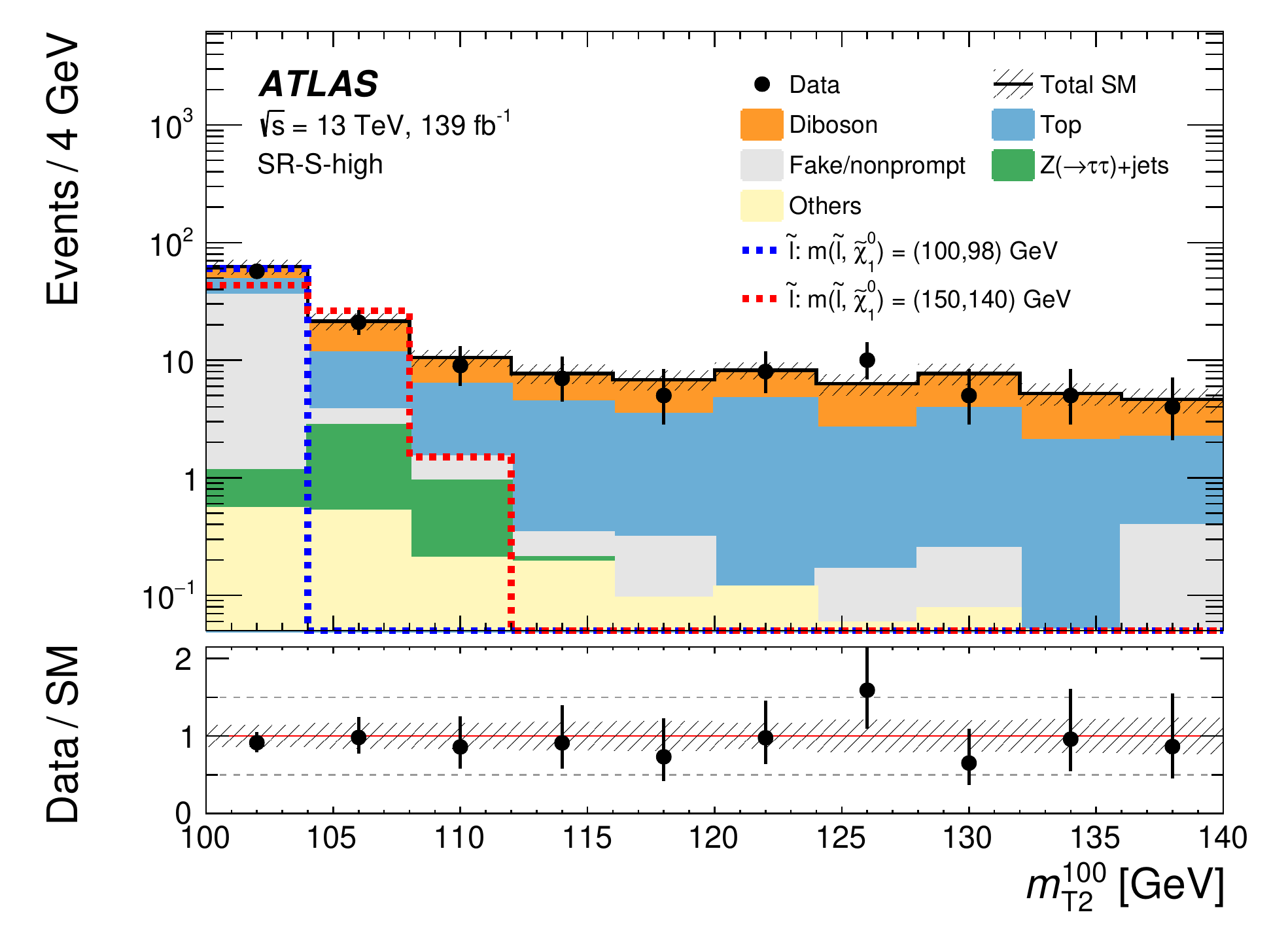}
\includegraphics[width=0.45\columnwidth]{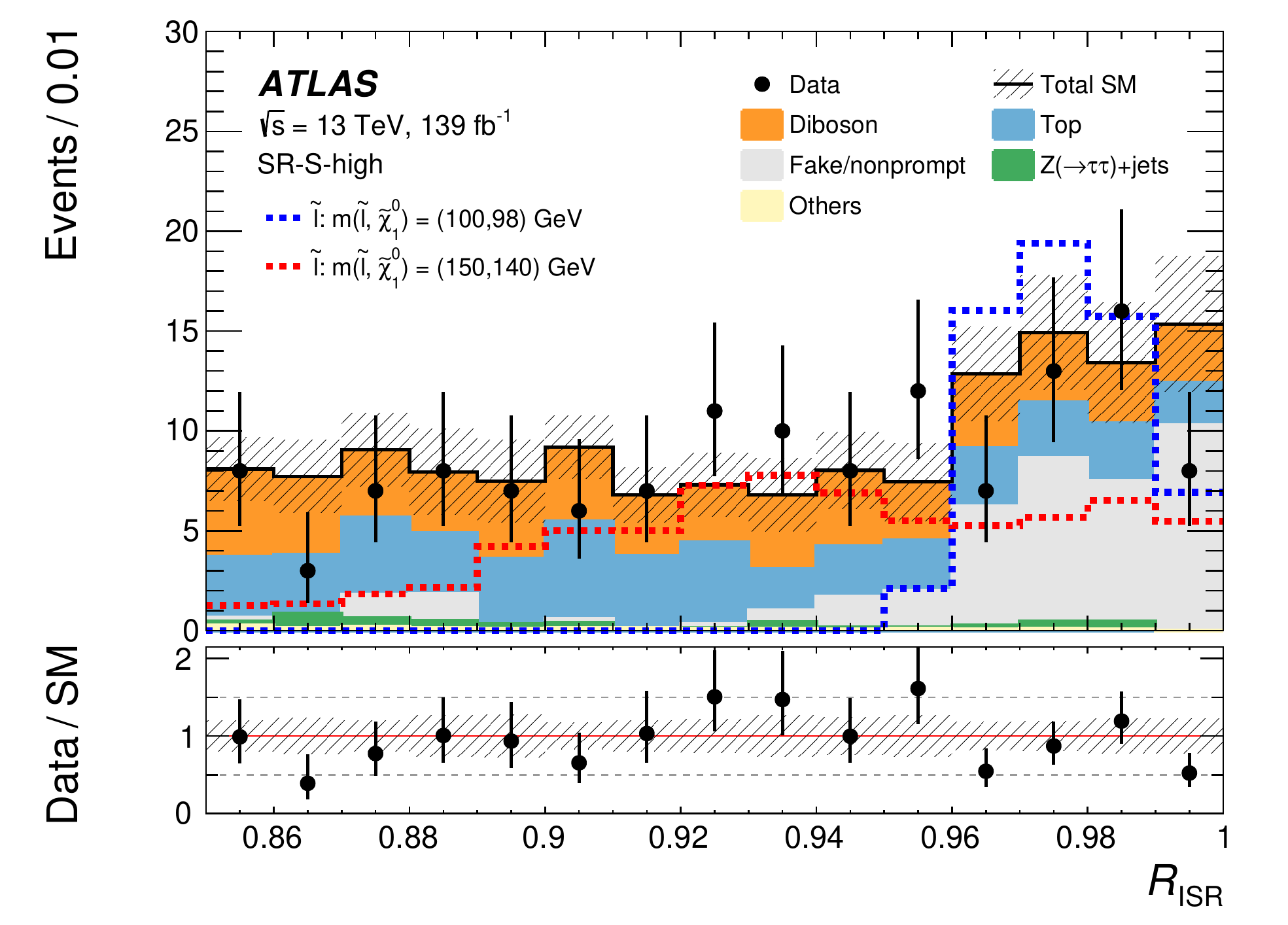}\\
\includegraphics[width=0.45\columnwidth]{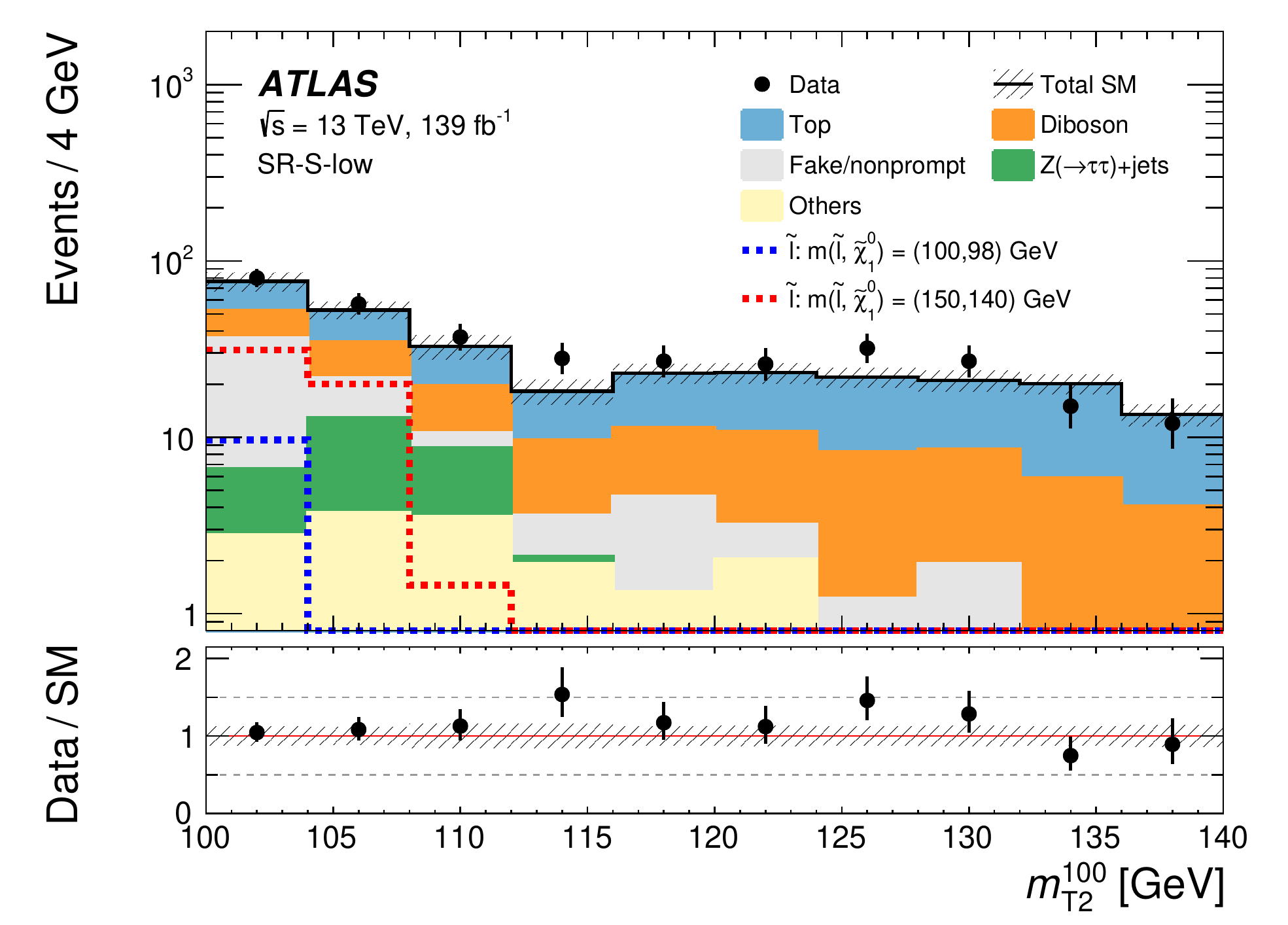}
\includegraphics[width=0.45\columnwidth]{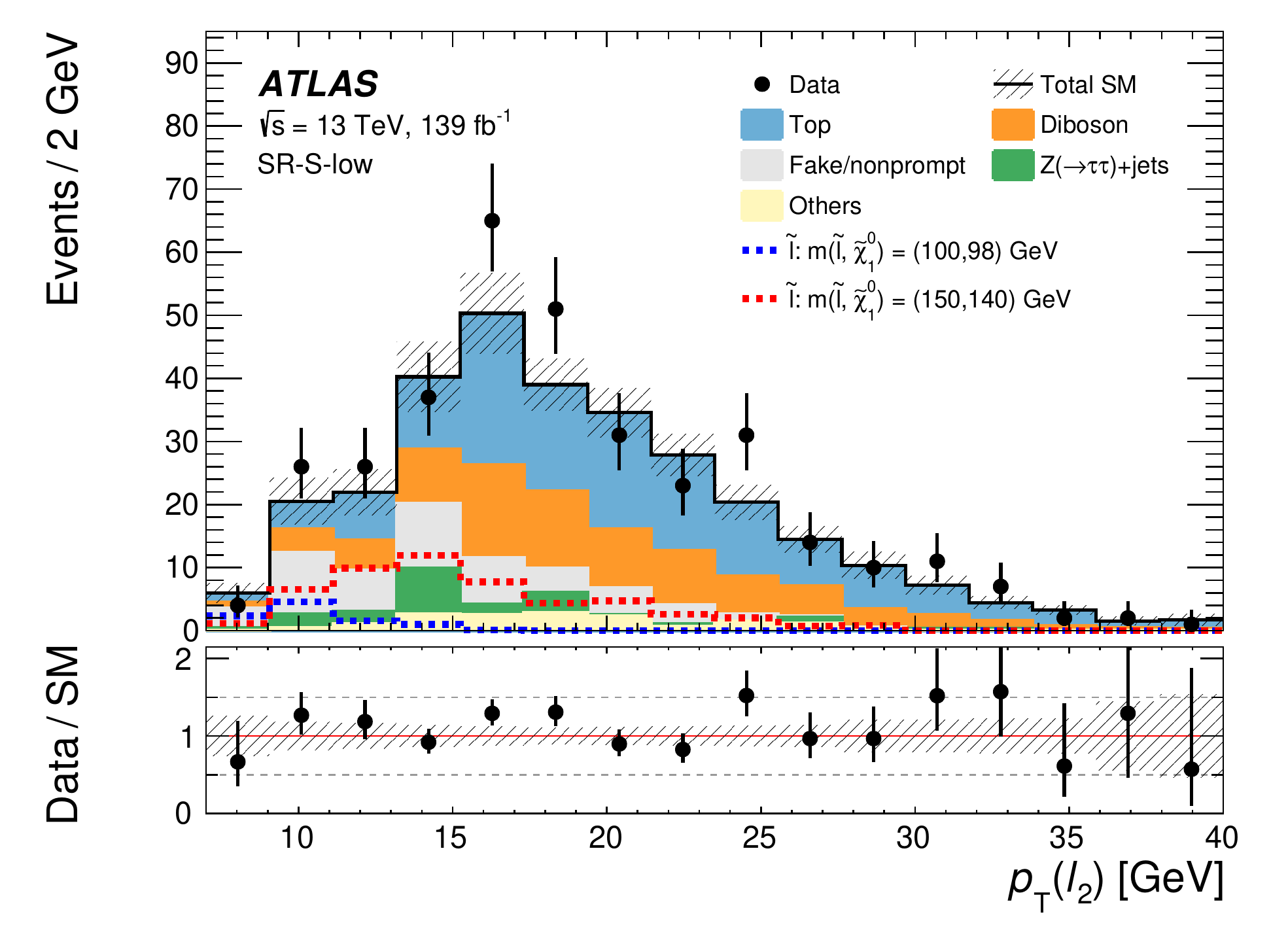}\\
\caption{Examples of kinematic distributions after the \CRBGonlyfit{} showing the data as well as the expected background in the signal regions sensitive to sleptons. The full event selection of the corresponding regions is applied, except for distributions showing blue arrows, where the requirement on the variable being plotted is removed and indicated by the arrows instead in the distributions. The first~(last) bin includes underflow~(overflow). The uncertainty bands plotted include all statistical and systematic uncertainties.}
\label{fig:results:sleptonSRkinematics}
\end{figure}

The exclusive signal regions are used to constrain specific SUSY models.
An exclusion fit extends a \CRBGonlyfit{} to include signal regions relevant for the model under study.
All regions are fit simultaneously with a parameter of interest corresponding to the signal strength, a factor that
coherently scales the signal yield across all regions.
In order to assess the stability of the exclusion fit,
a ``\SRCRBGonlyfit'' of the CRs and the exclusive signal regions is performed in which the signal strength is fixed to zero.
Comparisons of the data yields with background prediction in the $\mll$ and $\mtth$ bins of the SRs,
after the \SRCRBGonlyfit, are shown in Tables~\ref{tab:yields:higgsino:all}--\ref{tab:yields:slepton:all}
and Figure~\ref{fig:results:mllmt2}, with all deviations less than $2\sigma$.
Examples of kinematic distributions in the SRs after a \CRBGonlyfit{} are presented in
Figures~\ref{fig:results:ewkinoSRkinematics},  \ref{fig:results:vbfSRkinematics} and~\ref{fig:results:sleptonSRkinematics},
where good agreement between data and the background predictions is seen in both the shape and the normalization of the
discriminating variables.

 
\begin{table}[tbp]
\caption{Observed event yields and fit results using a \SRCRBGonlyfit\ for the exclusive electroweakino signal regions. Background processes containing fewer than two prompt leptons are categorized as `Fake/nonprompt'. The category `Others' contains rare backgrounds from triboson, Higgs boson, and the remaining top-quark production processes listed in Table 1. Uncertainties in the fitted background estimates combine statistical and systematic uncertainties.}
\maxsizebox*{\textwidth}{\textheight}{
\begin{tabular}{l l rrrrrrrr }
\toprule
\toprule
&SR bin [\GeV] &{[1,2]} & {[2,3]} &{[3.2,5]} & {[5,10]} & {[10,20]} & {[20,30]} & {[30,40]} & {[40,60]} \\
\toprule
\toprule
\vtlab{\large SR--E--high $ee$}{7}
&Observed  &&& $1$ & $16$ & $13$ & $8$ & $8$ & $18$ \\
\cmidrule{2-10}
\cmidrule{2-10}
&Fitted SM events &&& $0.7 \pm 0.4$ & $10.3 \pm 2.5$ & $12.1 \pm 2.2$ & $10.1 \pm 1.7$ & $10.4 \pm 1.7$ & $19.3 \pm 2.5$ \\
\cmidrule{2-10}
&Fake/nonprompt &&& $0.03^{+0.19}_{-0.03}$ & $6.6 \pm 2.7$ & $4.6 \pm 2.0$ & $4.0 \pm 1.5$ & $4.4 \pm 1.6$ & $6.7 \pm 2.3$ \\
&$t\bar{t}$, single top &&& $0.01^{+0.06}_{-0.01}$ & $0.59 \pm 0.27$ & $1.9 \pm 0.5$ & $1.6 \pm 0.4$ & $3.3 \pm 0.6$ & $6.4 \pm 0.9$ \\
&Diboson &&& $0.62 \pm 0.23$ & $1.4 \pm 0.5$ & $2.3 \pm 0.7$ & $2.5 \pm 0.7$ & $2.3 \pm 0.6$ & $5.4 \pm 1.3$ \\
&$Z(\to\tau\tau)$+jets &&& $0.06^{+0.29}_{-0.06}$ & $1.7 \pm 0.7$ & $2.6 \pm 1.2$ & $0.93 \pm 0.24$ & $0.04 \pm 0.04$ & $0.62 \pm 0.23$ \\
&Others &&& $0.000^{+0.004}_{-0.000}$ & $0.12 \pm 0.05$ & $0.74 \pm 0.18$ & $1.14 \pm 0.19$ & $0.29 \pm 0.07$ & $0.27 \pm 0.14$ \\
\bottomrule
\toprule
\vtlab{\large SR--E--high $\mu\mu$}{7}
&Observed & $5$ & $5$ & $0$ & $9$ & $23$ & $3$ & $5$ & $20$ \\
\cmidrule{2-10}
\cmidrule{2-10}
&Fitted SM events & $3.4 \pm 1.2$ & $3.5 \pm 1.3$ & $3.9 \pm 1.3$ & $11.0 \pm 2.0$ & $17.8 \pm 2.7$ & $8.3 \pm 1.4$ & $10.1 \pm 1.5$ & $19.6 \pm 2.3$ \\
\cmidrule{2-10}
&Fake/nonprompt & $2.4 \pm 1.2$ & $2.6 \pm 1.4$ & $1.9 \pm 1.0$ & $3.1 \pm 1.7$ & $6.0 \pm 2.8$ & $1.3 \pm 0.8$ & $2.0 \pm 0.9$ & $1.4 \pm 1.3$ \\
&$t\bar{t}$, single top & $0.01^{+0.06}_{-0.01}$ & $0.01^{+0.06}_{-0.01}$ & $0.09 \pm 0.07$ & $0.67 \pm 0.25$ & $2.0 \pm 0.5$ & $2.4 \pm 0.5$ & $3.7 \pm 0.9$ & $10.2 \pm 1.7$ \\
&Diboson & $0.92 \pm 0.32$ & $0.84 \pm 0.32$ & $0.9 \pm 0.4$ & $2.7 \pm 0.7$ & $3.1 \pm 0.8$ & $3.3 \pm 0.8$ & $3.6 \pm 0.8$ & $6.6 \pm 1.5$ \\
&$Z(\to\tau\tau)$+jets & $0.07^{+0.34}_{-0.07}$ & $0.06^{+0.34}_{-0.06}$ & $1.0 \pm 0.4$ & $3.9 \pm 0.9$ & $5.7 \pm 1.6$ & $0.31 \pm 0.25$ & $0.00^{+0.04}_{-0.00}$ & $0.31 \pm 0.16$ \\
&Others & $0.032^{+0.035}_{-0.032}$ & -- & $0.025 \pm 0.018$ & $0.66 \pm 0.33$ & $0.91 \pm 0.14$ & $1.10 \pm 0.18$ & $0.75 \pm 0.16$ & $1.06 \pm 0.09$ \\
\bottomrule
\toprule
\vtlab{\large SR--E--med $ee$}{7}
&Observed  &&& $0$ & $4$ & $11$ & $4$ \\
\cmidrule{2-10}
\cmidrule{2-10}
&Fitted SM events &&& $0.11 \pm 0.08$ & $5.1 \pm 1.6$ & $7.3 \pm 1.9$ & $2.2 \pm 0.9$ \\
\cmidrule{2-10}
&Fake/nonprompt &&& $0.000^{+0.016}_{-0.000}$ & $3.8 \pm 1.3$ & $6.9 \pm 2.0$ & $1.6 \pm 1.1$ \\
&$t\bar{t}$, single top &&& $0.00^{+0.05}_{-0.00}$ & $0.00^{+0.04}_{-0.00}$ & $0.01^{+0.06}_{-0.01}$ & $0.23^{+0.25}_{-0.23}$ \\
&Diboson &&& $0.10 \pm 0.05$ & $0.10 \pm 0.09$ & $0.28 \pm 0.26$ & $0.02^{+0.13}_{-0.02}$ \\
&$Z(\to\tau\tau)$+jets &&& $0.000^{+0.028}_{-0.000}$ & $1.2 \pm 1.2$ & $0.1^{+0.5}_{-0.1}$ & $0.3^{+0.6}_{-0.3}$ \\
&Others &&& $0.000^{+0.012}_{-0.000}$ & -- & -- & -- \\
\bottomrule
\toprule
\vtlab{\large SR--E--med $\mu\mu$}{7}
&Observed & $16$ & $8$ & $6$ & $41$ & $59$ & $21$ \\
\cmidrule{2-10}
\cmidrule{2-10}
&Fitted SM events & $14.6 \pm 2.9$ & $6.9 \pm 2.1$ & $6.2 \pm 1.9$ & $34 \pm 4$ & $52 \pm 6$ & $18.5 \pm 3.2$ \\
\cmidrule{2-10}
&Fake/nonprompt & $7.9 \pm 3.2$ & $4.8 \pm 2.1$ & $5.1 \pm 2.0$ & $27 \pm 5$ & $44 \pm 6$ & $18.2 \pm 3.2$ \\
&$t\bar{t}$, single top & $0.01^{+0.06}_{-0.01}$ & $0.01^{+0.06}_{-0.01}$ & $0.00^{+0.05}_{-0.00}$ & $0.12^{+0.13}_{-0.12}$ & $0.24 \pm 0.08$ & $0.14^{+0.19}_{-0.14}$ \\
&Diboson & $2.3 \pm 0.8$ & $0.9 \pm 0.4$ & $0.73 \pm 0.24$ & $1.9 \pm 0.7$ & $0.87 \pm 0.26$ & $0.13 \pm 0.07$ \\
&$Z(\to\tau\tau)$+jets & $3.8 \pm 1.8$ & $1.2 \pm 0.5$ & $0.3^{+0.6}_{-0.3}$ & $4.9 \pm 1.6$ & $6.1 \pm 2.1$ & $0.02^{+0.29}_{-0.02}$ \\
&Others & $0.5 \pm 0.4$ & $0.000^{+0.026}_{-0.000}$ & $0.036 \pm 0.015$ & $0.019 \pm 0.017$ & $0.9 \pm 0.6$ & -- \\
\bottomrule
\toprule
\vtlab{\large SR--E--low $ee$}{7}
&Observed &&& $7$ & $11$ & $16$ & $16$ & $10$ & $9$ \\
\cmidrule{2-10}
\cmidrule{2-10}
&Fitted SM events &&& $5.3 \pm 1.5$ & $8.6 \pm 1.8$ & $16.7 \pm 2.5$ & $15.5 \pm 2.6$ & $12.9 \pm 2.1$ & $18.8 \pm 2.2$ \\
\cmidrule{2-10}
&Fake/nonprompt &&& $1.6 \pm 1.1$ & $3.8 \pm 1.8$ & $6.2 \pm 2.2$ & $5.8 \pm 2.3$ & $4.2 \pm 1.8$ & $2.8 \pm 1.4$ \\
&$t\bar{t}$, single top &&& $0.015 \pm 0.006$ & $0.32 \pm 0.30$ & $2.8 \pm 0.6$ & $3.4 \pm 1.1$ & $4.5 \pm 0.9$ & $9.7 \pm 1.5$ \\
&Diboson &&& $1.3 \pm 0.6$ & $2.4 \pm 0.8$ & $3.0 \pm 0.7$ & $2.1 \pm 0.7$ & $2.4 \pm 0.7$ & $4.2 \pm 1.0$ \\
&$Z(\to\tau\tau)$+jets &&& $2.5 \pm 1.1$ & $1.8 \pm 0.7$ & $3.9 \pm 1.3$ & $2.8 \pm 1.0$ & $1.4 \pm 0.7$ & $0.07^{+0.20}_{-0.07}$ \\
&Others &&& $0.01^{+0.05}_{-0.01}$ & $0.20 \pm 0.05$ & $0.79 \pm 0.23$ & $1.3 \pm 0.8$ & $0.54 \pm 0.09$ & $2.10 \pm 0.34$ \\
\bottomrule
\toprule
\vtlab{\large SR--E--low $\mu\mu$}{7}
&Observed & $9$ & $7$ & $7$ & $12$ & $17$ & $18$ & $16$ & $44$ \\
\cmidrule{2-10}
\cmidrule{2-10}
&Fitted SM events & $15.4 \pm 2.4$ & $8.0 \pm 1.7$ & $6.5 \pm 1.6$ & $11.3 \pm 1.9$ & $15.6 \pm 2.3$ & $16.7 \pm 2.3$ & $15.3 \pm 2.0$ & $35.9 \pm 3.3$ \\
\cmidrule{2-10}
&Fake/nonprompt & $7.7 \pm 1.9$ & $0.3^{+0.6}_{-0.3}$ & $0.01^{+0.22}_{-0.01}$ & $2.6 \pm 1.3$ & $4.7 \pm 1.9$ & $2.8 \pm 1.6$ & $2.8 \pm 1.6$ & $4.9 \pm 2.3$ \\
&$t\bar{t}$, single top & $0.00^{+0.04}_{-0.00}$ & $0.26 \pm 0.07$ & $0.01^{+0.06}_{-0.01}$ & $1.2 \pm 0.5$ & $3.4 \pm 0.7$ & $5.1 \pm 1.5$ & $7.8 \pm 1.3$ & $18.9 \pm 2.7$ \\
&Diboson & $4.9 \pm 1.3$ & $2.7 \pm 0.7$ & $3.2 \pm 0.9$ & $3.8 \pm 0.9$ & $4.1 \pm 1.0$ & $3.7 \pm 0.9$ & $3.8 \pm 0.8$ & $7.8 \pm 1.6$ \\
&$Z(\to\tau\tau)$+jets & $2.0 \pm 0.7$ & $3.8 \pm 1.1$ & $2.7 \pm 1.2$ & $3.2 \pm 1.1$ & $2.0 \pm 1.2$ & $2.9 \pm 0.8$ & $0.01^{+0.27}_{-0.01}$ & $1.6 \pm 0.6$ \\
&Others & $0.8 \pm 0.5$ & $0.9 \pm 0.8$ & $0.52 \pm 0.24$ & $0.57 \pm 0.16$ & $1.32 \pm 0.18$ & $2.1 \pm 0.4$ & $0.94 \pm 0.11$ & $2.60 \pm 0.20$ \\
\bottomrule
\bottomrule
\end{tabular}
}
\label{tab:yields:higgsino:all}
\end{table}
 
\begin{table}[tbp]
\centering
\caption{Observed event yields and fit results using a \SRCRBGonlyfit\ for the exclusive electroweakino \oneleponetrack{} regions. All backgrounds are determined from the same-sign method.  Uncertainties in the fitted background estimates combine statistical and systematic uncertainties.}
\begin{tabular}{l rrrrrr }
\toprule
\toprule
SR bin [\GeV] & {[0.5,1.0]} & {[1.0,1.5]} & {[1.5,2.0]} & {[2.0,3.0]} & {[3.2,4.0]} & {[4.0,5.0]} \\
\midrule
Observed & $0$ & $8$ & $8$ & $24$ & $24$ & $16$ \\
\midrule
Fitted SM events & $0.5 \pm 0.5$ & $6.0 \pm 1.9$ & $7.6 \pm 2.1$ & $20.7 \pm 3.4$ & $24 \pm 4$ & $18.1 \pm 3.1$ \\
\bottomrule
\bottomrule
\end{tabular}
\end{table}
 
 
\begin{table}[tbp]
\caption{Observed event yields and fit results using a \SRCRBGonlyfit\ for the exclusive VBF signal regions. Background processes containing fewer than two prompt leptons are categorized as `Fake/nonprompt'. The category `Others' contains rare backgrounds from triboson, Higgs boson, and the remaining top-quark production processes listed in Table 1. Uncertainties in the fitted background estimates combine statistical and systematic uncertainties.}
\maxsizebox*{\textwidth}{\textheight}{
\begin{tabular}{l l rrrrrrr }
\toprule
\toprule
&SR bin [\GeV] &{[1,2]} & {[2,3]} &{[3.2,5]} & {[5,10]} & {[10,20]} & {[20,30]} & {[30,40]} \\
\toprule
\toprule
\vtlab{\large SR--VBF--low}{7}
&Observed  & $0$              & $0$              & $0$              & $6$              & $21$              & $14$              & $15$                     \\
\cmidrule{2-9}
\cmidrule{2-9}
& Fitted SM events    & $0.7 \pm 0.4$          & $0.47 \pm 0.25$          & $0.64 \pm 0.32$          & $4.9 \pm 1.2$          & $17.3 \pm 2.6$          & $12.5 \pm 1.8$          & $15.2 \pm 2.7$  \\
\cmidrule{2-9}
& $Z(\to\tau\tau)$+jets & $0.11_{-0.11}^{+0.22}$          & $0.17 \pm 0.12$          & $0.009_{-0.009}^{+0.018}$          & $1.8 \pm 0.7$          & $6.4 \pm 1.4$          & $5.7 \pm 1.3$          & $2.6 \pm 1.0$     \\
& Fake/nonprompt & $0.01_{-0.01}^{+0.05}$          & $0.01_{-0.01}^{+0.05}$          & $0.01_{-0.01}^{+0.05}$          & $1.5 \pm 1.0$          & $3.4 \pm 2.0$          & $0.01_{-0.01}^{+0.06}$          & $1.8_{-1.8}^{+2.5}$    \\
& Diboson & $0.57 \pm 0.29$          & $0.28 \pm 0.17$          & $0.35 \pm 0.20$          & $1.0 \pm 0.4$          & $2.8 \pm 1.1$          & $2.7 \pm 1.004$          & $4.0 \pm 1.4$    \\
& $t\bar{t}$, single top & $0.01_{-0.01}^{+0.04}$          & $0.01_{-0.01}^{+0.05}$          & $0.26 \pm 0.18$          & $0.55 \pm 0.27$          & $3.6 \pm 1.3$          & $3.1 \pm 0.7$          & $6.4 \pm 1.1$     \\
& Others & $0.007 \pm 0.007$          & $0.007 \pm 0.004$          & $0.01_{-0.01}^{+0.05}$          & $0.056 \pm 0.026$          & $1.0 \pm 0.4$          & $1.03 \pm 0.32$          & $0.37 \pm 0.13$    \\
\bottomrule
\toprule
\vtlab{\large SR--VBF--high}{7}
& Observed           & $0$              & $1$              & $1$              & $1$              & $6$              & $8$              & $9$             \\
\cmidrule{2-9}
\cmidrule{2-9}
& Fitted SM events       & $1.6 \pm 0.7$          & $0.8 \pm 0.6$          & $0.25 \pm 0.13$          & $0.9 \pm 0.5$          & $7.1 \pm 1.5$          & $8.5 \pm 2.2$          & $7.7 \pm 1.5$      \\
\cmidrule{2-9}
& $Z(\to\tau\tau)$+jets    & $0.009_{-0.009}^{+0.018}$          & $0.010_{-0.010}^{+0.021}$          & $0.012_{-0.012}^{+0.026}$          & $0.19_{-0.19}^{+0.29}$          & $1.7 \pm 0.8$          & $1.8 \pm 1.3$          & $0.27 \pm 0.09$      \\
& Fake/nonprompt     & $1.4 \pm 0.7$          & $0.7 \pm 0.6$          & $0.08_{-0.08}^{+0.11}$          & $0.3_{-0.3}^{+0.5}$          & $1.5 \pm 1.0$          & $1.4_{-1.4}^{+1.5}$          & $1.2 \pm 1.2$     \\
& Diboson      & $0.27 \pm 0.17$          & $0.13 \pm 0.11$          & $0.10 \pm 0.05$          & $0.37 \pm 0.19$          & $1.1 \pm 0.5$          & $1.8 \pm 0.7$          & $2.0 \pm 0.8$        \\
& $t\bar{t}$, single top       & $0.01_{-0.01}^{+0.05}$          & $0.01_{-0.01}^{+0.06}$          & $0.05_{-0.05}^{+0.09}$          & $0.01_{-0.01}^{+0.06}$          & $1.7 \pm 0.6$          & $0.9 \pm 0.6$          & $3.5 \pm 0.8$   \\
& Others & $--$          & $--$          & $--$          & $0.01_{-0.01}^{+0.02}$          & $1.2 \pm 0.4$          & $2.6 \pm 1.5$          & $0.57 \pm 0.21$            \\
\bottomrule
\bottomrule
\end{tabular}
}
\label{tab:yields:vbf:all}
\end{table}
 
 
\begin{table}[tbp]
\caption{Observed event yields and fit results using a \SRCRBGonlyfit\ for the exclusive slepton signal regions. Background processes containing fewer than two prompt leptons are categorized as `Fake/nonprompt'. The category `Others' contains rare backgrounds from triboson, Higgs boson, and the remaining top-quark production processes listed in Table 1. Uncertainties in the fitted background estimates combine statistical and systematic uncertainties.}
\maxsizebox*{\textwidth}{\textheight}{
\begin{tabular}{l l rrrrrrrr }
\toprule
\toprule
&SR bin [\GeV] & {[100,100.5]} & {[100.5,101]} & {[101,102]} & {[102,105]} & {[105,110]} & {[110,120]} & {[120,130]} & {[130,140]} \\
\toprule
\toprule
\vtlab{\large SR-S-high $ee$}{7}
&Observed & $3$ & $3$ & $9$ & $13$ & $9$ & $6$ & $8$ & $6$ \\
\cmidrule{2-10}
\cmidrule{2-10}
&Fitted SM events & $4.0 \pm 1.1$ & $3.6 \pm 1.0$ & $7.9 \pm 1.9$ & $13.2 \pm 2.1$ & $8.6 \pm 1.4$ & $5.7 \pm 1.0$ & $7.0 \pm 1.2$ & $6.8 \pm 1.1$ \\
\cmidrule{2-10}
&Fake/nonprompt & $2.7 \pm 1.1$ & $2.1 \pm 1.0$ & $5.6 \pm 1.9$ & $4.7 \pm 1.9$ & $0.2^{+0.5}_{-0.2}$ & $0.01^{+0.17}_{-0.01}$ & $0.01^{+0.17}_{-0.01}$ & $0.00^{+0.15}_{-0.00}$ \\
&$t\bar{t}$, single top & $0.8 \pm 0.4$ & $0.8 \pm 0.5$ & $0.8 \pm 0.4$ & $3.5 \pm 0.7$ & $4.5 \pm 1.2$ & $3.0 \pm 0.7$ & $3.9 \pm 0.9$ & $3.9 \pm 0.9$ \\
&Diboson & $0.42 \pm 0.16$ & $0.68 \pm 0.23$ & $1.4 \pm 0.4$ & $4.2 \pm 1.1$ & $2.4 \pm 0.7$ & $2.5 \pm 0.7$ & $3.0 \pm 0.8$ & $2.8 \pm 0.7$ \\
&$Z(\to\tau\tau)$+jets & $0.00^{+0.08}_{-0.00}$ & $0.00^{+0.18}_{-0.00}$ & $0.027 \pm 0.012$ & $0.38 \pm 0.16$ & $1.32 \pm 0.31$ & $0.00^{+0.12}_{-0.00}$ & $0.02^{+0.22}_{-0.02}$ & $0.00^{+0.19}_{-0.00}$ \\
&Others & $0.0 \pm 0.0$ & $0.06^{+0.11}_{-0.06}$ & $0.09 \pm 0.05$ & $0.43 \pm 0.32$ & $0.26 \pm 0.14$ & $0.2^{+0.5}_{-0.2}$ & $0.06^{+0.08}_{-0.06}$ & $0.05 \pm 0.05$ \\
\bottomrule
\toprule
\vtlab{\large SR-S-high $\mu\mu$}{7}
&Observed & $10$ & $3$ & $11$ & $12$ & $9$ & $11$ & $10$ & $8$ \\
\cmidrule{2-10}
\cmidrule{2-10}
&Fitted SM events & $11.0 \pm 2.2$ & $5.8 \pm 1.3$ & $8.6 \pm 1.6$ & $14.2 \pm 1.9$ & $10.0 \pm 1.5$ & $11.2 \pm 1.6$ & $11.5 \pm 1.5$ & $7.8 \pm 1.4$ \\
\cmidrule{2-10}
&Fake/nonprompt & $9.1 \pm 2.2$ & $3.0 \pm 1.1$ & $3.5 \pm 1.4$ & $2.4 \pm 1.2$ & $1.5 \pm 1.0$ & $0.7^{+0.8}_{-0.7}$ & $0.4^{+0.5}_{-0.4}$ & $0.19^{+0.33}_{-0.19}$ \\
&$t\bar{t}$, single top & $0.8 \pm 0.5$ & $1.5 \pm 0.5$ & $1.9 \pm 0.5$ & $4.4 \pm 0.8$ & $3.3 \pm 0.7$ & $5.9 \pm 1.1$ & $5.9 \pm 0.9$ & $3.9 \pm 1.3$ \\
&Diboson & $1.1 \pm 0.4$ & $1.2 \pm 0.4$ & $2.9 \pm 1.3$ & $6.7 \pm 1.7$ & $3.9 \pm 1.1$ & $4.2 \pm 1.0$ & $5.0 \pm 1.3$ & $3.7 \pm 0.9$ \\
&$Z(\to\tau\tau)$+jets & $0.00^{+0.19}_{-0.00}$ & $0.15 \pm 0.04$ & $0.22 \pm 0.19$ & $0.40 \pm 0.34$ & $1.03 \pm 0.34$ & $0.19 \pm 0.12$ & $0.00^{+0.19}_{-0.00}$ & $0.00^{+0.21}_{-0.00}$ \\
&Others & $0.000^{+0.019}_{-0.000}$ & $0.029 \pm 0.017$ & $0.09 \pm 0.05$ & $0.29 \pm 0.14$ & $0.32 \pm 0.22$ & $0.13 \pm 0.11$ & $0.15 \pm 0.12$ & $0.06 \pm 0.05$ \\
\bottomrule
\toprule
\vtlab{\large SR-S-low $ee$}{7}
&Observed & $8$ & $5$ & $15$ & $19$ & $30$ & $24$ & $32$ & $11$ \\
\cmidrule{2-10}
\cmidrule{2-10}
&Fitted SM events & $6.0 \pm 1.4$ & $5.3 \pm 2.1$ & $11.6 \pm 2.5$ & $22.9 \pm 3.3$ & $31 \pm 4$ & $23.3 \pm 3.0$ & $27.1 \pm 3.1$ & $16.8 \pm 2.1$ \\
\cmidrule{2-10}
&Fake/nonprompt & $2.4 \pm 1.2$ & $2.5 \pm 1.2$ & $4.4 \pm 2.0$ & $9.0 \pm 2.8$ & $5.7 \pm 2.7$ & $1.6^{+1.7}_{-1.6}$ & $3.4 \pm 2.3$ & $1.0 \pm 0.9$ \\
&$t\bar{t}$, single top & $2.3 \pm 0.9$ & $1.4 \pm 0.5$ & $2.2 \pm 0.7$ & $7.6 \pm 1.7$ & $9.6 \pm 1.7$ & $13.3 \pm 3.3$ & $16.4 \pm 3.0$ & $9.8 \pm 1.5$ \\
&Diboson & $1.1 \pm 0.6$ & $0.71 \pm 0.30$ & $2.4 \pm 0.8$ & $3.8 \pm 1.3$ & $6.9 \pm 2.1$ & $7.1 \pm 2.1$ & $6.2 \pm 2.0$ & $5.9 \pm 1.6$ \\
&$Z(\to\tau\tau)$+jets & $0.1^{+0.4}_{-0.1}$ & $0.6^{+2.0}_{-0.6}$ & $2.5 \pm 2.4$ & $0.7^{+1.5}_{-0.7}$ & $6.5 \pm 2.2$ & $0.01^{+0.26}_{-0.01}$ & $0.03^{+0.30}_{-0.03}$ & $0.000^{+0.032}_{-0.000}$ \\
&Others & $0.11 \pm 0.06$ & $0.17 \pm 0.15$ & $0.13 \pm 0.09$ & $1.8 \pm 0.9$ & $2.4 \pm 2.4$ & $1.3 \pm 1.2$ & $1.1 \pm 1.0$ & $0.042 \pm 0.034$ \\
\bottomrule
\toprule
\vtlab{\large SR-S-low $\mu\mu$}{7}
&Observed & $3$ & $6$ & $15$ & $23$ & $37$ & $44$ & $41$ & $28$ \\
\cmidrule{2-10}
\cmidrule{2-10}
&Fitted SM events & $5.2 \pm 1.1$ & $4.3 \pm 1.0$ & $12.8 \pm 1.8$ & $24.8 \pm 2.6$ & $38 \pm 5$ & $37.8 \pm 3.3$ & $36.0 \pm 3.4$ & $28.0 \pm 2.7$ \\
\cmidrule{2-10}
&Fake/nonprompt & $3.2 \pm 1.0$ & $0.9 \pm 0.7$ & $4.6 \pm 1.5$ & $5.6 \pm 1.8$ & $2.8 \pm 1.7$ & $3.8 \pm 2.0$ & $1.5 \pm 1.2$ & $0.00^{+0.10}_{-0.00}$ \\
&$t\bar{t}$, single top & $0.45 \pm 0.18$ & $2.0 \pm 0.5$ & $4.7 \pm 1.0$ & $9.1 \pm 1.6$ & $10.6 \pm 1.9$ & $21.2 \pm 2.9$ & $21.8 \pm 2.6$ & $20.2 \pm 2.7$ \\
&Diboson & $1.4 \pm 0.5$ & $1.02 \pm 0.34$ & $2.2 \pm 0.8$ & $6.7 \pm 1.9$ & $8.8 \pm 2.6$ & $9.4 \pm 2.6$ & $11.2 \pm 3.2$ & $7.5 \pm 2.2$ \\
&$Z(\to\tau\tau)$+jets & $0.09^{+0.16}_{-0.09}$ & $0.1^{+0.5}_{-0.1}$ & $0.9 \pm 0.4$ & $2.1 \pm 1.0$ & $13 \pm 5$ & $1.0 \pm 0.6$ & $0.02^{+0.29}_{-0.02}$ & $0.00^{+0.26}_{-0.00}$ \\
&Others & $0.032 \pm 0.026$ & $0.19 \pm 0.11$ & $0.37 \pm 0.19$ & $1.4 \pm 0.8$ & $2.6 \pm 1.5$ & $2.4 \pm 1.2$ & $1.5 \pm 0.8$ & $0.22 \pm 0.13$ \\
\bottomrule
\bottomrule
\end{tabular}
}
\label{tab:yields:slepton:all}
\end{table}
 
\FloatBarrier
 
\FloatBarrier
 
The CL$_\text{s}$ prescription is used to perform hypothesis tests of specific SUSY models.
The SRs defined using \mll{} are used for electroweakino models, while regions defined using \mtth{} are used for slepton models.
Exclusions at 95\% confidence level are presented in a two-dimensional plane with the horizontal axis given by the mass of the \chitz{}, and
the vertical axis defined by the difference in mass between the \chitz{} or slepton and the \chioz{}.
 
\begin{figure}[tbp]
\centering
\includegraphics[width=0.49\columnwidth]{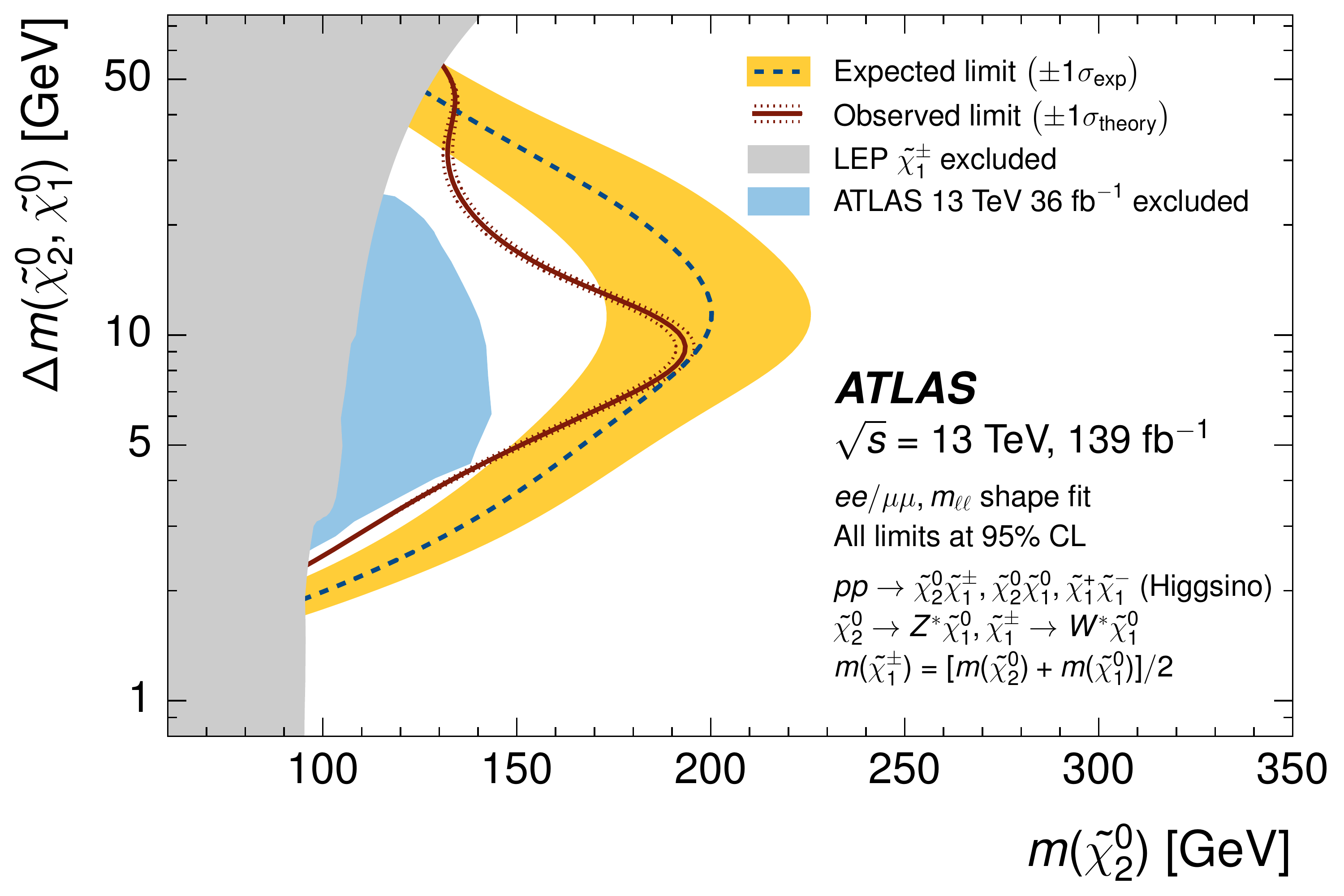}\\
\includegraphics[width=0.49\columnwidth]{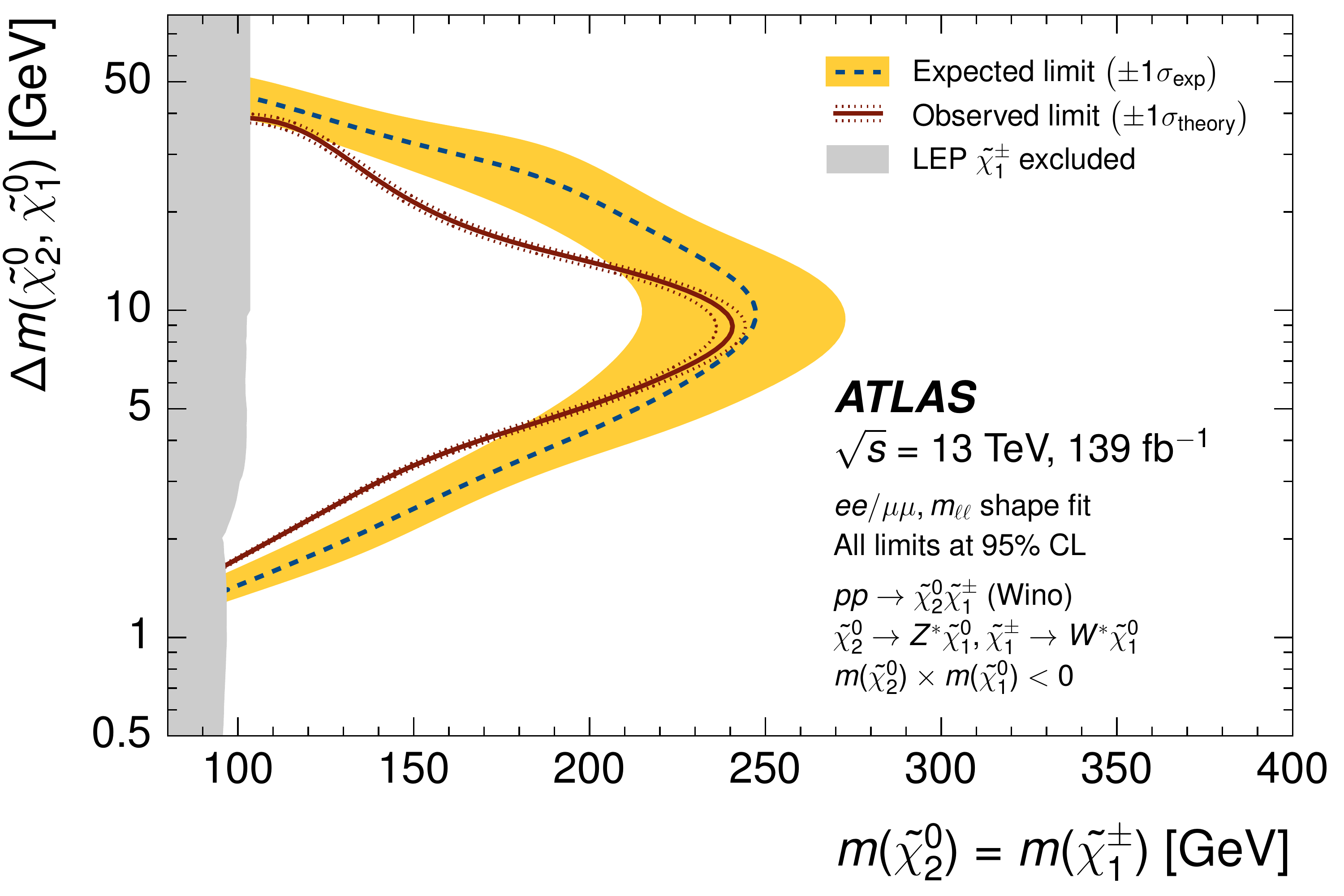}
\includegraphics[width=0.49\columnwidth]{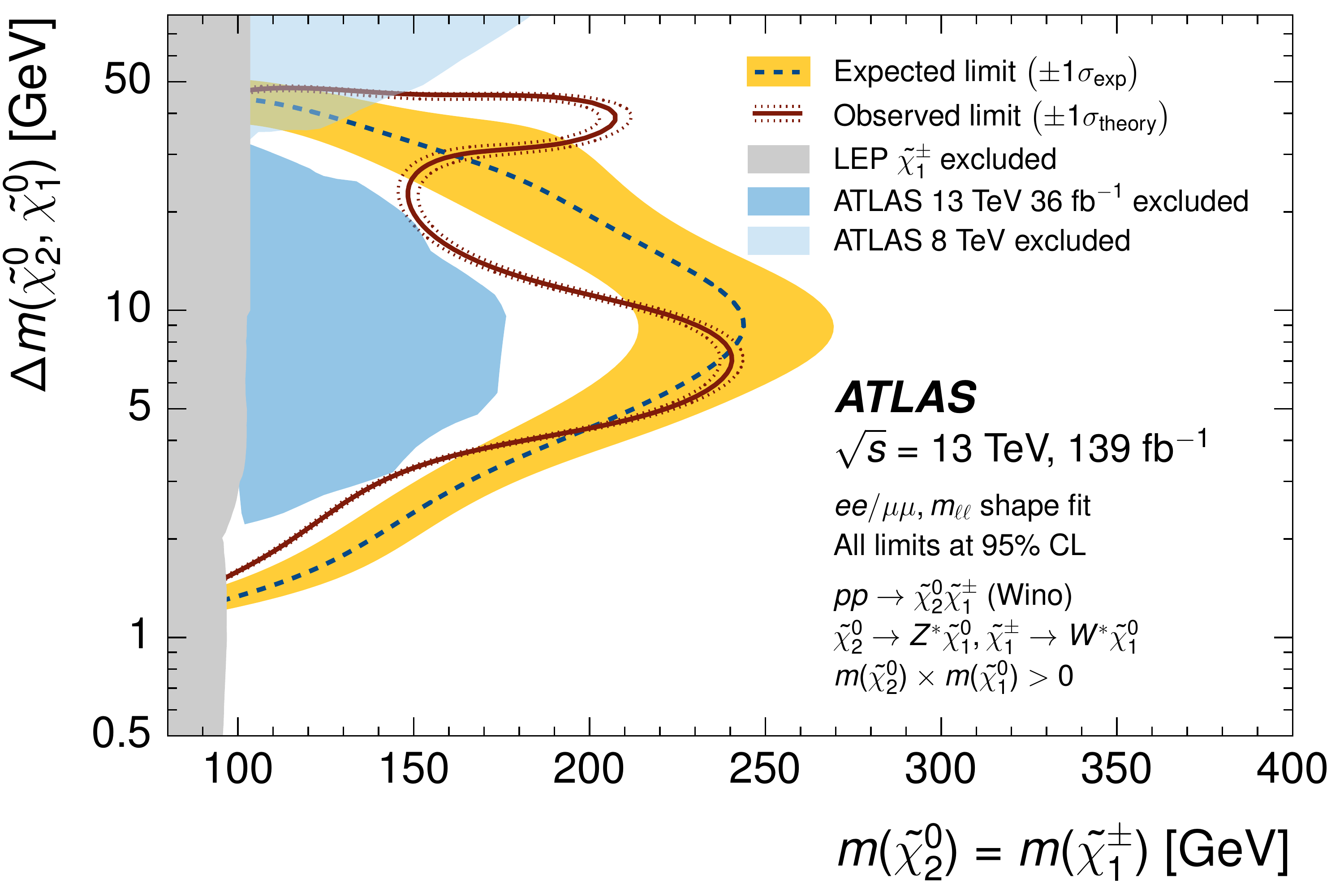}\\
\caption{Expected 95\% CL exclusion sensitivity (blue dashed line), with $\pm1\sigma_{\text{exp}}$ (yellow band) from experimental systematic uncertainties and statistical uncertainties on the data yields, and observed limits (red solid line) with $\pm1\sigma_{\text{theory}}$ (dotted red line) from signal cross-section uncertainties for simplified models of direct higgsino (top) and wino (bottom) production. A fit of signals to the $\mll$ spectrum is used to derive the limit, which is projected into the $\dm(\chitz,\chioz)$ vs.~$m(\chitz)$ plane. For higgsino production, the chargino $\chiopm$ mass is assumed to be halfway between the \chitz{} and \chioz{} masses, while $m(\chitz) = m(\chiopm)$ is assumed for the wino/bino model. Following the discussion in Section~\ref{sec:samples}, the $\mll$ shape in the wino/bino model depends on the relative sign of the $\chioz$ and $\chitz$ mass parameters. The bottom left plot assumes $m(\chioz) \times m(\chitz) < 0$, while $m(\chioz) \times m(\chitz) > 0$ is assumed on the bottom right. The gray regions denote the lower chargino mass limit from LEP~\cite{LEPlimits}. The blue regions indicates the limits from ATLAS searches at 8~\TeV~\cite{SUSY-2013-11,SUSY-2013-12} and at 13~\TeV{} with 36~\ifb~\cite{Aaboud:2017leg}.}
\label{fig:results:ewkinoexclusion}
\end{figure}

Exclusion contours for both wino and higgsino production are shown in Figure~\ref{fig:results:ewkinoexclusion}. Most of the exclusion power originates from the high-\met\ channel, with added sensitivity provided by the \oneleponetrack{} search at small mass splittings and by the low-\met\ channels at higher mass splittings.
The behavior of the observed exclusion contours at large $\dm(\chitz,\chioz)$ is due to the SM background expectation underestimating the data for events with $10<m_{\mu\mu}<20~\GeV$ in SR--E--high, while it overestimates for events with $20<m_{\mu\mu}<40~\GeV$ in the same signal region.
This is also visible in Figure~\ref{fig:results:mllmt2}, which shows the results of a \SRCRBGonlyfit{} assuming that no signal is present.  The lack of allowed contributions from signal processes in the SR-constrained fit reduces the significance of bin-by-bin deviations, while the presence of a signal normalization parameter in the exclusion fit allows for larger deviations from the background constraints.
When assuming wino production with $\mchitz\times\mchioz>0$, electroweakino masses of up to \WinoMaxLimit{} for mass splittings of \WinoMaxLimitDM{}
are excluded.
For electroweakino masses at the edge of LEP exclusions, mass splittings from \WinoMinDM{} to \WinoMaxDM{} are excluded. Assuming higgsino production, \chitz{} masses below \HiggsinoMaxLimit{} are excluded for mass splittings of \HiggsinoMaxLimitDM.  At the LEP bounds on \mchitz, mass splittings from \HiggsinoMinDM{} to \HiggsinoMaxDM{} are excluded. All observed limits are within 2$\sigma$ of the median expected limit.
 
\begin{figure}[tbp]
\centering
\includegraphics[width=0.49\columnwidth]{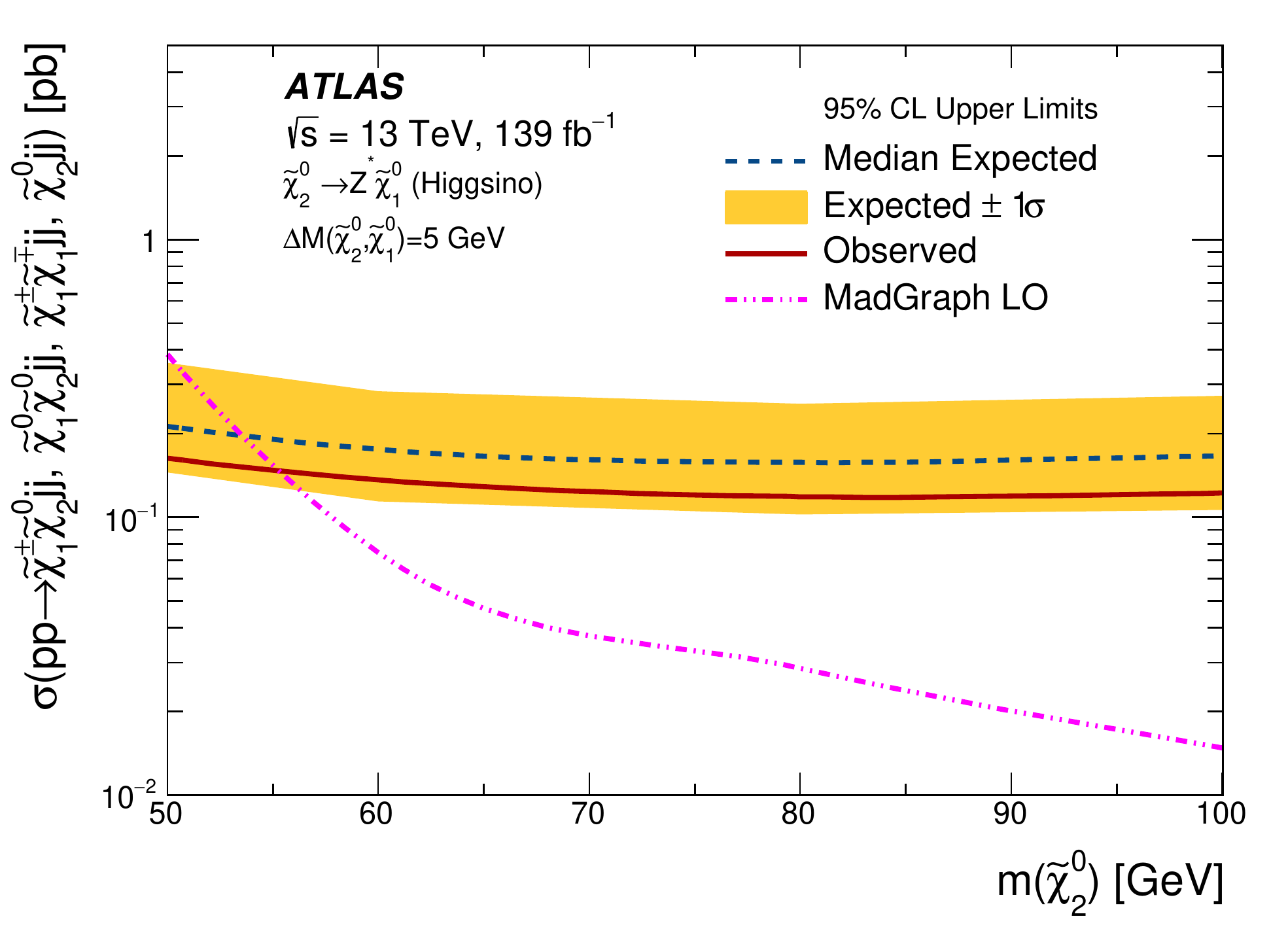}
\includegraphics[width=0.49\columnwidth]{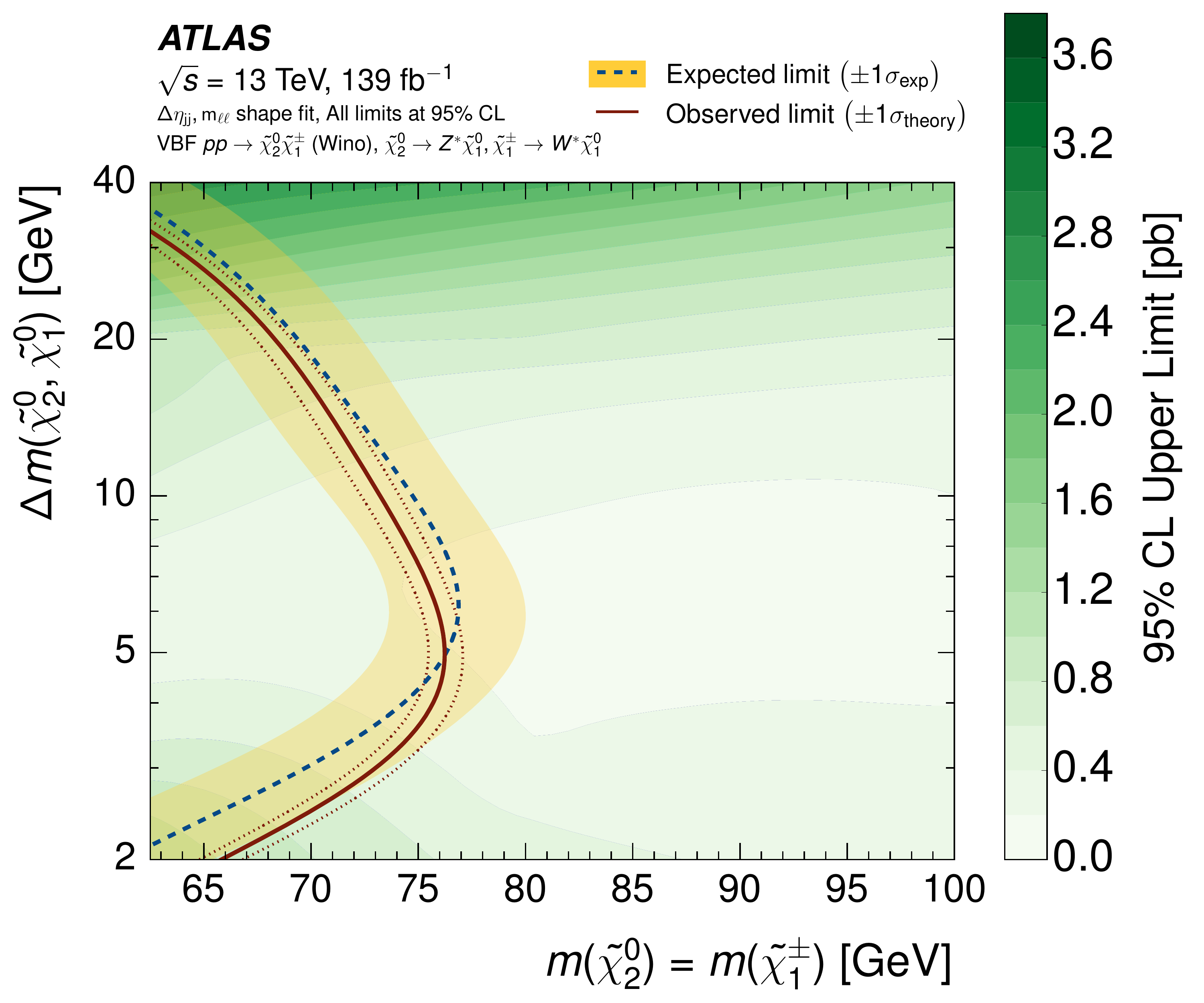}\\
\caption{Expected 95\% CL exclusion sensitivity (blue dashed line) and observed limits (red solid line) for simplified models of higgsino (left) and wino (right) production through VBF. A fit of signals to the $\mll$ spectrum in the VBF signal regions is used to derive the limit. On the left, the limit for higgsinos is shown as a function of $m(\chitz)$ for a mass splitting of $\dm(\chitz,\chioz)=5$~GeV (the chargino $\chiopm$ mass is assumed to be halfway between the \chitz{} and \chioz{} masses). The yellow band indicates $\pm1\sigma_{\text{exp}}$ from experimental systematic uncertainties and statistical uncertainties on the data yields.  On the right the limit for winos is projected into the $\dm(\chitz,\chioz)$ vs.~$m(\chitz)$ plane ($m(\chitz) = m(\chiopm)$ is assumed for the wino/bino model). The red dotted line indicates the $\pm1\sigma_{\text{theory}}$ from signal cross-section uncertainties and the colored map illustrates the 95\% CL upper limits on the cross-section.  The cross-section corresponds to the leading-order prediction from MG5\_aMC@NLO for the process $pp\to\chitz\chiopm jj$ including the parton-level requirements described in Section~\ref{sec:samples}.  The contour lines represent steps of 0.2 pb.}
\label{fig:results:vbfexclusion}
\end{figure}

Models containing electroweakinos produced through VBF processes are constrained using the VBF signal regions.
These constraints are shown in Figure~\ref{fig:results:vbfexclusion}.
The limits on VBF higgsino production cross-sections have a weak dependence on the mass splittings and are shown
assuming $\dm=5~\GeV$. Higgsinos with masses below \VBFHiggsinoMaxLimitDMfive{} are excluded for mass splittings of 5~\GeV.
Assuming VBF production of winos, electroweakino masses up to \VBFWinoMaxLimit{} for mass splittings of \VBFWinoMaxLimitDM{}
are excluded. For wino masses near half of the Higgs boson mass, mass splittings between \VBFWinoMinDM{} and \VBFWinoMaxDM{}
are excluded.
 
\begin{figure}[tbp]
\centering
\includegraphics[width=0.49\columnwidth]{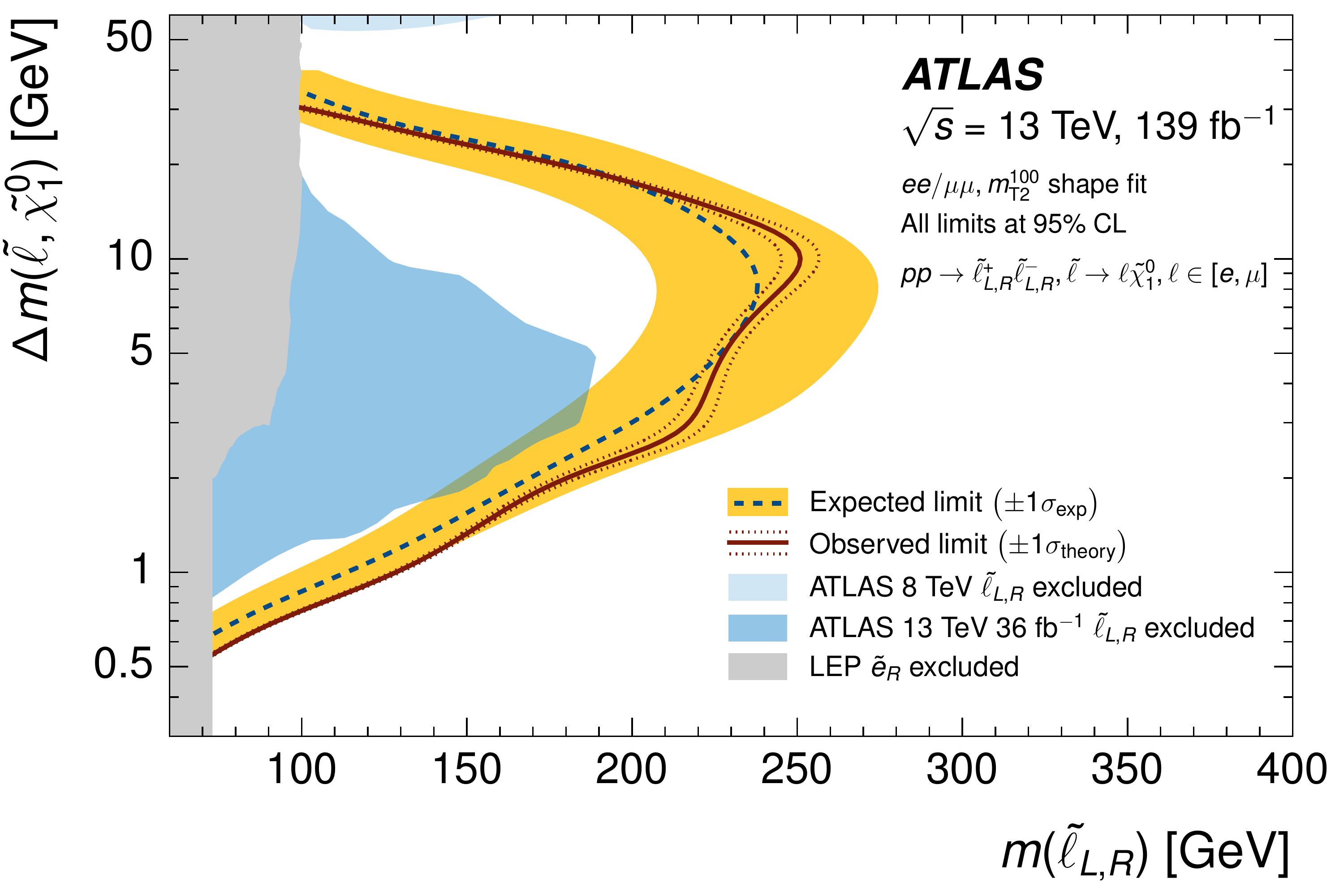}
\includegraphics[width=0.49\columnwidth]{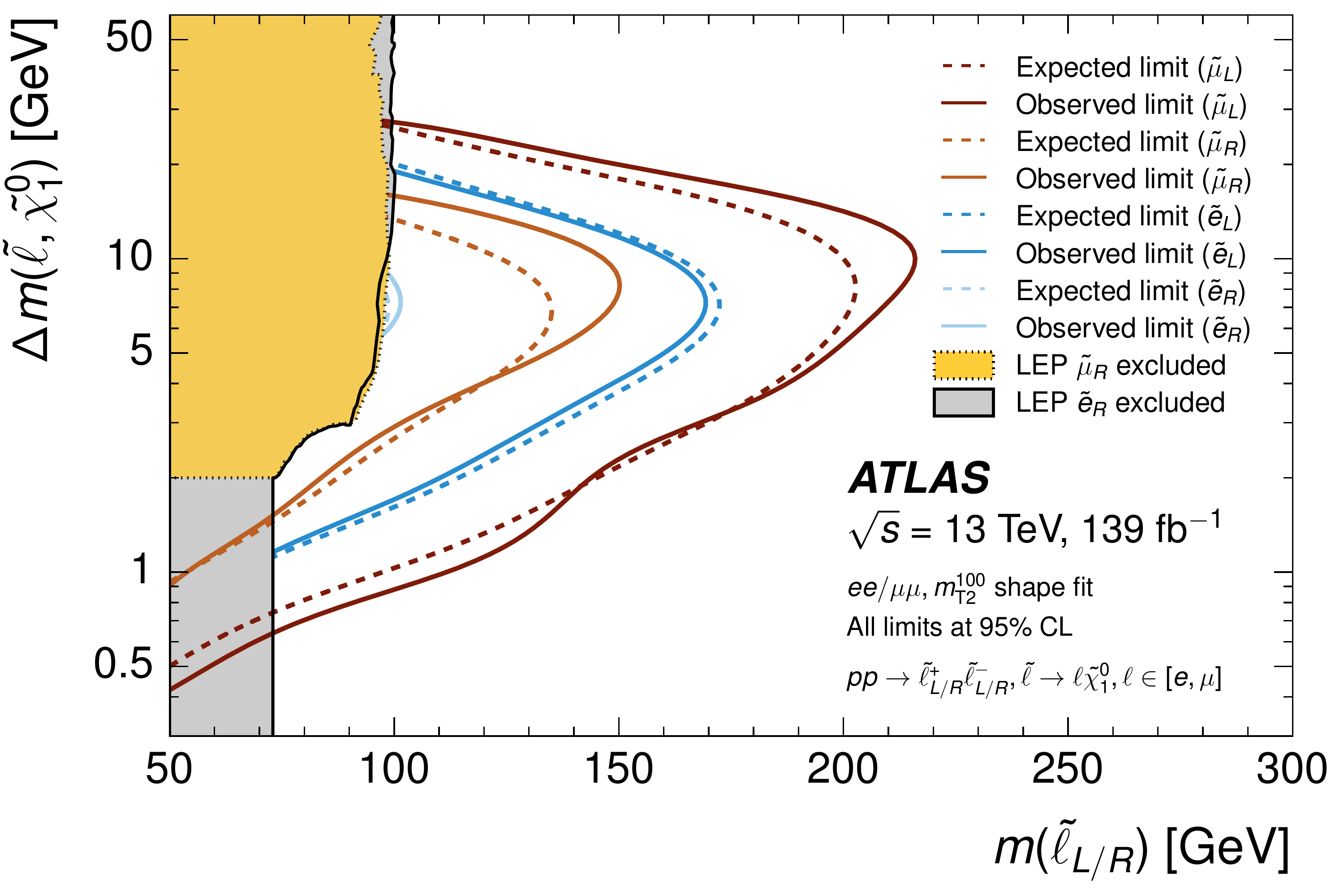}\\
\caption{Expected 95\% CL sensitivity (dashed lines) and observed limits (solid lines) for simplified models of direct slepton production. A fit of slepton signals to the $\mtth$ spectrum is used to derive the limits, which are projected into the $\dm(\slepton,\chioz)$ vs.~$m(\slepton)$ plane. Slepton $\slepton$ refers to the scalar partners of left- and right-handed electrons and muons. The gray region is the $\selectron_{R}$ limit from LEP~\cite{LEPlimits}. On the left, the sleptons are assumed to be fourfold mass degenerate with $m(\selectron_{L}) = m(\selectron_{R}) = m(\smuon_{L}) = m(\smuon_{R})$, the expected sensitivity (blue dashed line) is shown with with $\pm1\sigma_{\text{exp}}$ (yellow band) from experimental systematic uncertainties and statistical uncertainties on the data yields, the observed limit (red solid line) is shown with $\pm1\sigma_{\text{theory}}$ (dotted red line) from signal cross-section uncertainties, and the blue regions are the fourfold mass-degenerate slepton limits from ATLAS Run 1~\cite{SUSY-2013-11} and Run 2~\cite{Aaboud:2017leg}. On the right, no degeneracy is assumed for the masses of the sleptons and the limits are presented separately for $\selectron_{L}$, $\selectron_{R}$, $\smuon_{L}$ and $\smuon_{R}$.}
\label{fig:results:sleptons}
\end{figure}

Exclusion contours for light-flavor sleptons are shown in Figure~\ref{fig:results:sleptons}.
Assuming mass-degenerate selectrons and smuons, slepton masses below \SleptonMaxLimit{} are excluded for
mass splittings of \SleptonMaxLimitDM.
For sleptons with masses just above the LEP limits, mass splittings from \SleptonMinDM{} to \SleptonMaxDM{} are excluded.
Figure~\ref{fig:results:sleptons} also shows results where only the right/left-handed selectron or smuon is produced.
When producing these results, only $ee$ or $\mu\mu$ events in the SRs are considered.
Right-handed sleptons have smaller cross-sections than their left-handed counterparts, due to their different couplings to
the weak gauge fields~\cite{Martin:1997ns}.
Right-handed smuons are excluded up to \SmuonRMaxLimit{} for mass splittings of \SmuonRMaxLimitDM, while left-handed smuons are
excluded up to \SmuonLMaxLimit{} for mass splittings of \SmuonLMaxLimitDM.
Left-handed selectrons are excluded up to \SelectronLMaxLimit{} for mass splittings of \SelectronLMaxLimitDM{}.
Right-handed selectrons are excluded up to \SelectronRMaxLimit{} for mass splittings of \SelectronRMaxLimitDM{}.
 
\FloatBarrier

\FloatBarrier
\section{Conclusion}
Results of searches for electroweak production of supersymmetric particles in models with compressed mass spectra are presented,
using $\sqrt{s}=13~\TeV$ proton--proton collision data corresponding to \FullRunTwoLumi{} collected by the ATLAS experiment at
the CERN Large Hadron Collider.
Events with missing transverse momentum, two same-flavor, opposite-charge, low transverse momentum leptons, and hadronic activity
from initial-state radiation or characteristic of vector-boson fusion production are selected.
The data are found to be consistent with predictions from the Standard Model.
Assuming wino production, constraints at 95\% confidence level are placed on the minimum mass of the \chitz{} at \WinoMaxLimit{}
for a mass splitting of \WinoMaxLimitDM, and extend down to a mass splitting of \WinoMinDM{} at the LEP chargino mass limit
of 92.4~\GeV{}.
For higgsino production, the corresponding lower limits are at \HiggsinoMaxLimit{} at a mass splitting of \HiggsinoMaxLimitDM,
and extend down to a mass splitting of \HiggsinoMinDM{} at the LEP chargino mass limit.
Events consistent with the production of electroweak SUSY states through vector-boson fusion processes are used to constrain
wino/bino and higgsino models while assuming a vanishing $q\bar{q}$ fusion production cross-section.
Light-flavor sleptons are constrained to have masses above \SleptonMaxLimit{} for a mass splitting of \SleptonMaxLimitDM,
with constraints extending down to mass splittings of \SleptonMinDM{} at the LEP slepton limits (73~\GeV).
 

\section*{Acknowledgments}
\setcounter{section}{0}
\renewcommand{\thesection}{\Alph{section}}


We thank CERN for the very successful operation of the LHC, as well as the
support staff from our institutions without whom ATLAS could not be
operated efficiently.
 
We acknowledge the support of ANPCyT, Argentina; YerPhI, Armenia; ARC, Australia; BMWFW and FWF, Austria; ANAS, Azerbaijan; SSTC, Belarus; CNPq and FAPESP, Brazil; NSERC, NRC and CFI, Canada; CERN; CONICYT, Chile; CAS, MOST and NSFC, China; COLCIENCIAS, Colombia; MSMT CR, MPO CR and VSC CR, Czech Republic; DNRF and DNSRC, Denmark; IN2P3-CNRS and CEA-DRF/IRFU, France; SRNSFG, Georgia; BMBF, HGF and MPG, Germany; GSRT, Greece; RGC and Hong Kong SAR, China; ISF and Benoziyo Center, Israel; INFN, Italy; MEXT and JSPS, Japan; CNRST, Morocco; NWO, Netherlands; RCN, Norway; MNiSW and NCN, Poland; FCT, Portugal; MNE/IFA, Romania; MES of Russia and NRC KI, Russia Federation; JINR; MESTD, Serbia; MSSR, Slovakia; ARRS and MIZ\v{S}, Slovenia; DST/NRF, South Africa; MINECO, Spain; SRC and Wallenberg Foundation, Sweden; SERI, SNSF and Cantons of Bern and Geneva, Switzerland; MOST, Taiwan; TAEK, Turkey; STFC, United Kingdom; DOE and NSF, United States of America. In addition, individual groups and members have received support from BCKDF, CANARIE, Compute Canada and CRC, Canada; ERC, ERDF, Horizon 2020, Marie Sk{\l}odowska-Curie Actions and COST, European Union; Investissements d'Avenir Labex, Investissements d'Avenir Idex and ANR, France; DFG and AvH Foundation, Germany; Herakleitos, Thales and Aristeia programmes co-financed by EU-ESF and the Greek NSRF, Greece; BSF-NSF and GIF, Israel; CERCA Programme Generalitat de Catalunya and PROMETEO Programme Generalitat Valenciana, Spain; G\"{o}ran Gustafssons Stiftelse, Sweden; The Royal Society and Leverhulme Trust, United Kingdom.
 
The crucial computing support from all WLCG partners is acknowledged gratefully, in particular from CERN, the ATLAS Tier-1 facilities at TRIUMF (Canada), NDGF (Denmark, Norway, Sweden), CC-IN2P3 (France), KIT/GridKA (Germany), INFN-CNAF (Italy), NL-T1 (Netherlands), PIC (Spain), ASGC (Taiwan), RAL (UK) and BNL (USA), the Tier-2 facilities worldwide and large non-WLCG resource providers. Major contributors of computing resources are listed in Ref.~\cite{ATL-GEN-PUB-2016-002}.
 

\printbibliography

\clearpage 
 
\begin{flushleft}
{\Large The ATLAS Collaboration}

\bigskip

G.~Aad$^\textrm{\scriptsize 102}$,    
B.~Abbott$^\textrm{\scriptsize 129}$,    
D.C.~Abbott$^\textrm{\scriptsize 103}$,    
A.~Abed~Abud$^\textrm{\scriptsize 36}$,    
K.~Abeling$^\textrm{\scriptsize 53}$,    
D.K.~Abhayasinghe$^\textrm{\scriptsize 94}$,    
S.H.~Abidi$^\textrm{\scriptsize 167}$,    
O.S.~AbouZeid$^\textrm{\scriptsize 40}$,    
N.L.~Abraham$^\textrm{\scriptsize 156}$,    
H.~Abramowicz$^\textrm{\scriptsize 161}$
% ...
\bigskip
\\

% ...
$^{36}$CERN, Geneva; Switzerland.\\
% ...
$^{40}$Niels Bohr Institute, University of Copenhagen, Copenhagen; Denmark.\\
% ...
$^{53}$II. Physikalisches Institut, Georg-August-Universit\"{a}t G\"ottingen, G\"ottingen; Germany.\\
% ...
$^{94}$Department of Physics, Royal Holloway University of London, Egham; United Kingdom.\\
% ...
$^{102}$CPPM, Aix-Marseille Universit\'e, CNRS/IN2P3, Marseille; France.\\
$^{103}$Department of Physics, University of Massachusetts, Amherst MA; United States of America.\\
% ...
$^{129}$Homer L. Dodge Department of Physics and Astronomy, University of Oklahoma, Norman OK; United States of America.\\
% ...
$^{156}$Department of Physics and Astronomy, University of Sussex, Brighton; United Kingdom.\\
% ...
$^{161}$Raymond and Beverly Sackler School of Physics and Astronomy, Tel Aviv University, Tel Aviv; Israel.\\
% ...
$^{167}$Department of Physics, University of Toronto, Toronto ON; Canada.\\
% ...

\end{flushleft}


\end{document}